\documentclass[usenatbib,useAMS,twocolumn]{mnras}
\usepackage[dvipdfmx]{graphicx}
\usepackage[dvipdfmx]{color}
\usepackage{amsmath,amssymb}
\usepackage{epstopdf}
\usepackage{grffile}
\usepackage{times}
\usepackage{url}
\usepackage{bm}
\usepackage{xcolor}

\newcommand{\simgt}{\lower.5ex\hbox{$\; \buildrel > \over \sim \;$}}
\newcommand{\simlt}{\lower.5ex\hbox{$\; \buildrel < \over \sim \;$}}

\newcommand{\himpc}{{\hbox {$~h^{-1}$}{\rm ~Mpc}}}
\newcommand{\higpc}{{\hbox {$~h^{-1}$}{\rm ~Gpc}}}
\newcommand{\hmpci}{{\hbox {$~h{\rm ~Mpc}^{-1}$}}}

\newcommand{\vq}{\mathbf{q}}

\newcommand{\vr}{\mathbf{r}}

\newcommand{\vk}{\mathbf{k}}
\newcommand{\vn}{\mathbf{n}}

\newcommand{\fc}{f_{\mathrm{c}}}
\newcommand{\fracs}{f_{\mathrm{s}}}

\newcommand{\tfc}{\widetilde{f}_{\mathrm{c}}}
\newcommand{\tfs}{\widetilde{f}_{\mathrm{s}}}

\newcommand{\pshh}{P^S_{\mathrm{hh}}}

\newcommand{\tpscc}{\widetilde{P}^S_{\mathrm{cc}}}

\newcommand{\prss}{P^R_{\mathrm{ss}}}
\newcommand{\prggt}{P^R_{\mathrm{gg,2}}}

\newcommand{\tprcct}{\widetilde{P}^R_{\mathrm{cc},2}}
\newcommand{\tprcst}{\widetilde{P}^R_{\mathrm{cs},2}}

\newcommand{\be}{\begin{equation}}
\newcommand{\ee}{\end{equation}}
\newcommand{\bey}{\begin{eqnarray}}
\newcommand{\eey}{\end{eqnarray}}
\newcommand{\nn}{\nonumber}

\newcommand{\wt}{\widetilde}

\begin{document}
\title[Halo power spectrum reconstruction]{Reconstruction of halo power spectrum from redshift-space galaxy distribution:
cylinder-grouping method and halo exclusion effect}

\author[T. Okumura et al.]{
\parbox{\textwidth}{
Teppei Okumura,$^{1,2}$\thanks{tokumura@asiaa.sinica.edu.tw} Masahiro Takada,$^{1}$ Surhud More$^{1}$ and Shogo Masaki$^{3}$}
\vspace*{4pt} \\
$^{1}$ Kavli Institute for the Physics and Mathematics of the Universe (WPI), 
The University of Tokyo Institutes for Advanced Study, \\ The University of Tokyo, Kashiwa, Chiba 277-8583, Japan\\
$^{2}$ Institute of Astronomy and Astrophysics, Academia Sinica, P. O. Box 23-141, Taipei 10617, Taiwan \\
$^{3}$ NTT Secure Platform Laboratories, NTT Corporation, Tokyo 180-8585, Japan}

\date{\today} 
\pagerange{\pageref{firstpage}--\pageref{lastpage}} \pubyear{2016}

\maketitle
\label{firstpage}

\begin{abstract}
The peculiar velocity field measured by redshift-space distortions (RSD) in
galaxy surveys provides a unique probe of the growth of large-scale
 structure.
 However, systematic effects arise when
including satellite galaxies in the clustering analysis.  Since
 satellite galaxies tend to reside in massive halos with a
  greater halo bias, the inclusion boosts the clustering power. In
 addition, virial motions of the satellite galaxies cause a
 significant suppression of the clustering power due to nonlinear RSD
 effects.  We develop a novel method to recover the redshift-space power
 spectrum of halos from the observed galaxy distribution by minimizing
 the contamination of satellite galaxies.  The cylinder grouping method
 (CGM) we study effectively excludes satellite galaxies from a galaxy
 sample. However, we find that this technique produces apparent
 anisotropies in the reconstructed halo distribution over all the scales
 which mimic RSD. On small scales, the apparent anisotropic clustering
 is caused by exclusion of halos within the anisotropic cylinder used by
 the CGM. On large scales, the misidentification of different halos in
 the large-scale structures, aligned along the line-of-sight, 
into the same CGM group
causes the apparent anisotropic clustering via their
 cross-correlation with the CGM halos.  We
 construct an empirical model for the CGM halo power spectrum, which
 includes correction terms derived using the CGM window function at
 small scales as well as the linear matter power spectrum multiplied by
 a simple anisotropic function at large scales.  We apply this model to
 a mock galaxy catalog at $z=0.5$, designed to resemble
 SDSS-III BOSS CMASS galaxies, and find that our model
 can predict both the monopole and quadrupole power spectra of the host
 halos up to $k < 0.5\hmpci$ to within 5\%.
\end{abstract}


\section{Introduction} \label{sec:intro}

Observation of large-scale structure of the Universe through imaging
and spectroscopic surveys of galaxies is a powerful tool to probe the
origin of the accelerated expansion of the Universe, i.e., dark energy
or its alternative such as modified gravity theories \citep[e.g.,
see][for a review]{Weinberg:2013}.  Imaging surveys enable us to probe
the growth of cosmic structure through observations of the
gravitational lensing effect.  On the other hand, spectroscopic
surveys provide information on the expansion rate through baryon
acoustic oscillations (BAO).  Due to the fact that the distance to
each galaxy is measured by redshift, spectroscopic surveys also enable
us to probe the growth rate of the structure by measuring
redshift-space distortions (RSD) \citep{Kaiser:1987,Hamilton:1998}.
Combining imaging and spectroscopic surveys
is a powerful tool to study the origin of the
cosmic acceleration.

While the BAO scale can be precisely predicted, 
the constraints from RSD are known to be affected by
various systematic effects
\citep[e.g.,][]{Tinker:2006,Okumura:2011,Jennings:2011}. One of the most
serious systematic effects arises from
the nontrivial relationship between dark matter halos and galaxies,
especially due to the presence of ``satellite'' galaxies in the clustering
analysis.  Since satellite galaxies tend to reside in massive halos,
which are more biased tracers and rare objects, the massive halos are
upweighted in the clustering analysis, boosting the clustering amplitude
compared to the bias of typical halos hosting other majority of the
galaxies.  In addition, virial motions of satellite galaxies inside such
massive halos, known as the Finger-of-God (FoG) effect
\citep{Jackson:1972}, cause a significant modification of the
redshift-space clustering power at small scales. Thus, a direct approach
of comparing the measured galaxy power spectrum with theory requires a
detailed knowledge of how galaxies populate their dark matter halos as
a function of redshift, and perhaps their surrounding environments
\citep{Reid:2014}.

One way to connect theory and observations is an analytical approach
based on cosmological perturbation theory \citep{Bernardeau:2002} and
including an empirical FoG damping function
\citep{Peacock:1994,Scoccimarro:2004,Okumura:2012,Zheng:2016}.  While
halos can be straightforwardly identified in $N$-body simulations
\citep{Davis:1985,Behroozi:2013}, it is not straightforward to include
halo bias prescriptions into the perturbation theory in a
self-consistent manner, which generally requires  many parameters to
model halo bias at each order \citep[e.g.,][]{Saito:2014}.  Therefore,
\citet{Okumura:2015} proposed a hybrid method that combines
perturbation theory or $N$-body simulations with a halo model to both
model the galaxy clustering at large- and small-scales simultaneously
\citep{White:2001,Seljak:2001,Cooray:2002,Mohammed:2014,Seljak:2015}.

Another approach could be to modify the measurement method to
correspond closely to theory. There are several clustering analysis
methods in the literature which compress or filter out the impact of
high-density galaxy regions \citep{Tegmark:2006,Simpson:2013}.
\citet{Reid:2009} used the redshift space friends-of-friends (FOF)
percolation algorithm to group galaxies and demonstrated that the
groups can be used to reconstruct the halo power spectrum from the
distribution of luminous red galaxies (LRGs) in the
Sloan Digital Sky Survey (SDSS) \citep{Reid:2010}.  If the halo power spectrum is
recovered, we might not need to model the clustering signal from
satellite galaxies and can directly compare the measured power
spectrum with the theoretical predictions of halo clustering, based
either on perturbation theory and/or $N$-body simulations
\citep[e.g.,][]{Matsubara:2008a,Nishimichi:2011,Sato:2011,Reid:2011,Okumura:2012b,Nishizawa:2013,Vlah:2013}.
However, the accuracy of such a halo reconstruction method has not
been fully assessed, and is known to strongly depend on the cylinder
shape (or the linking length in redshift space) and the properties of
the galaxy sample: the number density, the host halo mass and the
satellite fraction. For example, since CMASS galaxy sample of the SDSS
BOSS survey \citep{Eisenstein:2011} has a higher number density than
that of LRGs by about factor of 3, a naive implementation of the FOF
algorithm leads to a much deeper percolation of groups, and causes a
large number of separate smaller halos to be grouped as one.

Hence the purpose of this paper is to revisit the halo reconstruction
method proposed by \citet{Reid:2009} and refine it to enable a more
accurate recovery of the halo power spectrum. Instead of the FOF
method, we adopt a similar but simpler cylinder grouping (CG) method.
We then include terms to correct
for the apparent anisotropic clustering caused by the anisotropic
shape of our cylinder as well as by the large-scale clustering of
halos misidentified by the CG method, extending the method in
\citep{van-den-Bosch:2013,Baldauf:2013}. We will carefully investigate
the accuracy and performance of the method using mock galaxy catalogs
that are constructed from $N$-body simulations.

The structure of this paper is as follows.  In
Section~\ref{sec:motivation} we first motivate the present work.  In
Section~\ref{sec:theory} we present our theoretical formalism to
reconstruct the power spectrum of halos from the observed distribution
of galaxies.  Section~\ref{sec:sim} describes $N$-body simulations,
the catalogs of halos and subhalos, and the mock galaxy catalogs which
we use to test our method.  In Section~\ref{sec:analysis} we show the
numerical results of the reconstructed halo power spectrum in redshift
space. We present a brief summary of our conclusions in
Section~\ref{sec:conclusion}.
In Appendix~\ref{sec:changing_pi}-\ref{sec:velocity_bias} we test how
the accuracy of our formalism changes if properties of target galaxy
sample and the cylinder size for the CG method are changed.  Unless
stated, throughout this paper we adopt the flat $\Lambda$CDM model
with a matter density $\Omega_{\rm m0} = 0.265$, baryon density
$\Omega_{\rm b0} = 0.0448$, the spectral index $n_s = 0.963$, the rms
density fluctuation in a sphere of radius $8~h^{-1}{\rm Mpc}$,
$\sigma_8=0.80$, and the Hubble parameter $h = 0.71$, which is
consistent with the WMAP cosmology \citep{Komatsu:2009}. 


\section{Motivation}\label{sec:motivation}

\begin{figure*}
\begin{center}
\includegraphics[width=85mm, bb=0 0 610 570]{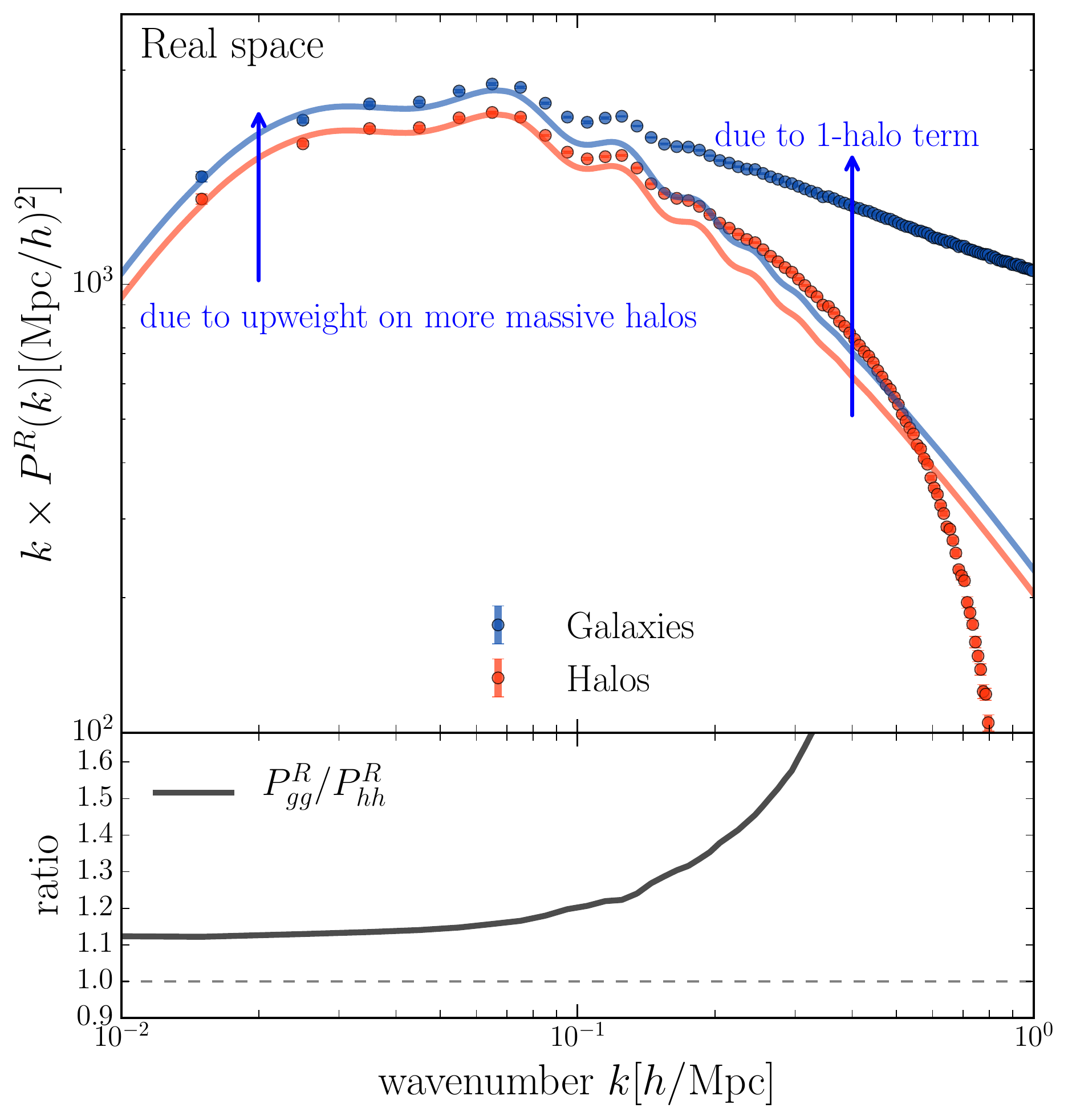}
 \includegraphics[width=85mm,bb=0 0 610 570]{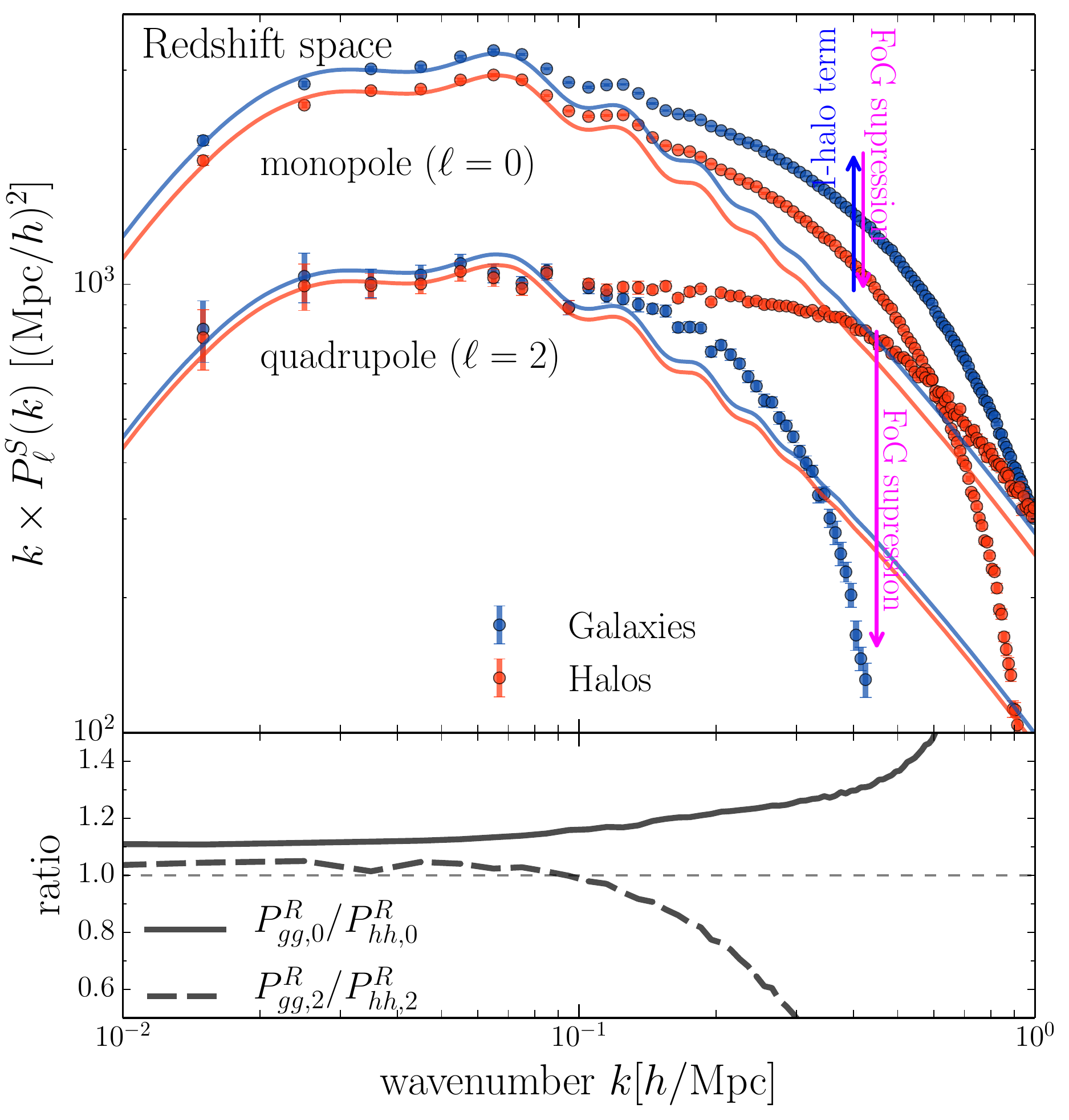}
 \caption{({\it Top}) An illustration of how including satellite galaxies in the
 clustering analysis affects the real-space (left) and redshift-space (right) power spectrum,
 which are computed using the mock catalogs of CMASS-like galaxies and
 their host halos at $z=0.5$. Errorbars are the standard error of the mean.
 The satellite galaxies alter the power
 spectrum amplitudes over all the scales, compared to the power spectrum of their host
 halos, due to the upweights on more massive halos with greater biases, 
 the 1-halo term clustering contribution and the
 Finger-of-God effect in redshift space. These effects need to be
 properly taken into account in order to extract the information on the
 underlying matter power spectrum from the measured galaxy power
 spectrum if including satellite galaxies in the analysis. 
({\it Bottom}) Ratios of the measured spectra of galaxies to those of their host halos. 
The deviation from unity is again caused by the satellite galaxies. 
 \label{fig:pkr_vs_pks}}
\end{center}
\end{figure*}
We begin by highlighting the impact of satellite galaxies on the
measured power spectrum in real and redshift space. In
Fig.~\ref{fig:pkr_vs_pks}, the solid curves correspond to the linear
power spectra in real space and redshift space computed using the
publicly available, CAMB code \citep{Lewis:2000}, where the amplitudes
are normalized so as to reproduce
the spectrum amplitude for halos or galaxies
at large scales $k=[0.01,0.05]\hmpci$. These power spectra were
calculated using 32 realizations of mock catalogs of CMASS-like
galaxies at $z=0.5$, which we will describe in Section~\ref{sec:sim}
below in detail. For the spectrum of host halos, we used halos in the
mock catalog that host at least one galaxy. Note that the galaxy power
spectra are from the exactly same catalog of host halos as used in the
halo power spectra, and the fraction of satellite galaxies is about
10\% in each realization. The satellite galaxies, albeit a small
fraction, affect the power spectrum over all scales as we will
describe below.

First, let us consider the real-space power spectrum, as illustrated
in the left panel of Fig.~\ref{fig:pkr_vs_pks}. On large length
scales, i.e., for small $k$, the galaxy power spectrum generally has a
greater amplitude than the halo power spectrum. This can be
interpreted as follows. Satellite galaxies tend to reside in more
massive halos, and such massive halos generally have a larger halo
bias with respect to the underlying matter distribution.  Thus, when
satellite galaxies are included in the power spectrum estimation, such
massive halos are counted multiple times, causing a boost in the power
spectrum amplitude.

On the other hand, the galaxy spectrum on small scales, i.e., $k\simgt
0.1 h/$Mpc, has a contribution arising from correlations between two
galaxies, either central-satellite galaxies or satellite-satellite
galaxies in the same halo, the so-called 1-halo term
\citep{Seljak:2000,Cooray:2002}.  The correlations are preferentially
from massive halos that can host multiple galaxies inside. The 1-halo
term causes a boost in the power spectrum amplitude than the halo
power spectrum. The 1-halo term has a complex scale dependence due
to various effects such as nonlinear clustering, the exclusion effect
and the nonlinear bias effects \citep{Peebles:1980, Smith:2007,
van-den-Bosch:2013, Nishizawa:2013, Baldauf:2013, Mohammed:2014,
Okumura:2015, Baldauf:2016}. The amount of the amplitude boost is
sensitive to the details of the 1-halo term (the satellite fraction,
the radial profile of satellite galaxies in the host halos, and the
mass distribution of halos hosting multiple galaxies). On the other
hand, the halo power spectrum arises from correlations between
different halos, the 2-halo term alone.

However, redshift space distortions (RSD) due to peculiar motions of
galaxies modify the shape of galaxy power spectrum in a complicated way,
as illustrated in the right panel of Fig.~\ref{fig:pkr_vs_pks}, where we
show the monopole $(P_0^S)$ and quadrupole $(P_2^S)$ power
spectra\footnote{The superscript $S$ denotes a quantity measured in
redshift space.} (see equation~\ref{eq:multipoles} below for the
definitions). The relative difference between the galaxy and halo
power spectra at small $k$ is similar to that in real space, although
the difference in the quadrupole power spectrum is smaller due to the
fact that it depends linearly on the galaxy bias (the monopole scales
with bias as $P_0^S\propto b^2$ in the linear regime, while the
quadrupole spectrum scales as $P_2^S\propto b$). At $k\simgt
0.1~h/$Mpc, the virial motion of galaxies inside their host halos
yields a significant suppression in the power for both the monopole
and quadrupole power spectra, which is known as the Finger-of-God
(FoG) effect \citep{Jackson:1972}. For the monopole spectrum, the FoG
smearing effect somehow compensates the boost due to the 1-halo term
contribution, making the overall shape closer to the halo power
spectrum. This is, however, just a coincidence and the net power
depends severely on the details of satellite galaxies in their host
halos. For the quadrupole power spectrum, the FoG suppression leads to
a much smaller amplitude in the power at $k\simgt 0.1~h$/Mpc than that
of the halo power spectrum.

Thus the difference between the power spectra of galaxies and their host
halos at $k\simgt 0.1~h$/Mpc seems to arise mainly from the satellite
galaxies, which would prevent from recovering the underlying halo power
spectrum. The purpose of this paper is to develop a method to
reconstruct the halo power spectrum from the observed distribution of
galaxies based on the CG halo reconstruction method.


\section{Formalism} \label{sec:theory}

We decompose galaxies into two populations, central and satellite
galaxies based upon their location in their host halos.  Accordingly,
as developed in \cite{Okumura:2015} \citep[also
see][]{Hikage:2012, van-den-Bosch:2013, Hikage:2013a}, the galaxy
power spectrum in redshift space can be decomposed into different
contributions as
\bey
P_{\rm gg}^S(\vk)=f_{\mathrm{c}}^2P^S_{\mathrm{cc}}(\vk) + 2\fc \fracs
P^S_{\mathrm{cs}}(\vk)
+ \fracs^2P^S_{\mathrm{ss}}(\vk) , \label{eq:pgg_decom}
\eey
where $P^S_{\mathrm{cc}}$, $P^S_{\mathrm{ss}}$ and $P^S_{\mathrm{cs}}$
are the auto power spectra of central and satellite galaxies, and
their cross spectrum, respectively, $\fc$ and $\fracs$ are the
fractions of central and satellite galaxies to the total number of
galaxies, respectively and satisfy $\fc+\fracs=1$, while $\vk$ is the
wavevector. We define the direction cosine of the angle between $\vk$
and the line-of-sight direction as $\mu\equiv
\hat{\vn}\cdot\hat{\vk}$, where $\hat{\vn}$ is the unit vector along
the line of sight. Note that, throughout this paper, we assume a
distant observer approximation. The expression for the corresponding
real-space power spectrum, denoted by the superscript $R$, can be
obtained by taking the transverse modes; $P^R_{\mathrm{gg}}(k) =
P^S_{\mathrm{gg}}(k,\mu=0)$.  Considering the multipole expansion in
terms of the Legendre polynomials ${\cal L}_\ell(\mu)$ as
\be P_\ell^S(k) = \frac{2l+1}{2}\int^1_{-1} d\mu P^S(\vk) {\cal L}_\ell(\mu), \label{eq:multipoles} \ee
we have
 \be
P^S_{\mathrm{gg},\ell}(k)=\fc^2P^S_{\mathrm{cc},\ell}(k) + 2\fc\fracs
P^S_{\mathrm{cs},\ell}(k) + \fracs^2P^S_{\mathrm{ss},\ell}(k) .
\label{eq:ps_break}
\ee
Assuming statistical isotropy in the plane of the sky, the odd-number
multipole spectra vanish. On the other hand, the statistical isotropy
of the real-space power spectrum implies that only the monopole power
spectrum is non-vanishing, i.e., $P^R_{\mathrm{gg},\ell}(k)=0$ with
$\ell\ge 2$.

In linear theory, the power spectrum for biased objects ${\rm x}(={\rm
g,c, etc.})$  can be described as \citep{Kaiser:1987}
\be
P_{\rm xx,{\rm lin}}^S(\vk) 
= (b_{\rm x}+f_{\rm g}\mu^2)^2 P^R_{\rm lin}(k), \label{eq:kaiser}
\ee
where 
$b_{\rm x}$ is the linear bias parameter, $f_{\rm g}$ is the linear
growth rate\footnote{ The linear growth rate $f_{\rm g}$ should not be
confused with the fraction of an given sample ${\rm x}$, $f_{\rm x}$,
and should be distinguished.} defined as $f_{\rm g} = d \ln D / d \ln
a$ and $P^R_{\rm lin}$ is the linear power spectrum of total matter in
real space. By considering the multipole expansion
(equation~\ref{eq:multipoles}), only the lowest three moments
(monopole, quadrupole and hexadecapole) have non-zero values in linear
theory, and they can be written as \citep{Hamilton:1992,Cole:1994}:
\bey
P_{\rm xx,0}^S(k) &=& \left(b_{\rm x}^2 + \frac{2}{3}b_{\rm x}f_{\rm g}
+ \frac{1}{5}f_{\rm g}^2\right) P^R_{\rm lin}(k), \label{eq:kaiser0}\\
P_{\rm xx,2}^S(k) &=& \left(\frac{4}{3}b_{\rm x}f_{\rm g} +
\frac{4}{7}f_{\rm g}^2\right) P^R_{\rm lin}(k),  \label{eq:kaiser2} \\
P_{\rm xx,4}^S(k) &=& \frac{8}{35}f_{\rm g}^2 P^R_{\rm lin}(k).  \label{eq:kaiser4}
\eey


\subsection{Cylinder grouping (CG) method for halo reconstruction}
\label{sec:cgm}

If central galaxies reside at the center and have the same velocity of the host
halos, the power spectrum of central galaxies is equivalent to the halo power
spectrum of host halos, $\pshh$:
\begin{equation}
 P_{\mathrm{hh}}^S(\vk)=P_{\mathrm{cc}}^S(\vk).
  \label{eq:phh=pcc}
\end{equation}
Thus, in such an ideal case, we can reconstruct the halo power spectrum
from the catalog of central galaxies, which is a goal of this
paper. However, in practice it is very difficult to identify central galaxies
from observations. Therefore we need to consider the effects of possible
contamination on the halo reconstruction.

In this paper, we use the ``cylinder grouping method'' (hereafter, CG or CGM) to
minimize the satellite galaxy contamination. We implement this method as
follows:
\begin{itemize}
 \item[(i)] Make a ranked list of galaxies in  descending order of
	    their luminosities or stellar masses.
 \item[(ii)] Mark the top most galaxy in the ranked list as central and remove
        it from the list. Find any other galaxies within a cylinder around the
        central galaxy, defined by a base radius $\Delta r_\perp$, and height
        $\Delta r_\parallel$,  respectively perpendicular and parallel to the line of sight, 
        and mark them as satellite galaxies and remove them from the
        list as well.
 \item[(iii)] Repeat step (ii) until the entire list is exhausted.
\end{itemize}
We hereafter refer to the galaxies marked as centrals as ``CGM central
galaxies'' or ``CGM halos'', while we refer to the galaxies removed by the
grouping method as ``CGM satellite galaxies''. We denote CGM central galaxies
that do not have any satellite galaxy in their cylindrical region as ``CGM
isolated systems''. On the other hand, we denote CGM central galaxies that have
satellite galaxy(ies) in the cylinder region, as ``CGM multiple systems''.

Unless otherwise stated, we assume $\Delta r_\perp=1.5~\himpc$ and $\Delta
r_\parallel=15~\himpc$ as our fiducial parameters for the size of the cylinder.
A cylindrical shape is appropriate to group galaxies that are displaced along
the line-of-sight due to RSD. The peculiar velocities of satellite galaxies
within massive halos can displace them by up to $\sim \mbox{a few }10\himpc$. Hence
$\Delta r_\parallel$ needs to be sufficiently large to cover a RSD effect due to
virial motions of galaxies. in a massive halo.

Our method is based on a similar concept of the redshift space
friends-of-friends method \citep{Berlind:2006}, also used in the literature for
the halo reconstruction \citep{Reid:2009,Hikage:2013a}, but these methods are
different in some of their technical and implementation aspects. An advantage of
the cylinder grouping method compared to the redshift space friends-of-friends
method is that the different CGM cylinders are not linked or connected. All the
CGM central galaxies have exactly the same cylinder-shape territory region; no
pair of the CGM central galaxies are closer in the separation  than the cylinder
size. 

After removing CGM satellite galaxies, we {\it naively} expect that the
power spectrum of the host halos can be reconstructed from the measured power
spectrum of the central galaxies: 
\begin{equation}
 \pshh(\vk)\simeq \tpscc(\vk),
  \label{eq:pshh_naive}
\end{equation}
where tilde denotes a quantity for objects identified by the CG method, thus
$\tpscc$ is the power spectrum of the CGM central galaxies. Note that, it is
plausible that some of the galaxies labeled centrals are in reality
off-centered, so the power spectrum of central galaxies can differ from the halo
power spectrum \citep{Hikage:2012,Hikage:2013a}. We will discuss this issue
later in Appendix \ref{sec:off-centering}. 

Any method that groups observed galaxies together is subject to the following
problems:
\begin{itemize}
 \item[(i)] {\it Group merging}: The CG method could mistakenly group together
       different halos into the same CGM halo. 
 \item[(ii)] {\it Group fragmentation}: The CG method could mistakenly fragment a
     single halo into different CGM halos. 
 \item[(iii)] {\it Central/satellite misidentification}: The CG method could
     mistake a central galaxy to be a satellite, and vice versa.
\end{itemize}
Of these, the first problem is quite significant, as the merging of groups, can
preferentially happen in a high density region, or along a filament fortuitously
oriented along the line-of-sight direction. This can result in an apparent {\it
large scale anisotropy}, because correlations of the
misidentified halos with other majority of CGM halos are removed from the clustering
analysis.
Furthermore the grouping method is not able to identify
halos separated by the cylinder size, and thus suffer from a {\it cylindrical
exclusion effect}. The anisotropic shape of the cylinder thus can introduce 
anisotropic effects in the measured power spectrum of the CGM halos. These
anisotropic effects need to be corrected for before the measured power spectra
of galaxies can be used to constrain cosmological parameters.

When central and satellite galaxies are identified based on the CG method, 
the decomposition equation of the power spectrum (equation \ref{eq:pgg_decom}) 
needs to be modified, as 
\bey
P_{\rm gg}^S(\vk)=\wt{f}_{\mathrm{c}}^2\wt{P}^S_{\mathrm{cc}}(\vk) 
+ 2\wt{f}_{\mathrm{c}}\wt{f}_{\mathrm{s}} \wt{P}^S_{\mathrm{cs}}(\vk)
+ \wt{f}_{\mathrm{s}}^2\wt{P}^S_{\mathrm{ss}}(\vk), \label{eq:pgg_decom_cgm}
\eey
where $\wt{f}_{\mathrm{c}}+\wt{f}_{\mathrm{s}}=1$. 
Note that this equation holds, independently of whether or not the CG method 
correctly identifies centrals and satellites. 


\subsection{Connecting the power spectrum of CGM halos to the halo power
  spectrum}\label{sec:exclusion}

Similarly to equation (\ref{eq:pgg_decom_cgm}), one can write down the 
decomposition of the halo power spectrum based on the CG method, as 
\bey
P_{\rm hh}^S(\vk)=
\wt{g}_{\mathrm{c}}^2\wt{P}^S_{\mathrm{cc}}(\vk) 
+ 2\wt{g}_{\mathrm{c}}\wt{g}_{\mathrm{\bar{c}}} \wt{P}^S_{\mathrm{c\bar{c}}}(\vk)
+ \wt{g}_{\mathrm{\bar{c}}}^2\wt{P}^S_{\mathrm{\bar{c}\bar{c}}}(\vk) \nn \\
-2\wt{g}_{\mathrm{c}}\wt{g}_{\mathrm{\bar{s}}} \wt{P}^S_{\mathrm{c\bar{s}}}(\vk)
-2\wt{g}_{\mathrm{\bar{c}}}\wt{g}_{\mathrm{\bar{s}}} \wt{P}^S_{\mathrm{\bar{c}\bar{s}}}(\vk)
+ \wt{g}_{\mathrm{\bar{s}}}^2\wt{P}^S_{\mathrm{\bar{s}\bar{s}}}(\vk)
 , \label{eq:phh_decom_cgm}
\eey
where the subscript $\mathrm{\bar{c}}$ denotes galaxies which are centrals but grouped together with 
another central(s) living in the more massive halo(s) thus labeled as satellites, while
galaxies with $\mathrm{\bar{s}}$ are satellites but failed to be identified by the CGM 
thus labeled as centrals.  
$\wt{g}_\mathrm{x}$ is similar to $\wt{f}_\mathrm{x}$ but the fraction to the total number of halos,
$\wt{g}_\mathrm{x}=\wt{N}_\mathrm{x}/N_\mathrm{h}=\wt{N}_\mathrm{x}/N_\mathrm{c}$,
thus $\wt{g}_\mathrm{c}+\wt{g}_\mathrm{\bar{c}}-\wt{g}_\mathrm{\bar{s}}=1$. 
Note again that, with tilde, the subscript c denotes galaxies identified by the CGM as centrals, 
thus some of them are satellites in practice. 
Because $\wt{g}_\mathrm{c}\simeq 1 \gg \wt{g}_\mathrm{\bar{c}} > \wt{g}_\mathrm{\bar{s}}$
as we will see in Section \ref{sec:cgm_rec},  
by keeping major terms, we have, 
\bey
P_{\rm hh}^S(\vk)\simeq 
\wt{P}^S_{\mathrm{cc}}(\vk) 
+ 2 \left\{ \wt{g}_\mathrm{\bar{c}} \wt{P}^{S1h}_{\mathrm{c\bar{c}}}(\vk) -  \wt{g}_\mathrm{\bar{s}} \wt{P}^{S1h}_{\mathrm{c\bar{s}}}(\vk) \right\} \nn \\
+ 2 \left\{ \wt{g}_\mathrm{\bar{c}} \wt{P}^{S2h}_{\mathrm{c\bar{c}}}(\vk) - \wt{g}_\mathrm{\bar{s}} \wt{P}^{S2h}_{\mathrm{c\bar{s}}}(\vk) \right\},
 \label{eq:phh_decom_cgm_approx}
\eey 
where we explicitly write the contributions of $\wt{P}^S_{\mathrm{c\bar{c}}}$ at 1-halo and 2-halo regimes with the superscripts $1h$ and $2h$, respectively.
The purpose of this paper is to reconstruct the halo power spectrum in redshift space, namely to model the total contributions from the terms of $\wt{P}_{\mathrm{c\bar{c}}}$ and $\wt{P}_{\mathrm{c\bar{s}}}$.

Instead of the naive reconstruction (equation~\ref{eq:pshh_naive}), 
here we propose the following model to relate the power spectrum of CGM
halos to that of underlying halos:
\bey
P_{\rm hh}^S(\vk)\simeq \wt{P}^S_{\rm cc}(\vk) +
\left\{ W(\vk)+ \left[ W(\vk)*\wt{P}^R_{\rm cc,0}(k)\right](\vk)
\right\} \nn \\
+ \Delta P^{S}(\vk). 
\label{eq:baldauf_mod}
\eey
In the following we explain each term on the r.h.s. and 
how they intend to correct for the systematic effects described in the 
previous subsection.

The terms in the curly brackets arise from the Fourier transform of the 
correlation function with the exclusion effect \citep{van-den-Bosch:2013,Baldauf:2013}.
They correspond to the contribution of the 1-halo terms in equation (\ref{eq:phh_decom_cgm_approx}), 
and correct for the exclusion effect and
the apparent anisotropy due to the cylinder shape, as discussed in the previous
subsection. We extend the formalism in \citet{van-den-Bosch:2013} and \citet{Baldauf:2013} 
to account for a cylindrical window exclusion effect.
The function $W(\vk)$ is the Fourier transform of the cylinder-shaped window,
expressed as
\be
W(\vk) = 2\frac{J_1(k_\perp \Delta r_\perp)}{k_\perp \Delta r_\perp}\frac{\sin(k_\parallel
\Delta r_\parallel)}{k_\parallel \Delta r_\parallel}V_W,
\label{eq:window_analytic}
\ee
where $V_W$ is the comoving volume of cylinder, $V_W=2\pi (\Delta
r_\perp)^2\Delta r_\parallel$ 
(see \S~\ref{sec:result_ps} for the shape of the window function).  Note that
the effective height of cylinder is $2\times \Delta r_\parallel$ as centers of
two CGM groups have to be separated by more than  $\Delta r_\parallel$ for both
positive and negative radial directions.  The first term in the brackets
corrects for the exclusion effect of CGM halo. The second term in the brackets
primarily corrects for an apparent anisotropic clustering due to the convolution
of the window function and the underlying halo power spectrum, assuming that the
halo power spectrum is a smooth function at scales down to the virial radii of
host halos, a few Mpc at maximum:
\bey 
\left[ W(\vk)*\wt{P}_{\rm cc,0}^R(k)\right](\vk) &=&
\int\!\!\frac{\mathrm{d}^3\vq}{(2\pi)^3}~W(\vq)\wt{P}_{\rm cc,0}^R(|\vk-\vq|)\nn\\
&=&\int\!\! \mathrm{d}^3\vr~ {\cal W}(\vr)\wt{\xi}_{\rm cc}^R(\vr)e^{i\vk\cdot\vr}, \label{eq:window_convolution} 
\eey
where ${\cal W}(\vr)$ is the cylinder window in real space.

The last term in equation~(\ref{eq:baldauf_mod}) is intended to correct for an
apparent anisotropic clustering due to the misidentified CGM halos. The
misidentification of CGM halos occurs if central galaxies in different halos are
grouped together by the CGM method and misidentified as satellite galaxies in
the same CGM halo, or if satellite galaxies in the same halo are fragmented into 
different halos. These two cases correspond to $\wt{P}_{\mathrm{c\bar{c}}}^{S2h}$
and $\wt{P}_{\mathrm{c\bar{s}}}^{S2h}$ in equation (\ref{eq:phh_decom_cgm_approx}), 
respectively.
This misidentification preferentially occurs in an
overdensity region in large-scale structure or if filamentary structure of
large-scale structure is aligned to the line-of-sight direction. As a
result, correlations of the misidentified halos in different overdensity regions
with other majority of CGM halos are removed from the clustering analysis. 
Since this occurs in the 2-halo regime, 
we employ a minimal model to describe this effect; we assume that the contaminating
effect is proportional to the linear power spectrum of total matter
with an coefficient function including the distortion due the cylinders anisotropies, 
$\Delta P^S(\vk)=\alpha(\mu)P^R_{\rm lin}(k)$. 
The linear power spectrum $P_{\rm lin}^R(k)$ can be computed using the public code
such as CMBFAST \citep{Seljak:1996} or CAMB \citep{Lewis:2000} for a given
cosmological model.
If we apply this model in real space and adopt an isotropic group finder such as spheres 
rather than cylinders, the coefficient is just the square of the effective bias of the two 2-halo terms,
$\alpha(\mu)=b_{\rm eff}^2$. 
The $\mu$ dependence is determined by both RSD and nature of
misidentified halos. We will investigate the functional form of our model and its performance 
in detail in Section \ref{sec:analysis}.

\section{Mock catalog of redshift galaxy survey}
\label{sec:sim}

In order to test the methodology developed in this paper and assess
its performance, we use a mock catalog of galaxies that is constructed
from $N$-body simulations for $\Lambda$CDM model. Here we describe the
details of the construction of these catalogs.


\subsection{$N$-body simulations and halo/subhalo catalogs} 
We use the 32 realizations of cosmological simulations presented by
\citet{Sato:2011}. Each realization evolves the matter density
distribution as sampled by $1024^3$ $N$-body particles in a box size
of $1~\higpc$. The adopted cosmological parameters are described in
Section~\ref{sec:intro}.  We use the snapshots at redshifts $z=0.5$
and $2$ for each realization, to construct a mock catalog. 
We use the friends-of-friends (FoF) group finder
\cite[e.g.,][]{Davis:1985} with a linking length of $0.2$ in units of
the mean interparticle spacing to create a catalog of halos from the
simulation output and use the {\it SubFind} algorithm
\citep{Springel:2001} to identify subhalos within each halo.  In this
paper, we use halos and subhalos with equal to or more than 20
particles.  Each particle in a halo region is assigned either to a
smooth component of the parent halo or to a satellite subhalo, where
the smooth component contains the majority of $N$-body particles in
the halo region.  Hereafter we call the smooth component a central
subhalo and call the subhalo(s) satellite subhalo(s).  For each
subhalo, we estimate its mass by counting the bounded particles, which
we call the subhalo mass ($M_{\rm sub}$).  We store the position and
velocity data of particles in halos and subhalos at different
redshifts.  To estimate the virial mass ($M_{\rm halo}$) for each
parent halo, we apply the spherical overdensity method to the FoF
halo, where the spherical boundary region is determined by the
interior virial overdensity, $\Delta_{\rm vir}$, relative to the mean
mass density \citep{Bryan:1998}.  The overdensity $\Delta_{\rm
vir}\simeq 242$ at $z=0.5$ for the assumed cosmological model.

\begin{figure*}
\begin{center}
\includegraphics[width=85mm,bb=0 0 610 460]{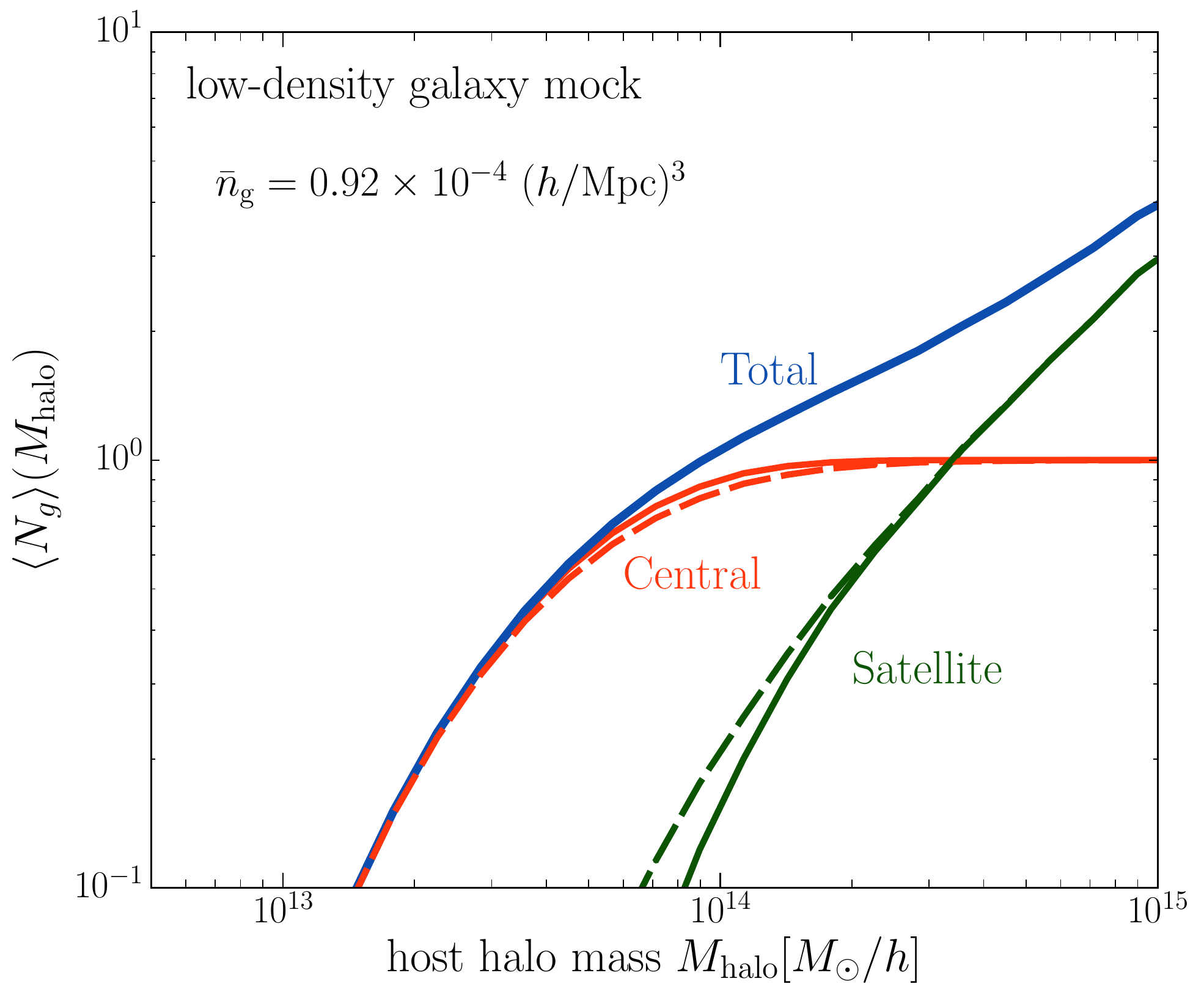}
\includegraphics[width=85mm,bb=0 0 610 460]{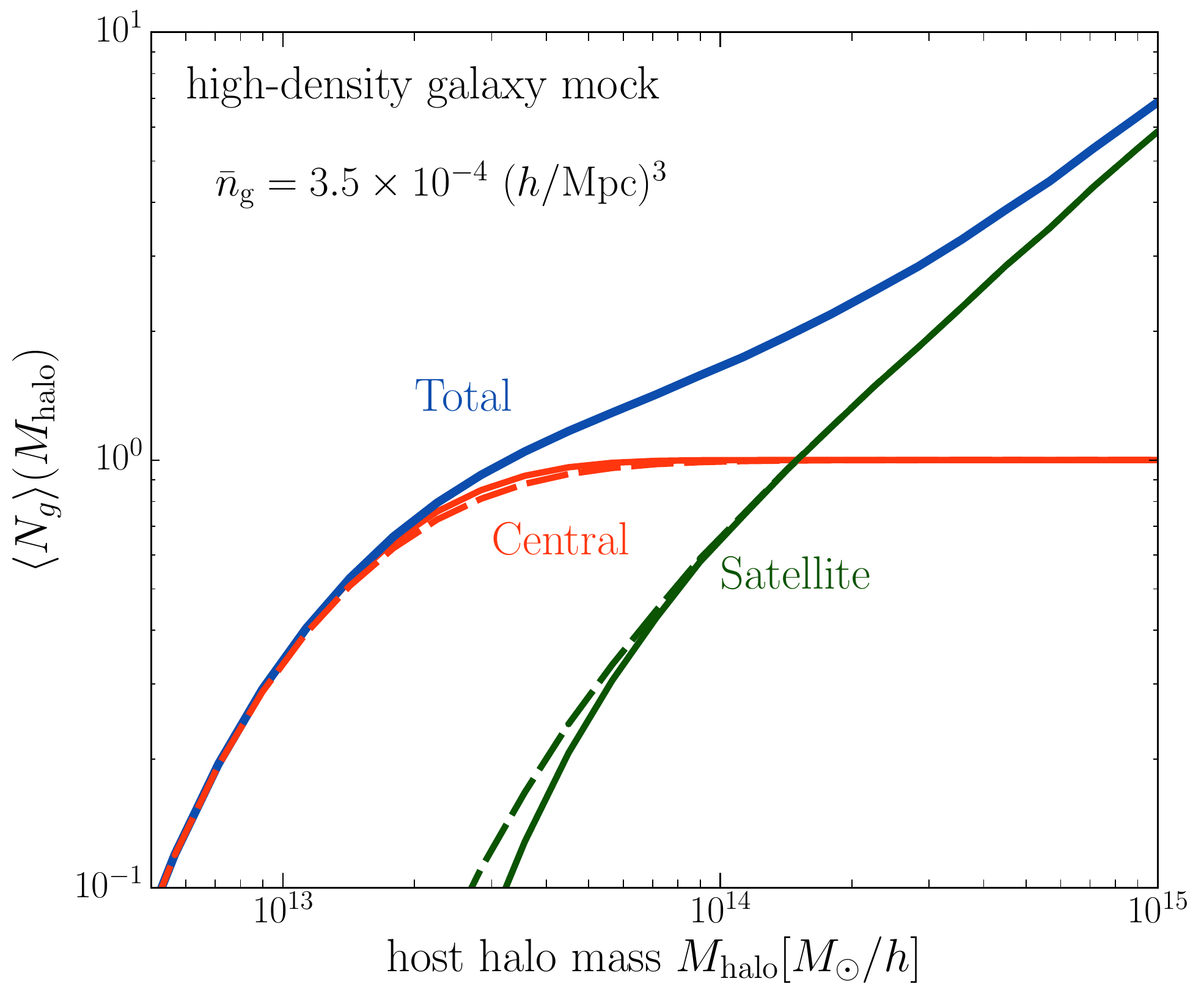}
 \caption{The halo occupation distribution (HOD) for galaxies in the
 mock catalog, constructed from $N$-body simulations (see text for
 details).  The left and right panels show the HODs for the
 ``low-density'' and ``high-density'' galaxy mock catalogs that we will
 use in this paper.  The upper bold solid curve in each panel is the
 total HOD, while the thin solid curves are the HODs for the central and
 satellite galaxies. Note that, if only a single galaxy of interest
 resides in the host halo, we refer to the galaxy as ``central galaxy''
 in this paper, even if the galaxy resides in a satellite subhalo in the
 parent halo. The dashed curves show the HODs, if we count the
 off-centered galaxy as a satellite galaxy.
 }
\label{fig:hod_cmass_offcentering}
\end{center}
\end{figure*}


\subsection{Abundance matching method in Masaki et al. (2013):
 populating galaxies into subhalos}
  
To populate mock galaxies into halos in the simulation realization, we
employ the abundance matching method described in \citet{Masaki:2013}.
In brief, the method assumes that simulated halos at $z=2$ are
progenitors for early-type galaxies at low redshift, for which we
adopt $z=0.5$ in this paper. We label the most tightly bound
particles in each progenitor halo as ``star particles'' as a building
block of the early-type galaxy.  We then label ``galaxies'' at the
target low redshift, by identifying the subhalos containing a majority
of the star particles in descending order of masses of the progenitor
halos until the comoving number density of the matched subhalos
becomes comparable to a target number density for the type of galaxies
we may be interested in \citep[see][for details]{Masaki:2013}.  Note that
building blocks (``star particles'') in the different progenitor halos
can merge together in the same subhalo at the low redshift.  Thus the
only free parameter in this method is the number density of progenitor
halos at $z=2$. In order to somewhat take into account stochastic
nature of galaxy formation, we employ a similar model to the halo
occupation distribution (HOD) \citep{Cooray:2002,Zheng:2005} including
a scatter around the halo mass threshold:
\begin{equation}
 p(M;z=2)=\frac{1}{2}\left[
1+{\rm erf}\left(\frac{\ln M -\ln M_{\rm th}}{\sqrt{2}\sigma}\right)		     
\right],
\end{equation}
where $M_{\rm th}$ is a parameter to model the mass threshold for halo
selection, and we fix the value of $\sigma=0.5$. Note that, $p(M)=0$
at $M\ll M_{\rm th}$, while $p(M)=1$ at $M\gg M_{\rm th}$. According
to the above probability in each mass bin, we randomly select
progenitor halos in each mass bin from the halo catalog of a given
simulation realization. The mass threshold parameter $M_{\rm th}$ is
determined so that the resulting mock galaxy sample has a desired
number density. In this paper, we consider two mock catalogs that
differ in the target number density: $\bar{n}_g \simeq 3.4\times
10^{-4}$ or $9.2\times 10^{-5}~(\hmpci)^3$ which we will refer to as
the ``high-density'' or ``low-density'' galaxy sample, respectively.
The mass threshold parameter for these samples are $M_{\rm th}\simeq
3.1\times 10^{12}~h^{-1}M_\odot$ and $7.7\times
10^{12}~h^{-1}M_\odot$, respectively. We employed the fixed value of
$M_{\rm th}$ for different realizations for simplicity, since the
resulting number density of galaxies in different realizations is
different only by less than 1\%.  The number density of the
high-density galaxy mock is comparable with that for CMASS galaxies in
the BOSS survey \citep{Eisenstein:2011}, while the low-density sample
is comparable with the number density for the LRG galaxies
\citep{Eisenstein:2005,Reid:2009} or the CMASS galaxy subsample with
higher stellar mass threshold \citep[e.g.,][]{More:2015}.

  \begin{table*}
   Mock galaxy catalogs\\
  \begin{tabular}{l|c|c|c|c|c|c|c|c|c|c|c||}\hline\hline
   Mock  & $\bar{n}_{\rm g}$& $N_{\rm g}$
	       & $f_{\rm sat}^{\rm gal}$
		       & single system   &
   \multicolumn{6}{|c|}{multiple systems $[\%]$} \\ 
   & $[10^{-4}(\hmpci)^3]$&
	       & $[\%]$
		       & $[\%]$ & 2 & 3 & 4 & 5 & 6 & $\ge 7$
		   \\ \hline
   high-dens &$3.4$ & $344,296$
	       & $9.59$
   & 91.75 
   & 6.64 
   & 1.14 
   & 0.30 
   & 0.10 
   & 0.04 
   & 0.03 
   \\ \hline
   low-dens & $0.92$& $92,362$
	       & $7.27$
   & 93.44 
   & 5.59 
   & 0.76 
   & 0.16 
   & 0.04 
   & 0.01 
   & 0.006
   \\ \hline\hline
  \end{tabular}
   \caption{Details of mock galaxy catalogs, the ``high-density'' or
   ``low-density'' mock catalogs we use in this paper (see text for
   details). $\bar{n}_{\rm g}$ is the mean number density of galaxies,
   $N_{\rm gal}$ is the total number of galaxies in the simulation
   volume, $V=1~(\higpc)^3$, and $f_{\rm sat}^{\rm gal}$ is the
   fraction of satellite galaxies.
   The single system is the halo hosting a single galaxy inside, and the
   multiple systems are the halos hosting multiple galaxies.  The
   numbers for these systems are the fraction of each system among all
   the host halos. All the numbers are the average values from the 32
   realizations of the mock catalogs.  \label{tab:mocks}}
  \end{table*}

The main purpose of this paper is not to develop a sophisticated mock catalog
of galaxies to reproduce observed properties of galaxies in any survey,
but rather to use such a mock catalog to test the validity and performance of
the halo reconstruction method. In this regard there are several
advantages of this method.  First, the mock catalog naturally consists
of both central and satellite galaxies, rather than assuming the two
populations as done in the HOD method. Secondly, each galaxy has its
own peculiar motion, including both virial motion inside the host halo
and the bulk motion of its host halo \citep[the motion of central
galaxy might be different from the center-of-mass motion of $N$-body
particles in the host halo as studied in] []{Masaki:2013}. Thirdly,
this mock generally predicts that some fraction of halos do not host a
central galaxy in its central subhalo, that is the off-centering
effect. This also means that, even if a given halo hosts a single
galaxy inside, the galaxy might be off-centered, or equivalently
reside in a satellite subhalo \citep[see][for the similar
discussion]{Hikage:2012,Masaki:2013}.


\subsection{HOD and clustering properties of mock galaxies}

Now we discuss properties of the mock galaxies in the context of
large-scale structure.  The left and right hand panels of
Fig.~\ref{fig:hod_cmass_offcentering} show the HOD for the galaxies in
the ``high-density'' and ``low-density'' mock catalogs, respectively.
Here we show the averaged HODs of the 32 realizations.  The method we
used allows us to compute the HODs directly from the mock catalog rather
than assuming a particular functional form. The figure shows the central
and satellite HODs. In our default definition, if a halo hosts a single
galaxy inside and even if the galaxy resides in a satellite subhalo of
the host halo, we include the galaxy in the central HOD. For comparison
we also show, by the dashed curve, the central HOD obtained only by
counting the galaxies residing in the central subhalos. The HOD for the
high-density galaxy mock extends to lower halo mass. Qualitatively
speaking the shapes of the HODs for our mock catalog appear to be
similar to those for the CMASS and LRG galaxies
\citep{White:2011,More:2015}.

Table~\ref{tab:mocks} summarizes the properties of the mock galaxy
catalogs. The fractions of satellite galaxies to all the galaxies are
about 9.6\% and 7.3\% for the high- and low-density galaxy mocks,
respectively. The table also shows fractions of ``single'' or
``multiple'' halo systems among all the host halos where the single
systems are the halos hosting a single galaxy inside and the multipole
systems are the halos hosting multiples galaxies.

Next we discuss clustering properties of these mock galaxies.  We
measure power spectra of given galaxy samples using a standard fast
Fourier transform (FFT) method, following the previous work of
\citet{Okumura:2012, Okumura:2012b, Okumura:2015}.  We first define
the density field by using a grid of $1024^3$ cells with cloud-in-cell
interpolation. When measuring the redshift-space density field, the
positions of objects are displaced along the line-of-sight direction
according to their peculiar velocities before they are assigned to the
grid.  We then perform an FFT to convert the density field of sample
``${\rm x}$'' to Fourier space, $\delta_{\rm x}(\vk)$.  The power
spectrum $P_{\rm xx}^S(\vk)$ is obtained by averaging $|\delta_{\rm
x}^2(\vk)|$ over finite bins in $\vk$. We assume the shot noise to be
Poisson, and hence subtract $1/\bar{n}_{\rm x}$ from the measured auto
power spectrum.  In case of the cross power spectrum of samples
``${\rm x}$'' and ``${\rm y}$'', $P_{\rm xy}^S(\vk)$, the two fields
need to be multiplied.  The aliasing effect arising from the
cloud-in-cell mass assignment method is corrected for using the
formula of \citet{Jing:2005}.  The errorbars are estimated from the
standard deviation of the 32 realizations, and the scatter is divided
by the square root of the number of the realizations, i.e.,
$\sqrt{32}$, to show the error of the mean.

\begin{figure}
\begin{center}
\includegraphics[width=85mm,bb=0 0 610 460]{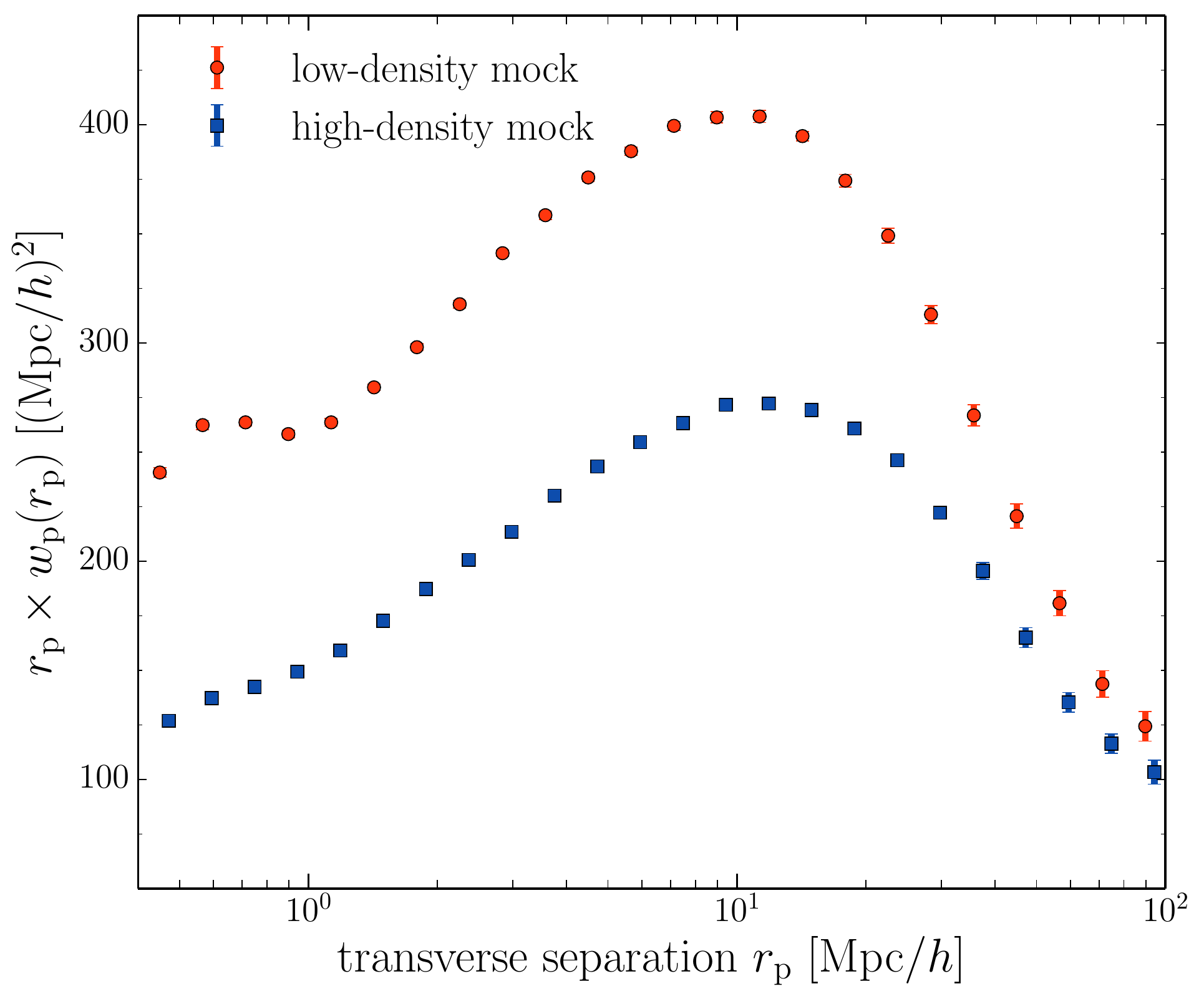}
 \caption{Projected correlation function, $r_p\times w_p(r_p)$, measured
 from the mock catalogs. The errorbars are the rms of the mean estimated
 from the 32 realizations each of which has volume of $1~(\higpc)^3$.
} \label{fig:wp}
\end{center}
\end{figure}

Fig.~\ref{fig:pkr_vs_pks} shows the real- and redshift-space power
spectra for the ``low-density'' galaxy mock. Even if the fraction of
satellite galaxies is only 7\% (see Table~\ref{tab:mocks}), the galaxies
cause a large difference from the underlying halo power spectra, as
discussed in Sec.~\ref{sec:motivation} in detail.

\begin{figure*}
\begin{center}
\includegraphics[width=85mm,bb=0 0 610 460]{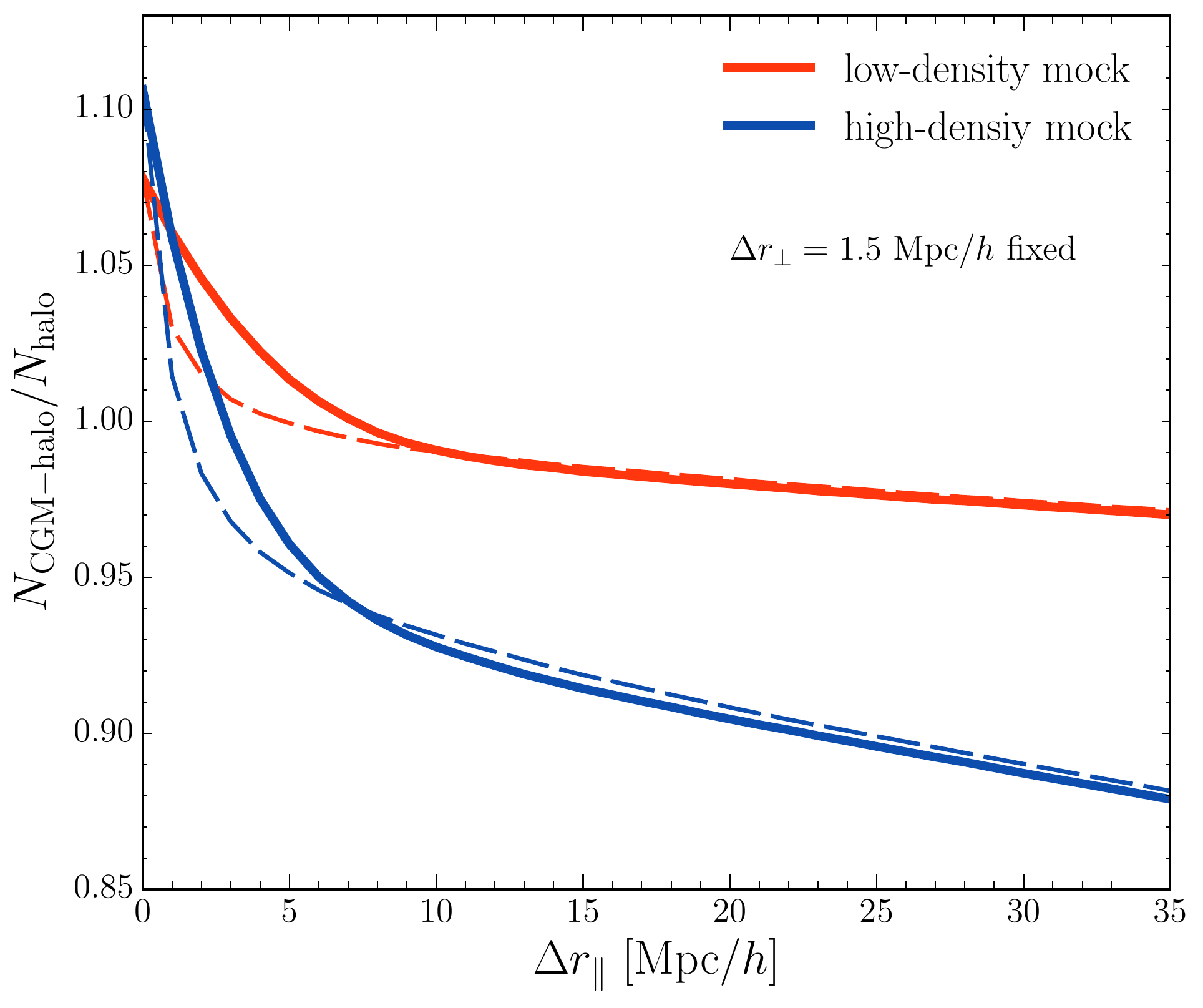}
\includegraphics[width=85mm,bb=0 0 610 460]{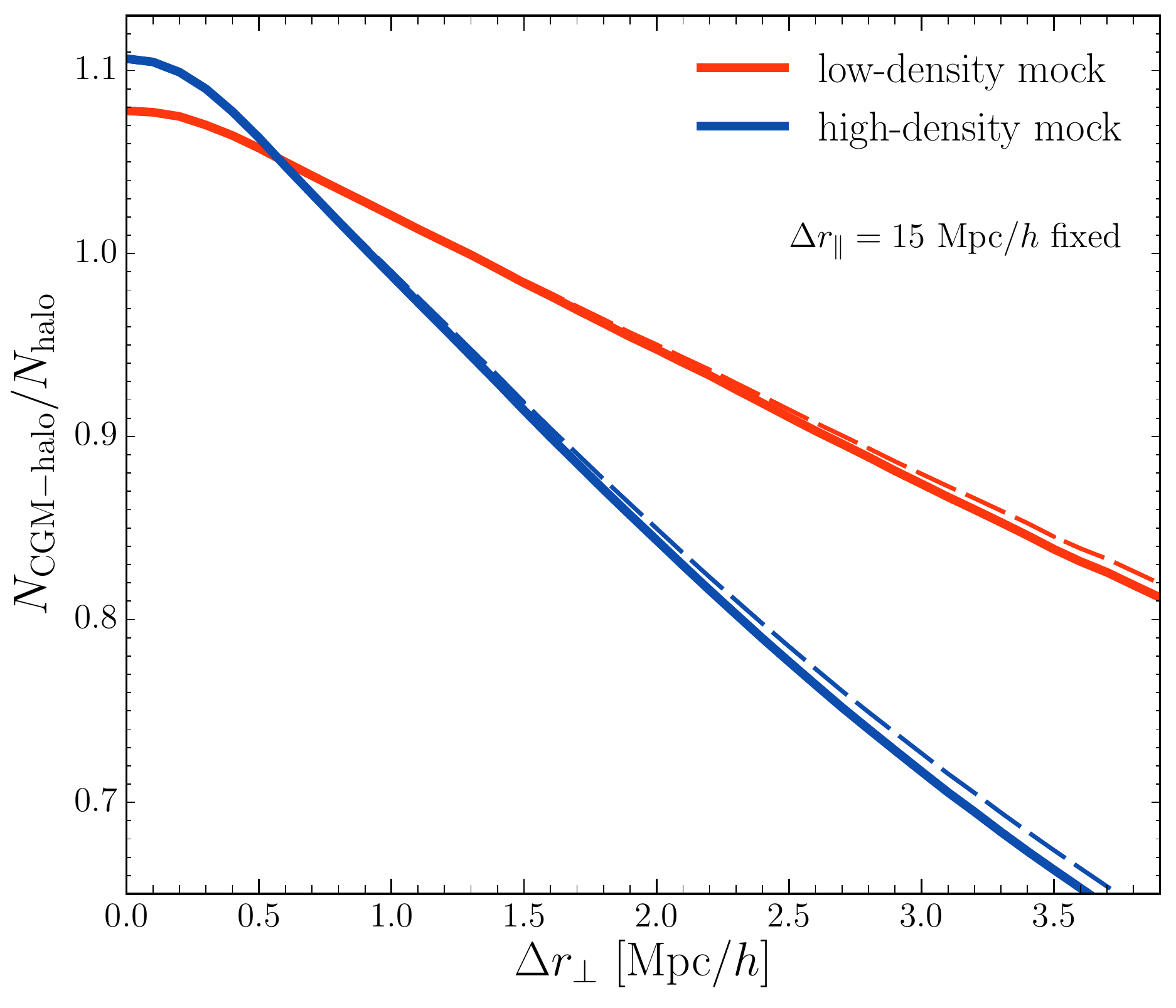}
 \caption{The number of halos identified using the cylinder-grouping
 method (CGM) relative
 to the true number. The left panel shows the result as a function of
 the height of the cylinder, $\Delta r_\parallel$, with its base circle
 radius being fixed to $\Delta r_\perp=1.5\himpc$, while the right panel
 as a function of $\Delta r_\perp$ with the height being fixed to
 $\Delta r_\parallel=15\himpc$.  The red and blue lines are for the low-
 and high-density mocks, respectively.  The dashed lines are obtained
 when the cylinders are identified in real space, while the solid lines
 are when identified in redshift space.  }
 \label{fig:dist_cylinder_galaxy}
\end{center}
\end{figure*}

The projected correlation function can be obtained by projecting the
redshift-space correlation function along the radial direction:
\be
w_p(r_p)=\int^{r_{\parallel,{\rm max}}}_{-r_{\parallel,{\rm max}}}\mathrm{d}r_\parallel \xi^s(r_\perp,r_\parallel),
\ee
where we choose $r_{\parallel, {\rm max}}=100~\himpc$. 
In Fig.~\ref{fig:wp} we show the projected correlation functions
measured from our mock samples. Roughly speaking, the projected
correlation functions measured from the low- and high-density mocks 
are similar to 
the SDSS CMASS measurements in \citet{More:2015}.

  \begin{table*}
  CGM halo catalogs\\
  \begin{tabular}{l|c|c|c|c|c|c|c|c|c|c|c|c||}\hline\hline
   Mock  & $N_{\rm h}$ &  cylinder   & $\tilde{N}_{\rm h}$ & $\tilde{f}_{\rm sat}^{\rm gal}$  & single system   &
   \multicolumn{6}{|c|}{multiple systems $[\%]$} \\ 
   & & identification&         & $[\%]$	       & $[\%]$ & 2 & 3 & 4 & 5 & 6 & $\ge 7$
		   \\ \hline
   high-dens & $311,277$ &real space 
   & 286,080 (91.9\%)
   & 16.91
   & 84.06
   & 12.65
   & 2.48
   & 0.58
   & 0.16
   & 0.04
   & 0.02 \\
   & & redshift space 
   & 284,681 (91.5\%)
   & 17.32
   & 83.67
   & 12.90
   & 2.59
   & 0.61
   & 0.16
   & 0.05
   & 0.02
   
   \\ \hline
   low-dens & 
   85,661  & real space & 84,394 (98.5\%)
   & 8.63		 
   & 91.74
   & 7.27
   &  0.85
   & 0.12
   &  0.02
   &  0.005
   &  0.001
   \\   
  & & redshift space & 
     84,257 (98.4\%)
   & 8.77		
   & 91.58
   & 7.41
   &  0.87
   & 0.12
   &  0.02
   &  0.004
   &  0.001
   \\ \hline\hline
  \end{tabular}
   \caption{Details of CGM halos for the low- and
   high-density mock catalogs, compared to Table~\ref{tab:mocks}.
   Here the CGM-halos are the halos reconstructed by placing the
   cylinder around each of more massive galaxies and by then removing
   satellite galaxies within a cylinder around the central galaxy. 
   (see text for the details). We employed
   $\Delta r_\perp=1.5~\himpc$ and $\Delta r_\parallel=15~\himpc$ for the radius and height of the CGM cylinder, and included
   the redshift space distortion of galaxies when implementing the CGM
   method in redshift space. The single CGM-halos are the isolated
   systems, and the multipole CGM-halos are the systems containing
   satellite galaxies in the same CGM halo. 
   The satellite fraction $\tilde{f}^{\rm gal}_{\rm
   sat}$ is the fraction of the CGM satellite galaxies
   among all the galaxies.
   \label{tab:mocks_cgm}}
  \end{table*}

\begin{figure*}
\begin{center}
\includegraphics[width=85mm,bb=0 0 610 570]{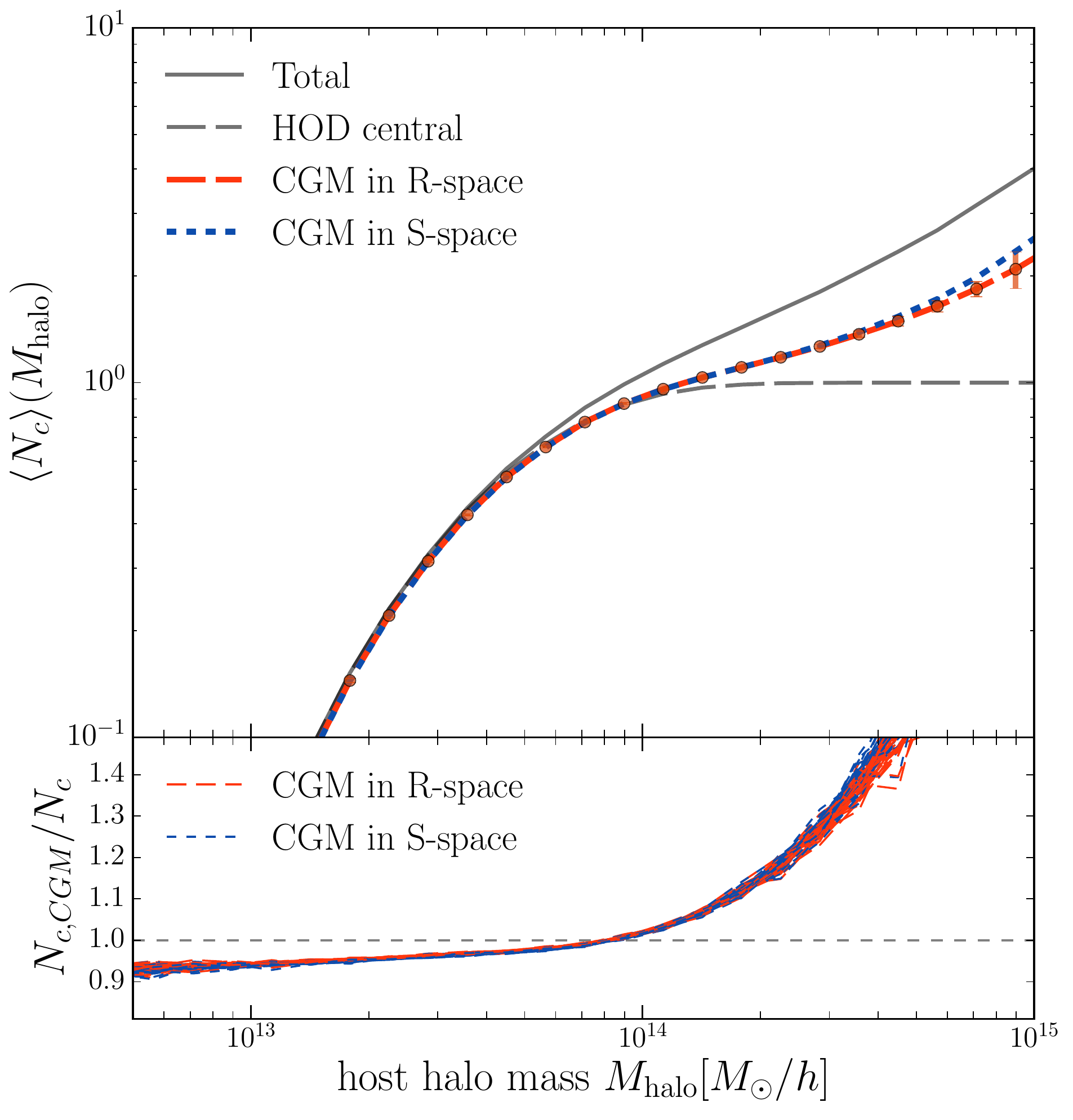}
\includegraphics[width=85mm,bb=0 0 610 570]{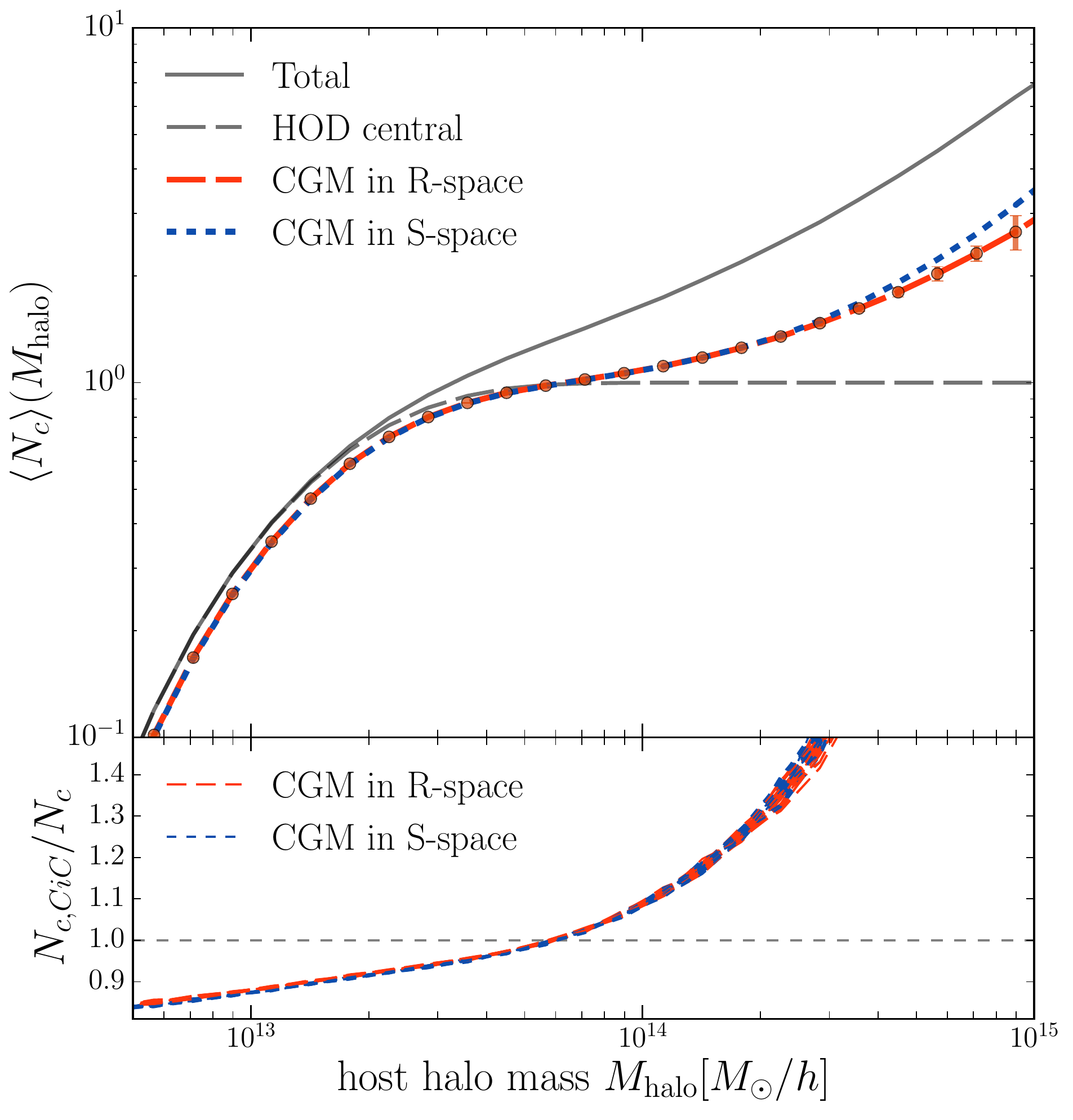}
\caption{({\it Top}) The HOD for central galaxies identified by the CG
 method for the low- and high-density mock catalogs in the left and
 right panels, respectively.  The red-dashed curves are obtained when
 the cylinders are identified in real space, while the blue-dotted curves
 are when identified in redshift space. The central HOD becomes larger
 than unity for massive halos because multiple galaxies in a massive
 halo can be misidentified as different CGM halos. 
 The gray solid and dashed curves are the total and central HOD which are
 the same as the blue and red solid curves in
 Fig.~\ref{fig:hod_cmass_offcentering}, respectively.  ({\it Bottom})
 The ratio of the CGM central galaxies to the true central galaxies. The
 ratio being less than unity for low-mass halos means that the galaxies
 in different halos are mis-grouped into one halo. 
\label{fig:hod_cic}}
\end{center}
\end{figure*}

\section{Numerical analysis} \label{sec:analysis}

In this section we show the main results of this paper, i.e. the
reconstruction of the halo distribution and power spectra from the galaxy
distribution using the mock catalogs.


\subsection{Applying the CGM halo reconstruction to mock galaxy catalogs}\label{sec:cgm_rec}

In order to perform the CGM halo reconstruction, we need to specify the
parameters that characterize the cylinder around each galaxy, the height
($\Delta r_\parallel$) and the base-circle radius ($\Delta r_\perp$), which are 
defined parallel and perpendicular to the line of sight, respectively. They are
arbitrary and need to be chosen so that galaxies which are in the same halo are
properly grouped into the same CGM halo.  In particular, the value of $\Delta
r_\parallel$ needs to be carefully tuned because the inferred radial separation
between galaxies is affected by their peculiar velocities due to virial motions,
i.e the FoG effect.  For the LRG sample at $z\sim 0.3$, which has a number
density of $10^{-4}~(\hmpci)^3$, \citet{Reid:2009} adopted $(\Delta
r_\perp,\Delta r_\parallel) = (0.8,20) [\himpc]$ to implement their halo
reconstruction.

In this paper we adopt $(\Delta r_\perp,\Delta r_\parallel) = (1.5,15) [\himpc]$
as our default parameters.  The left panel of
Fig.~\ref{fig:dist_cylinder_galaxy} shows the ratio of the number of halos
identified by the CG technique to the true number of halos as a function of
$\Delta r_\parallel$, for a fixed value of $\Delta r_\perp=1.5\himpc$.  The red
line corresponds to the low-density mock galaxy sample (see
Table~\ref{tab:mocks}), while the blue line corresponds to the high-density mock
sample. In the ideal case where equation (\ref{eq:phh=pcc}) holds, this ratio
becomes unity, which requires a fine tuning of the parameters, $(\Delta
r_\perp, \Delta r_\parallel)$.
However, an {\em apparent} perfect reconstruction, i.e. the ratio being unity,
could happen by chance, if the CGM halos contain both misidentified halos
(identifying satellite galaxies as different halos) and different halos
mis-grouped and identified as one halo. Thus the perfect reconstruction cannot be
realized in practice. 

The solid and dashed curves
show the results for the cases where cylinders are identified by the
positions of galaxies in redshift and real space, respectively.  The
discrepancy between the two results seen at small $\Delta r_\parallel$
for both high and low density samples implies that the length of the
cylinder is not long enough to capture the line-of-sight shifts of the
galaxies due to RSD.  The right panel of
Fig.~\ref{fig:dist_cylinder_galaxy} is the same as the left panel but
shows the ratios as a function of $\Delta r_\perp$, fixing $\Delta
r_\parallel=15\himpc$.  Since the angular positions are not affected
by RSD, the results where cylinders are defined in real space and
redshift space are almost the same, and the number of the CGM halos
decreases linearly with $\Delta r_\perp$.  For the low-density mock
catalog which has the number density $\bar{n}_{\rm g}\simeq
10^{-4}~(\hmpci)^3$, similar to that of LRGs, the choices of $\Delta
r_\perp=1.5~\himpc$ and $\Delta r_{\parallel}=15~\himpc$ seem
reasonable. For the high-density mock, the smaller size cylinder seems
better, but we in this paper employ the same parameters for
simplicity.  In Appendix~\ref{sec:changing_pi}, we will show how the
size of the cylinder affects the final results, by adopting $\Delta
r_\parallel = 30\himpc$. 

Table~\ref{tab:mocks_cgm} gives a summary of the halo catalogs
reconstructed from the CG method using the fiducial parameters
$(\Delta r_\perp,\Delta r_\parallel)=(1.5, 15)~\himpc$. The number of
the CGM halos is smaller than that of the underlying true halos,
because the CG method incorrectly connects different halos. For the
high-density mock, this problem is worse.  Compared to
Table~\ref{tab:mocks}, the fractions of the single or multiple
systems, which contain single or multiple galaxies in the CGM halos,
respectively, are modified.  These are unavoidable limitations of the
CG method, and below we will carefully study the impact of these
effects on the reconstructed halo power spectra. 
For the low- and high-density mocks, 
$(\wt{g}_\mathrm{c}, \wt{g}_\mathrm{\bar{c}}, \wt{g}_\mathrm{\bar{s}})=(0.985,0.0335,0.0187)$ 
and
$(0.919,0.0951,0.0141)$, respectively, 
thus equation (\ref{eq:phh_decom_cgm_approx}) is a good approximation
especially for our main, low-density sample.

The top left panel of Fig.~\ref{fig:hod_cic} shows the HOD for central
galaxies with lower density identified by the CG method.  The
red-dashed curve is obtained when the cylinders are identified in real
space, while the blue-dotted curve is when identified in redshift space.
The central HOD becomes larger than unity at the host halo mass of
$M_{\rm halo} > 10^{14} (h^{-1}M_\odot)$. This is mainly caused by
misidentification of different (central and satellite) galaxies in
massive halos as different CGM halos, that are separated along the
transverse direction to the light of sight direction rather than the
radial direction. We confirmed this; if we change the radius of cylinder
base circle to $\Delta r_\perp=3~\himpc$ from the fiducial value $\Delta
r_\perp=1.5~\himpc$, the central HOD becomes below unity up to more
massive halos. In other words, the CG method properly removes satellite
galaxies along the radial direction that are a dominant source of the
FoG effect. One can thus see
such a misidentification is a minor effect which occurs only in the
rare, massive halos.  The gray solid and dashed curves are the same as
the blue and red solid curves at the left panel of
Fig.~\ref{fig:hod_cmass_offcentering}, respectively, and shown here for
comparison.  The bottom left panel of Fig.~\ref{fig:hod_cic} shows the
ratio of the central galaxies identified by the CGM to the true central
galaxies.  The ratio being less than unity for low-mass halos with
$M_{\rm halo} < 10^{14} (h^{-1}M_\odot)$ means that the galaxies in
different halos are mis-grouped into one halo.
This is the dominant
effect which we will try to correct for in the following subsections.
The right panel of Fig.~\ref{fig:hod_cic} is the same as the left panel
but for the CGM central galaxies for the high-density mock.  The fact
that the deviation from unity at low $M_{\rm halo}$ at the bottom panel
than that in the low density sample implies that the larger correction
is required to reconstruct the halo spectrum for the higher density
sample.

\begin{figure*}
\begin{center}
\includegraphics[width=85mm,bb=0 0 610 460]{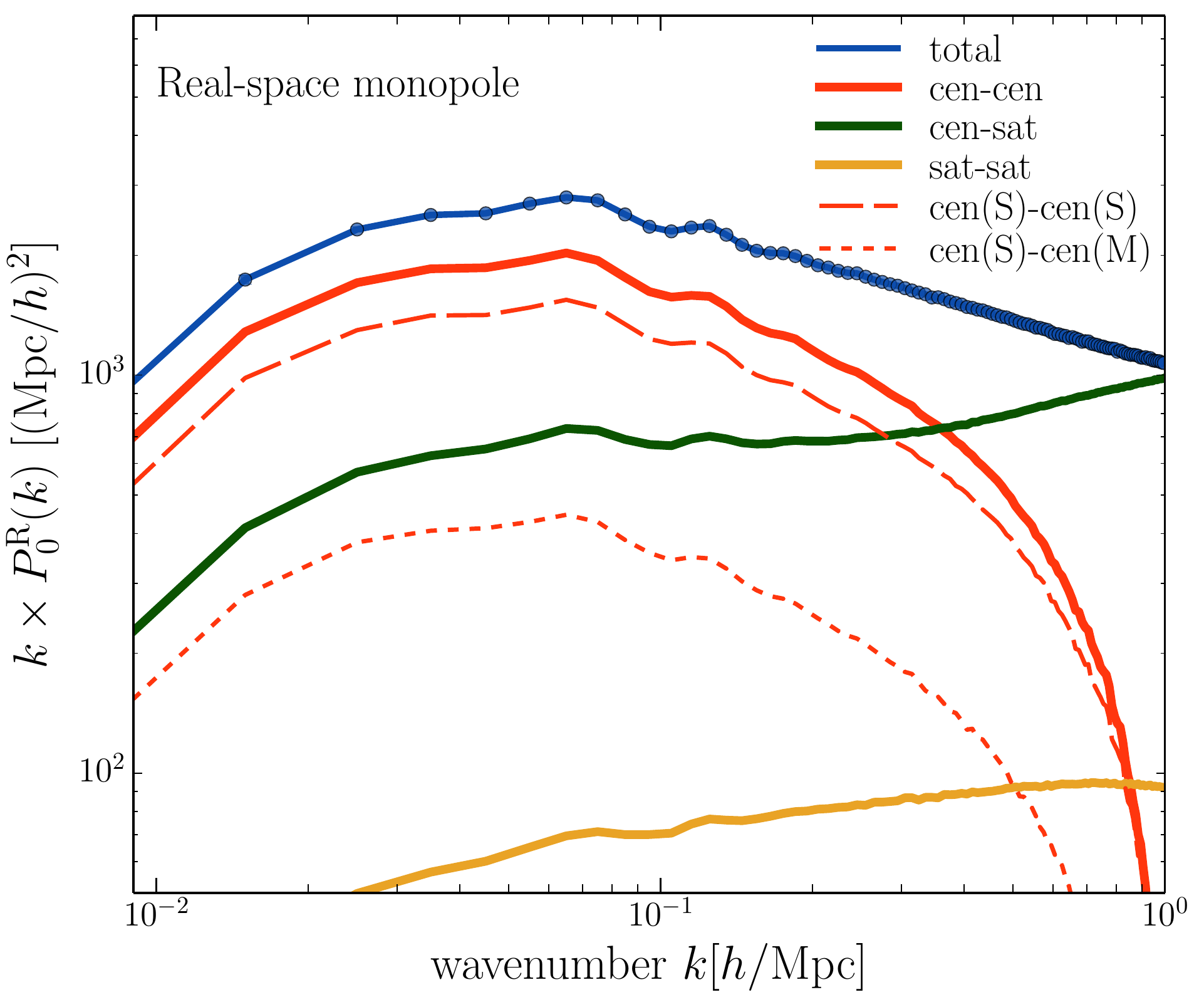}
\includegraphics[width=85mm,bb=0 0 610 460]{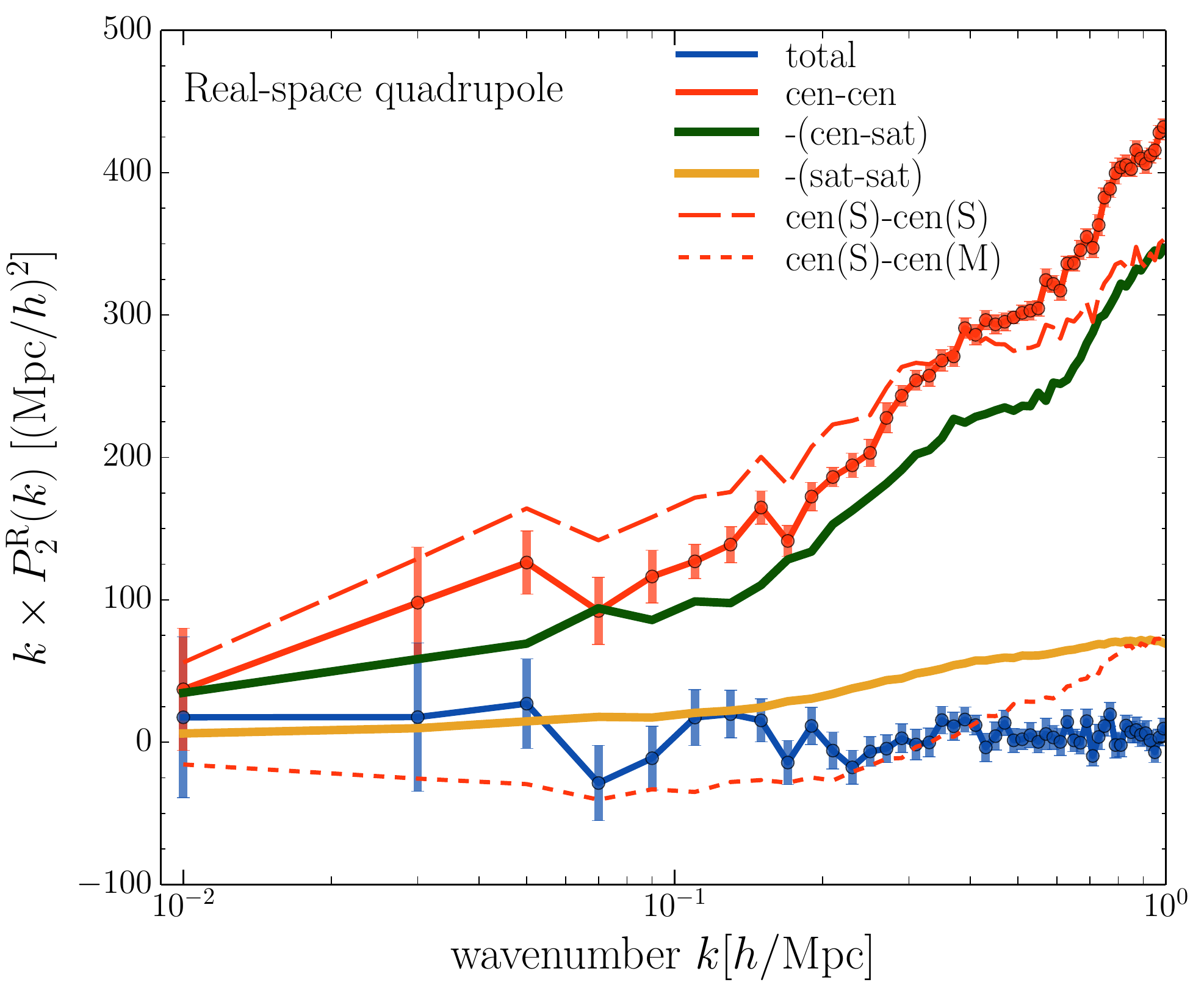}
 \caption{{\it Left panel}:
 Contributions of the decomposed galaxy samples by the
 CG method to the monopole power spectrum of the whole
galaxies in real space. The red and yellow solid curves are the contribution from the
central galaxies and satellite galaxies, denoted as $\wt{P}^R_{\rm
cc,0}$ and $\wt{P}^R_{\rm ss,0}$, respectively, while the green curve is
from their cross-spectrum $\wt{P}^R_{\rm cs,0}$. The power of
$\wt{P}_{\rm cc,0}$ is further decomposed to the three, and two of them
 are displayed; the long- and short-dashed curves show
 the correlation of the centrals in the single systems, denoted as cen(S)-cen(S), 
 and the cross-correlation of centrals in
 the single system and multiple system, cen(S)-cen(M), respectively. 
 For illustrative purpose, the errorbars are shown only for the
total power spectrum $P_{\rm gg,0}$. {\it Right}: The similar to the
left panel, but for the quadrupole spectrum. Note that the total power
is equal to a null signal as shown by the blue curve because of the
cosmological principle. Thus the errorbars are added here to the main
component, the quadrupole of the central galaxies
$\wt{P}^R_{\rm cc,2}$. Note that $\wt{P}^R_{\rm cs,2}$ and
$\wt{P}^R_{\rm ss,2}$ are multiplied by $(-1)$.}
\label{fig:pk_decon_r_9.2e-5}
\end{center}
\end{figure*}

\begin{figure}
\begin{center}
\includegraphics[width=85mm,bb=0 0 610 460]{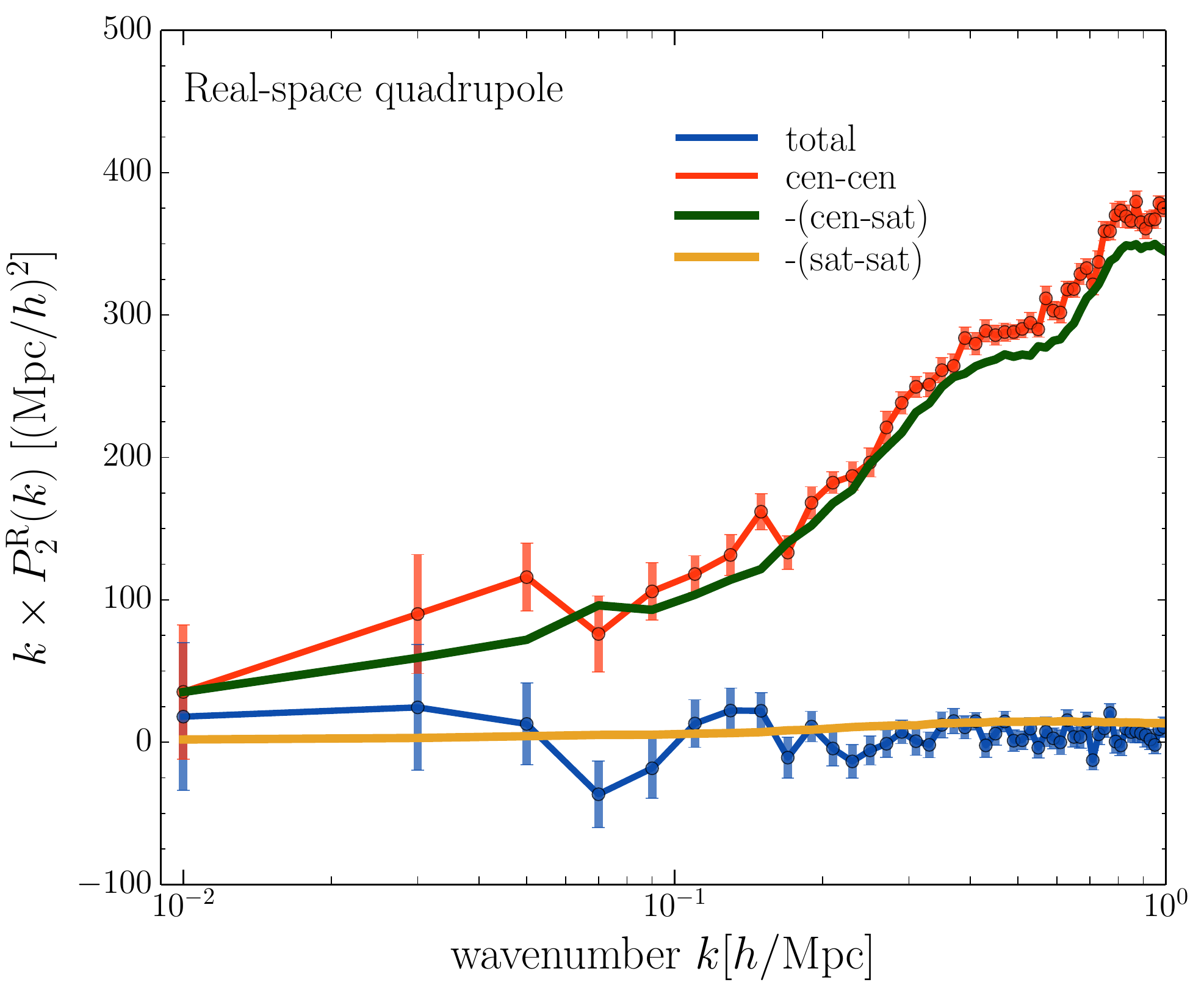}
 \caption{Same as the right panel of Fig. \ref{fig:pk_decon_r_9.2e-5} but 
 the CGM technique applied to the halo catalog.
 }
\label{fig:pk_decon_r_9.2e-5_halo}
\end{center}
\end{figure}

To conclude, the CGM single and
multiple halo systems are different from the underlying true systems of
halos given in Table~\ref{tab:mocks}, due to the imperfect grouping
by the CG method. Some CGM central galaxies are satellite galaxies in
larger more massive halos. Similarly, some CGM satellite galaxies are
in reality central galaxies, either due to group misidentification, or
a central-satellite misclassification.

\begin{figure*}[bt]
\begin{center}
\includegraphics[width=85mm,bb=0 0 610 460]{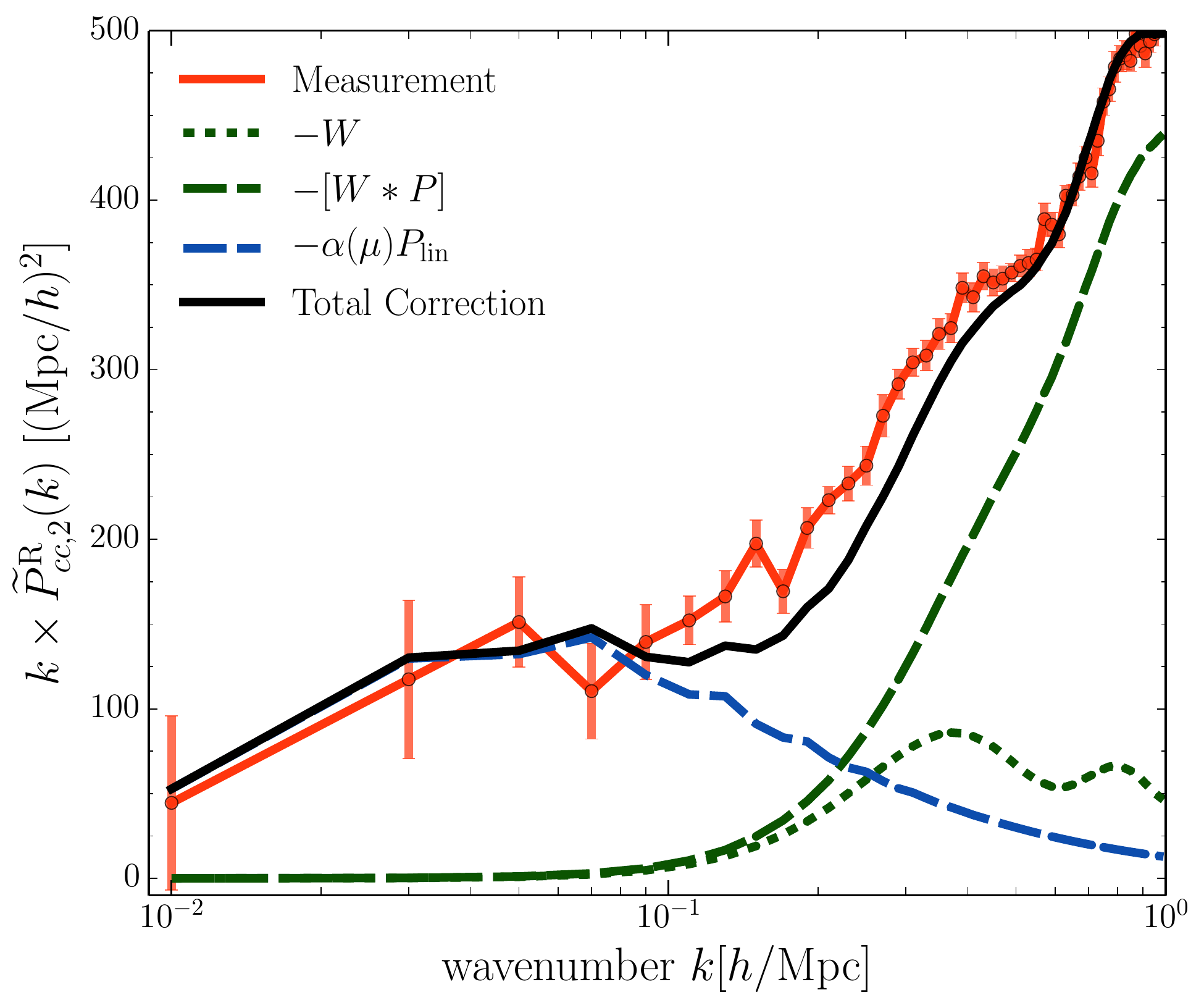}
 \includegraphics[width=85mm,bb=0 0 610 460]{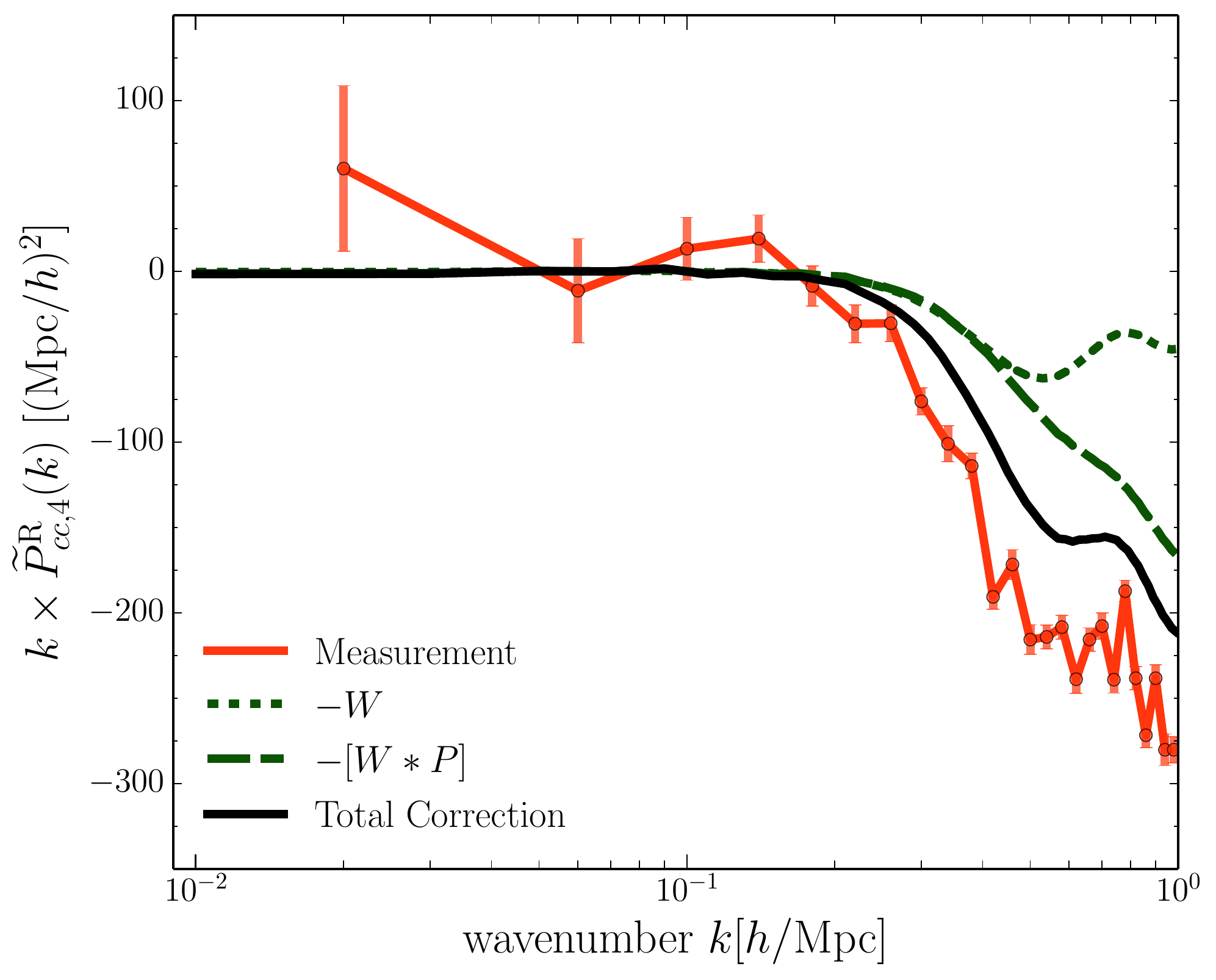}
 \caption{Correction terms of apparent anisotropic clustering of CGM
 halos; the terms modeling the anisotropic cylinder window and its exclusion
 effect (the 2nd and 3rd terms on the r.h.s. of
 equation~\ref{eq:baldauf_mod}) and the term correcting for the
 large-scale, anisotropic clustering due to misidentified CGM halos that
 tend to occur in large-scale structures aligned with the line-of-sight
 direction (the 4th term in equation~\ref{eq:baldauf_mod}). The data
 points with errorbars are the quadrupole (left panel) and hexadecapole
 (right) power spectra of CGM halos in {\em real space}, similar to 
 the right panel of Fig.~\ref{fig:pk_decon_r_9.2e-5}, which should
 vanish if the CGM halo reconstruction is perfect. The dotted and
 dashed green-curves are the terms involving the cylinder window, multiplied
 by $(-1)$. The dashed blue curve is the apparent large-scale
 clustering of misidentified CGM halos, multiplied by $(-1)$.
 The black curves are the sum of the correction terms,
 designed to cancel the power spectrum of CGM halos. 
 } \label{fig:pk_quad_window_r_9.2e-5}
\end{center}
\end{figure*}
%


\subsection{Decomposed power spectra of CGM galaxies}
\label{sec:result_ps}

As we have described, the CG halo reconstruction method has its limitations,
which makes it difficult to obtain a perfect reconstruction of the underlying
halos that host galaxies of interest.
To better understand the halo reconstruction obtained by the CG method, in this
subsection we consider the power spectrum decomposed into different correlation
contributions between the central and satellite galaxies in the CGM halos, as
given in equation~(\ref{eq:ps_break}). 
In the following we focus only on the low-density mock catalog.  

In the left panel of Fig.~\ref{fig:pk_decon_r_9.2e-5}, we first study
the monopole power spectra of the galaxies in real space. The circle
symbols are the monopole spectrum for all the galaxies in the mock
catalog, the same spectrum shown in the left panel of
Fig.~\ref{fig:pkr_vs_pks}.  The bold red and yellow curves are
respectively the spectra of the CGM ``central'' and ``satellite''
galaxies, while the bold green curve is their cross spectrum.
Again note that the CGM satellites contain misidentified halos as we
have discussed. The power spectrum of the galaxies is
given by the sum of the central and satellite power spectra, weighted by
their fractions (see equation~\ref{eq:ps_break}). The power spectrum of
the CGM central galaxies has lower amplitudes at large and small scales
(small and large $k$ bins) than the galaxy power spectrum. This is
because more massive halos are weighted low and the 1-halo term
contribution is mitigated.  However, the power spectrum of CGM halos
still does not correspond to that of the underlying halos.

In the right panel of Fig.~\ref{fig:pk_decon_r_9.2e-5} we show the
anisotropic component, the quadrupole spectrum in real space.  The
quadrupole of the whole galaxy sample is equal to zero because the
real-space galaxy distribution is isotropic, as shown with the blue
points. The real-space power spectrum of the CGM central galaxies,
however, has a non-zero quadrupole as shown by the red solid curve with
the errorbars.  Because the total power $\prggt=0$, 
the sum of the other two power spectra, the
cross power spectrum of the CGM halos and satellites and the auto
spectrum of the satellites, has the same amplitude as the power of the
CGM central galaxies but the opposite sign. The dashed curve shows that
the apparent quadrupole power arises mainly from the correlation between
central galaxies in single CGM halos. The contribution from the $\prss$
is negligibly small because the amplitude is suppressed by the square of
satellite galaxy fraction, $\tfs^2$, where $\tfs\sim 0.1$ for our mock
galaxy catalog, thus $\tfc \tprcct\simeq -2\tfs\tprcst$.  In particular,
since the CGM satellites include misidentified halos in the overdensity
region, the apparent quadrupole power at large scales can be understood
to arise from cross-correlations of the misidentified halos with other
majority of the central CGM galaxies.  We also measured the multipoles of the decomposed
galaxies in redshift space.  The trend of each spectrum is more or less
the same as shown in Fig.~5 of \cite{Okumura:2015}, thus we do not show
the redshift-space spectra of the decomposed CGM galaxies in this paper
and we refer readers to \cite{Okumura:2015}.

In order to see if the large-scale quadrupole is really caused by the 
misidentified halos, we consider to apply the CG method to the host halo distribution. 
In this case there is no satellite galaxy misidentified as a halo center 
($\wt{g}_{\rm \bar{s}\bar{s}}=0$), so the 
equation (\ref{eq:phh_decom_cgm}) is simplified as, 
\bey
P_{\rm hh}^S(\vk)&=&
\wt{g}_{\mathrm{h}}^2\wt{P}^S_{\mathrm{hh}}(\vk) 
+ 2\wt{g}_{\mathrm{h}}\wt{g}_{\mathrm{\bar{h}}} \wt{P}^S_{\mathrm{h\bar{h}}}(\vk)
+ \wt{g}_{\mathrm{\bar{h}}}^2\wt{P}^S_{\mathrm{\bar{h}\bar{h}}}(\vk) \nn \\
&\simeq& 
\wt{P}^S_{\mathrm{hh}}(\vk) 
+ 2\wt{g}_{\mathrm{\bar{h}}} \wt{P}^S_{\mathrm{h\bar{h}}}(\vk),
\eey
where $\wt{g}_{\mathrm{h}}=1-\wt{g}_{\mathrm{\bar{h}}}=0.967$ when
we adopt the same cylinder size, $(\Delta r_\perp,\Delta r_\parallel)=(1.5,15)\himpc$, 
for the host halos of the low-density sample. 
We measure the real-space quadrupole spectra of the halos and show them 
in Fig.~\ref{fig:pk_decon_r_9.2e-5_halo}.
The contributions of 
$\wt{P}^S_{\mathrm{hh}}$, $\wt{P}^S_{\mathrm{h\bar{h}}}$ and 
$\wt{P}^S_{\mathrm{\bar{h}\bar{h}}}$ to the total quadrupole are shown as the red, green and yellow lines, respectively. 
The contribution of $\wt{P}^S_{\mathrm{\bar{h}\bar{h}}}$ is suppressed by $\wt{g}_{\rm \bar{h}}^2 \sim 10^{-4}$ and  negligibly small, as expected, 
thus $\wt{P}^S_{\mathrm{hh},2} \simeq -2 \wt{g}_{\mathrm{\bar{h}}} \wt{P}^S_{\mathrm{h\bar{h}},2}$
because the real-space quadrupole of halos is zero. 
Even for this case where there is no satellite, 
one can see a clear non-zero signal of the quadrupole at small-$k$,
whose amplitude is similar to the quadrupole of the CGM halos 
identified from the galaxies as shown in the right panel of Fig.~\ref{fig:pk_decon_r_9.2e-5}.
This confirms that the large-scale anisotropy is indeed due to the central galaxies 
misidentified as satellite galaxies. 

In the next subsection, we test how well
equation~(\ref{eq:baldauf_mod}) can be used to reproduce the
anisotropy introduced even in the real space power spectrum of CGM
halos.

\subsection{Corrections for the CGM anisotropies and exclusion effects}\label{sec:corrections}

The two terms in equation (\ref{eq:baldauf_mod}), which involve the
window function $W$, correct for the apparent anisotropy due to the
shape of the cylinder. Since we know the shape of the cylinders to be
placed around each galaxy, we can analytically compute the Fourier
transform of the window function used in the halo reconstruction (i.e.
equation~\ref{eq:window_analytic}). 

The left panel of Fig.~\ref{fig:pk_quad_window_r_9.2e-5} compares the
model prediction of equation~(\ref{eq:baldauf_mod}) with the {\em
real-space} quadrupole power spectrum of CGM central galaxies. The
dashed and dotted curves show the quadrupole power spectra arising
from the halo exclusion effect via the anisotropic window function:
\begin{eqnarray}
 &&\hspace{-2em}W_2(k)\equiv
  5\int_0^1\!\mathrm{d}\mu~W(\vk){\cal L}_2(\mu),  \nonumber\\
 &&\hspace{-2em} \left[W(\vk)* \wt{P}_{\rm cc,0}^R(k)\right]_2(k)\nn\\
 && 
\ \ \ \  \equiv   5\int_0^1\!\mathrm{d}\mu~\left[W(\vk)*\wt{P}_{\rm cc,0}^R(k)\right](\vk){\cal L}_2(\mu), 
\label{eq:w_p2}
\end{eqnarray}
where we have used the monopole power spectrum of CGM halos as $P_{\rm
cc}(k)$ to estimate the convolution term.  Note that these
terms have negative values. The figure shows that the sum of these two
terms can explain some fraction of the apparent quadrupole power
spectrum (the data points with errorbars), and especially reproduce
the qualitative behavior of the quadrupole power; an increasing power
with increasing $k$ at $k\simgt 0.1~\hmpci$. However, since the size
of a cylinder is a few 10~Mpc$/h$ at most and its window
$|W(\vk)|\rightarrow 0$ for small $k$-limit, the above two terms
cannot explain the apparent quadrupole at large length scales,
$k\simlt 0.1~\hmpci$.

We argue that the small-$k$ apparent anisotropy in the CGM halo clustering can
be ascribed to misidentified halos in the CG method. The
misidentification tends to occur in an overdensity region that is
aligned with the line-of-sight direction, e.g., a filament. Hence as
implied from Figs.~\ref{fig:pk_decon_r_9.2e-5} and \ref{fig:pk_decon_r_9.2e-5_halo}, 
the apparent anisotropy at large scales are mainly from 
cross-correlations of the misidentified CGM satellites in the cosmic web with 
the CGM central galaxies. We assume that the large-scale apparent anisotropy 
seen in real space is proportional to the linear matter power spectrum with the anisotropic factor: 
\begin{equation}
 \Delta P^R(\vk)=\alpha(\mu) P^R_{\rm lin}(k),
  \label{eq:delta_pr}
\end{equation}
with
\begin{equation}
 \alpha(\mu)=\alpha_{\mu^0}-\alpha_{\mu^2}\mu^2, \label{eq:alpha_mu}
\end{equation}
where $\alpha_{\mu^0}$ and $\alpha_{\mu^2}$ are free parameters.  Thus
we assume that the apparent anisotropic clustering
can be explained by with
$O(\mu^2)$
correction term.
The monopole and quadrupole power spectra
are given by
\begin{eqnarray}
\Delta P_0^R(k)&=&\left( \alpha_{\mu^0} - \frac{\alpha_{\mu^2}}{3}
						 \right)P^R_{\rm lin}(k), \label{eq:Delta_p0r} \\
 \Delta P_2^R(k)&=&-
\frac{2\alpha_{\mu^2}}{3} P^R_{\rm lin}(k)\,. \label{eq:Delta_p2r} 
\end{eqnarray}
%

\begin{figure*}
\begin{center}
\includegraphics[width=85mm,bb=0 0 610 570]{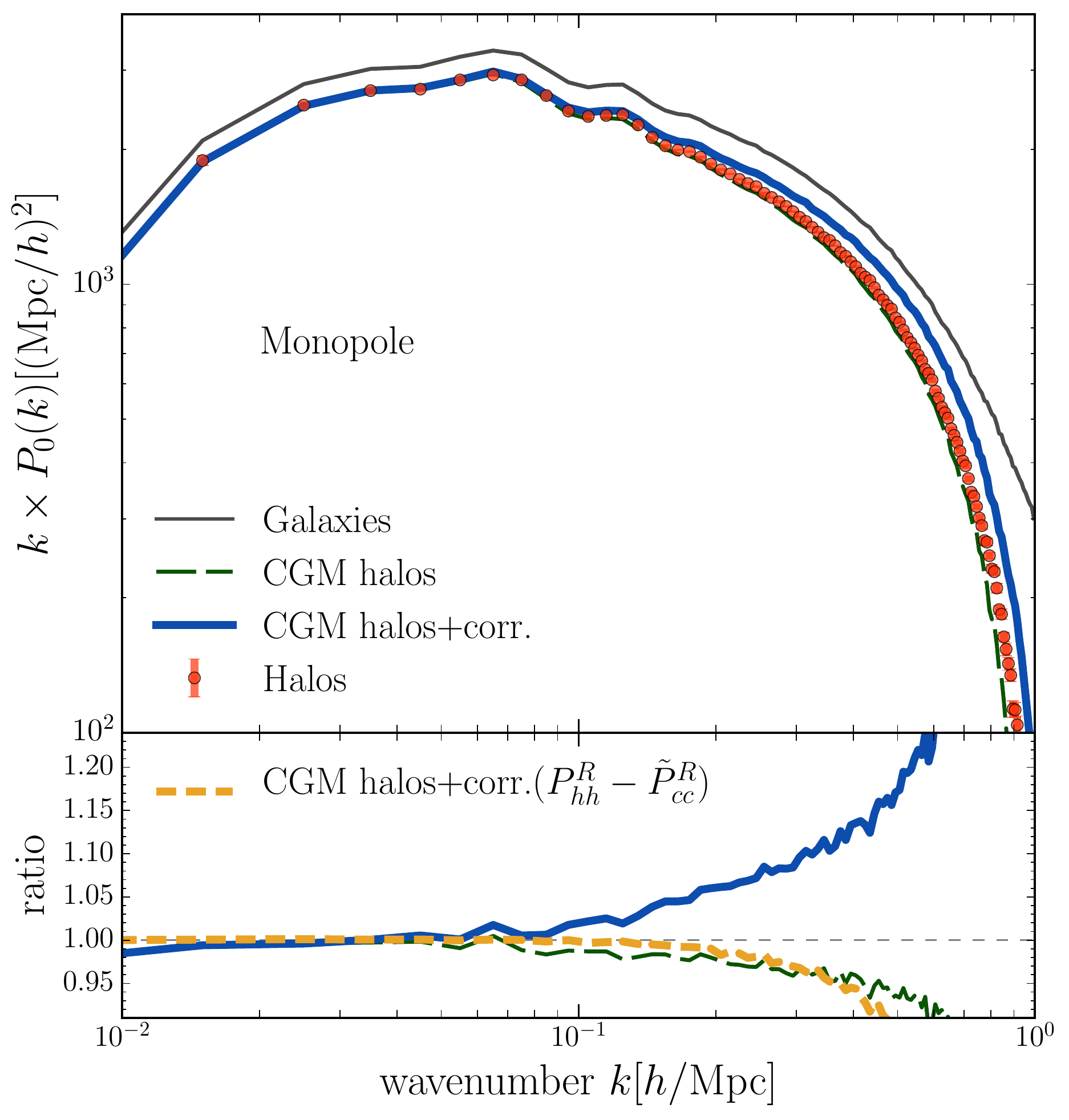}
\includegraphics[width=85mm,bb=0 0 610 570]{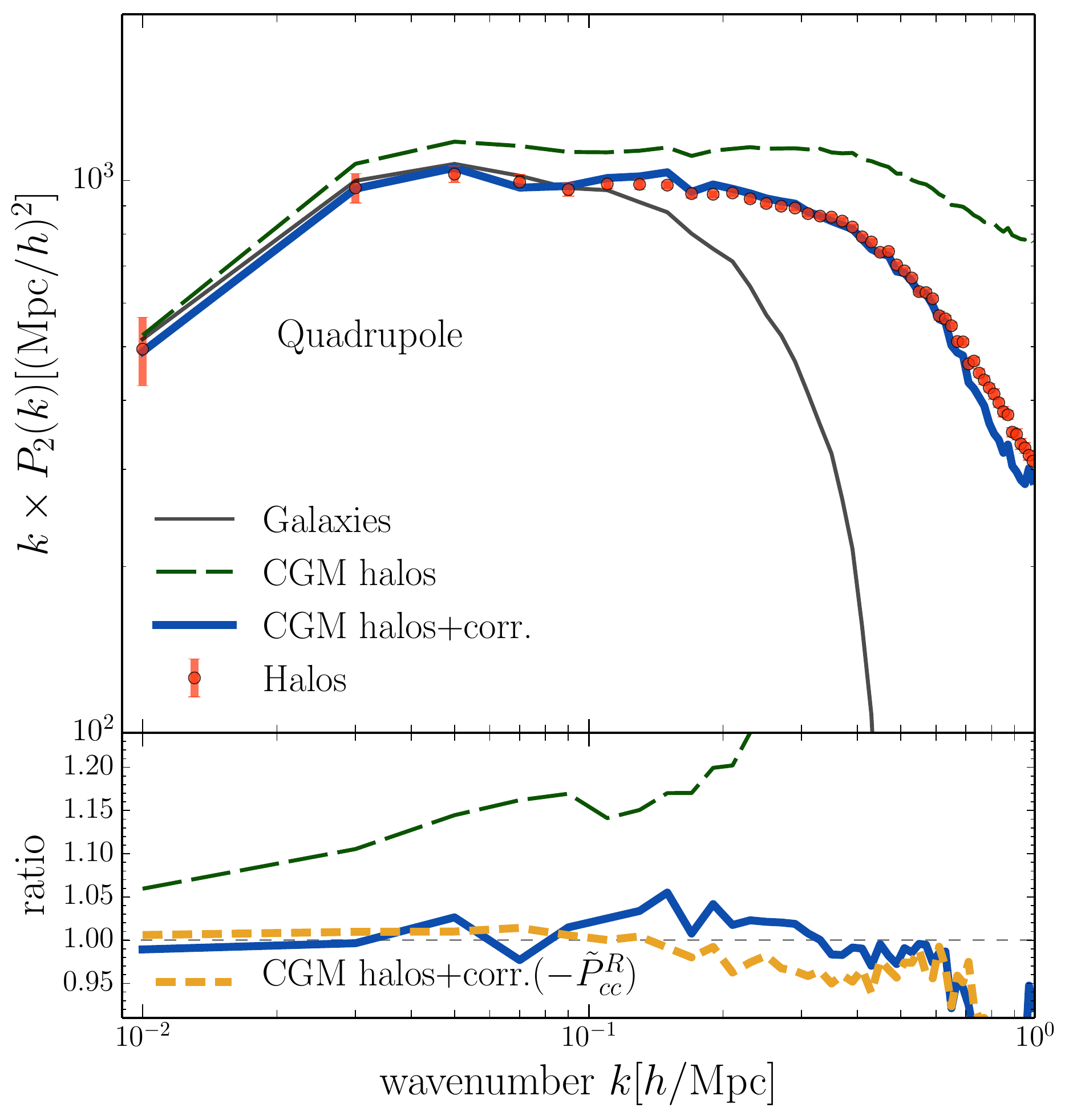}
 \caption{({\it Upper panels}) Reconstructed monopole (left) and quadrupole (right) power spectra of halos 
 in redshift space, which are given by a sum of the
 power spectrum of CGM halos and the correction terms
 (equation~\ref{eq:baldauf_mod}).  
 The bold solid curve is the result
 for our method for the monopole (left panel) and quadrupole (right)
 spectra, compared with the redshift-space spectra of the
 underlying true host halos (red points). We assumed the same model in real space,
 given in Fig.~\ref{fig:pk_quad_window_r_9.2e-5}, to
  compute the correction terms, hence did not introduce any
 additional free parameter to compute the model prediction. 
For comparison, the thin, green dashed
curve shows the results for CGM halos without the correction terms, and the
 gray solid curves are the spectra of mock galaxies. 
 ({\it Lower panels}) Ratios of the power spectrum of reconstructed halos to that of true halos. 
The green and blue curves are for the spectrum without and with the correction, 
corresponding to the curves with the same color in the upper panels.
 The yellow, dashed curve is similar to the blue solid line but the result 
 with the large-scale correction term determined by the non-parametric way.
 }
 \label{fig:pk_s_9.2e-5_corr_f}
\end{center}
\end{figure*}

The blue dashed curve in the left panel of
Fig.~\ref{fig:pk_quad_window_r_9.2e-5} shows the best-fitting model of
equation~(\ref{eq:Delta_p2r}), where we have estimated the best-fitting
parameter $\alpha_{\mu^2}$ by fitting equation~(\ref{eq:Delta_p2r}),
multiplied by negative sign ($-1$), to the quadrupole spectrum of CGM
halo power spectrum, over the range $k\in[0.01, 0.05]\hmpci$. We
obtain $\alpha_{\mu^2}=0.503$ for our fiducial model. The bold, black
curve in the figure shows the total power of the window terms
(equation~\ref{eq:w_p2}) and the correction power spectrum
(equation~\ref{eq:Delta_p2r}) which fairly well reproduces the overall
shape of the apparent quadrupole power spectrum of CGM halos, although
there is a small but residual systematic offset. Note that the physical RSD effect
caused by coherent peculiar motions of halos is much greater than the
apparent power that we discussed here, by about factor of 10 (see Fig.~\ref{fig:pkr_vs_pks}).
The result for a larger cylinder window $(\Delta r_\perp,\Delta
r_\parallel)=(1.5,30) [\himpc]$, instead of our fiducial values
$(\Delta r_\perp,\Delta r_\parallel)=(1.5,15) [\himpc]$, will be
presented in Appendix~\ref{sec:changing_pi} and
Fig.~\ref{fig:pk_quad_window_r_9.2e-5_pi30}.

The simple form for $\alpha(\mu)$ can be justified using the right
hand panel of Fig.~\ref{fig:pk_quad_window_r_9.2e-5}, which shows the
apparent hexadecapole power of the real-space CGM halo spectrum, that
arises due to the anisotropic shape of the cylinder. If the
correlation of misidentified CGM halos produced anisotropies
of order higher than $\mu^2$, it would generate a non-zero
hexadecapole on large scales. However, the measured CGM halo
hexadecapole is consistent with a null signal at $k<0.2\hmpci$. This
implies $\Delta P_4^R=0$, thus provides a justification of the truncation of 
the coefficients, $\alpha_{\mu^{2L}}=0$ ($L\geq 2$).
Note that large scatter at different $k$ bins is due to the discrete
summation over $\mu$ to compute the hexadecapole spectrum for a given
Fourier resolution. The black solid curve is the sum of the two window
terms multiplied by $(-1)$, and it explains the measured real-space
hexadecapole very well.


\subsection{Power spectra of CGM halos in redshift space}\label{sec:cgm_halo_power}

We now show the main results of this paper; we study how the
CGM-constructed halo power spectrum, with the correction terms in
equation~(\ref{eq:baldauf_mod}), can reproduce the redshift-space power
spectrum of the underlying halos. In this subsection we consider the
low-density mock sample, and show the results for the high density
sample in Appendix~\ref{sec:high_density}.

We first discuss the correction terms for the CGM-halo spectrum in
redshift space. For the window terms, $W$ and $[W*P]$, we can directly
compute the multipole power spectra as presented in
Fig.~\ref{fig:pk_quad_window_r_9.2e-5}.  For the apparent anisotropic
clustering on large scales, we simply assume that the apparent
anisotropy arises from the intrinsic distribution of galaxies in
large-scale structure, compared to the CGM technique. That is, we 
assume that the same correction term (equation~\ref{eq:delta_pr}), derived for the
real-space quadrupole spectrum of CGM halos in
Fig.~\ref{fig:pk_quad_window_r_9.2e-5}, can correct for the large-scale apparent
clustering even in redshift space:
\begin{eqnarray}
\Delta P^S(\vk)&\simeq&\Delta P^R(\vk). \label{eq:Delta_ps}
\end{eqnarray}
This term involves two free parameters for the fiducial cosmological
model: $\alpha_{\mu^0}$ and $\alpha_{\mu^2}$ for the $\alpha$ function.
We now estimate the best-fitting $\alpha_{\mu^0}$ in a similar manner
to our real space analysis such that the model matches
the residuals between the real-space monopole power spectra of the true halos and
the CGM halos on large scales, $0.01<k<0.05\hmpci$. We obtain the best-fitting
parameters $\alpha_{\mu^0}=0.246$ (and $\alpha_{\mu^2}=0.503$).
Thus, to test our model for the redshift-space halo power spectrum, we
do not introduce any additional free parameter. 

In Fig.~\ref{fig:pk_s_9.2e-5_corr_f}, we compare the monopole and
quadrupole power spectra of halos in redshift space with the model
prediction (equation~\ref{eq:baldauf_mod}).  To compute these
redshift-space power spectra, we use the peculiar velocities of halos
hosting the CGM central galaxies for simplicity; in other words we
ignore possible internal motion of the CGM central galaxy
inside the host halo. We will discuss the FoG effect of CGM central 
galaxies in Appendix~\ref{sec:off-centering}.  

\begin{figure}
\begin{center}
 \includegraphics[width=85mm,bb=0 0 610 570]{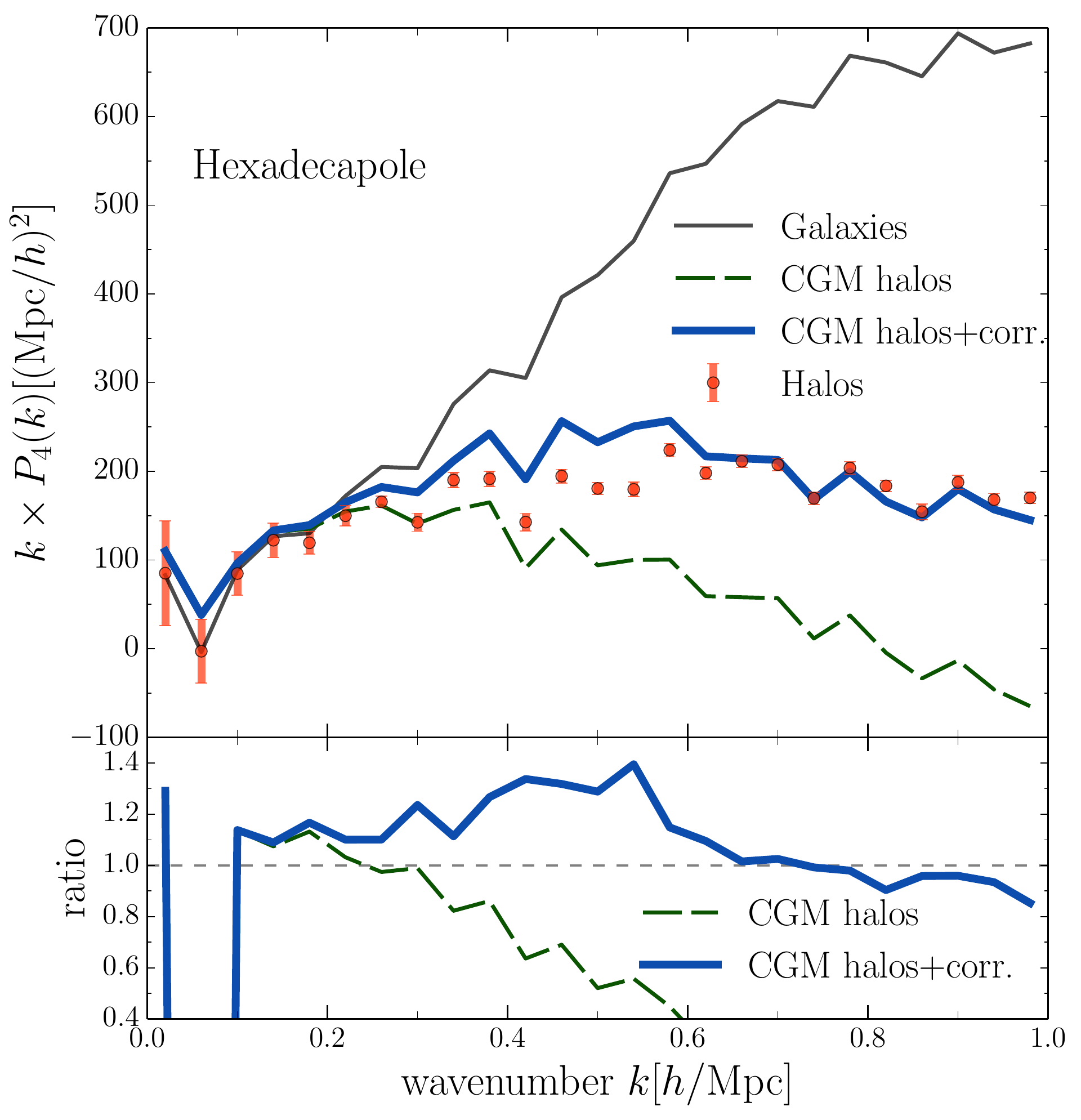}
 \caption{Similar to the previous figure, but for the hexadecapole power spectrum. 
}
\label{fig:p4_s_9.2e-5_corr_f}
\end{center}
\end{figure}

Let us first focus on the quadrupole power spectrum (right panel),
which is important in order to measure the RSD effect. The upper thin, green
dashed curve shows the spectrum for the CGM halos, i.e. the spectrum
obtained from a naive implementation of the CGM halo reconstruction.
This spectrum leads to a systematic overestimation of the underlying
halo power spectrum over all the wavenumbers plotted; the
deviation is about 15\% in power at $k\simeq 0.1~\hmpci$ and becomes
$\simgt 20\%$ at $k\simgt 0.2~\hmpci$, as seen in the bottom panel. 
On the other hand, the bold blue curve is the result for the quadrupole power spectrum of
equation~(\ref{eq:baldauf_mod}) including the CGM window correction
(equation~\ref{eq:w_p2}) and the large-scale correction
(equation~\ref{eq:Delta_ps}). This is now in excellent agreement with
the power spectrum of the underlying halos to within $\sim5$\% in the
power over all the $k$ range, even up to $k\simeq 0.5~\hmpci$.  This
implies that the model we develop in this paper can predict the
quadrupole power spectrum of CGM selected halos in both real and
redshift space simultaneously. This agreement is a huge improvement
compared to the quadrupole power spectrum of galaxies.

The upper left panel of Fig.~\ref{fig:pk_s_9.2e-5_corr_f} shows similar
results for the monopole power spectrum. Our model improves the
agreement with the spectrum of halos at small $k$, once the large
scale correction is included. However, the agreement in the weakly
nonlinear regime of $k\simgt 0.1~\hmpci$ is not as good as that for
the quadrupole spectrum. This could be ascribed to a residual shot
noise contribution that is caused by the difference between the number
densities of the true halos and the CGM halos (see
Table~\ref{tab:mocks_cgm}). The discrepancy at $k>0.1\hmpci$ can be
further improved by adding an additional free parameter to model the
residual shot noise \cite[e.g.,][]{Beutler:2014}.  The ratio of the
monopole and quadrupole power spectra at each $k$ bin provides a way
to measure the scale-dependent RSD effect: the Kaiser factor in the
linear regime of $k\simlt 0.1~\hmpci$ and the nonlinear RSD effect at
the larger $k$ bins. These signals are direct predictions of the
$\Lambda$CDM structure formation model, and can be used to test theory
of gravity at the respective scales. The results in Fig.
~\ref{fig:pk_s_9.2e-5_corr_f} show that our model should enable us to
measure the RSD effect to within 5\% accuracy for scales up to
$k\sim$ a few $0.1~\hmpci$, and also the assumption that the large-scale 
correction to the power spectrum in redshift space is the same as 
that in real space (equation \ref{eq:Delta_ps}) is reasonable.

We further examine the hexadecapole power spectrum in
Fig.~\ref{fig:p4_s_9.2e-5_corr_f}. The hexadecapole spectrum has a
smaller power than the quadrupole for the $\Lambda$CDM model as
predicted by the Kaiser formula. However, the nonlinear RSD effect can
to some extent enhance the power \citep{Scoccimarro:2004,Taruya:2010}.
In addition, the FoG effect causes a significant power in the
hexadecapole spectrum as shown by the upper solid curve  \citep[also
see][]{Hikage:2013}. Hence, the hexadecapole spectrum at least gives a
useful consistency check of the measurement and/or the systematic
effects. The figure shows that our model including the correction
terms nicely reproduces the results from the mock catalog, over the
range of $k$ we consider.  As is the case for the real-space
hexadecapole, implementing the CG technique does not produce the
apparent anisotropy in the redshift-space hexadecapole at
$k<0.2~\himpc$. It thus implies $\Delta P^S_4 = \Delta P^R_4 =0$ and
provides another justification of the ansatz we used in equation
(\ref{eq:Delta_ps}).

In the above analysis we have determined the large-scale corrections by
parameterizing $\alpha(\mu)$ using equation (\ref{eq:alpha_mu}).  For the
quadrupole, we have used the black line in the left panel of
Fig.~\ref{fig:pk_quad_window_r_9.2e-5} as the total correction to the CGM halo
spectrum.
Now, we present an alternative non-parametric way to correct the CGM halo
spectrum. We use the difference between the observed monopole and quadrupole of
the real space CGM halo power spectrum to those of the underlying halo power
spectrum as a correction for the monopole and quadrupole of the CGM halo power
spectrum in redshift space. This corresponds to using the red points for the
quadrupole correction. The procedure corresponds to the following model:
\bey
P_{\rm hh,0}^S(k) &\simeq& \wt{P}_{\rm cc,0}^S(k) + P_{\rm hh,0}^R(k) - \wt{P}_{\rm cc,0}^R(k),\\
P_{\rm hh,2}^S(k) &\simeq& \wt{P}_{\rm cc,2}^S(k) - \wt{P}_{\rm cc,2}^R(k).
\eey
The results of using such a non-parametric correction for the monopole and
quadrupole spectra are shown using the yellow dashed lines in the lower left and
right panels of Fig.~\ref{fig:pk_s_9.2e-5_corr_f}, respectively.  While there
is a slight improvement for the monopole compared to the parametric case
presented by the blue solid line, the accuracy for the quadrupole is similar to
the one in the parametric case. The agreement confirms that the
large-scale apparent
anisotropic clustering arises mainly from the intrinsic distribution of
galaxies. 

Note that the correction term for the large-scale anisotropy is modeled using 
the real-space multipole power spectra (equation \ref{eq:Delta_ps}), which are not an observable in 
real galaxy surveys. Although the contribution of the large-scale anisotropy from the mis-identified halos 
is only about 15\% and 20\% of the RSD quadrupole at $k=0.15\hmpci$ and $0.2\hmpci$, respectively, 
the cosmological information would be reduced if the corrections parameters $\alpha_{\mu^0}$ and $\alpha_{\mu^2}$ 
were treated as free parameters and marginalized over. 
Instead, we propose to use the mock catalog in order to derive priors on
the correction $\alpha(\mu)$ in real galaxy surveys
(see Fig.~\ref{fig:pk_quad_window_r_9.2e-5}).
This procedure will be tested in detail and applied to the real data in our future work.

Note also that the accuracy of our model to reconstruct the halo power spectrum depends on the choice of the cylinder size, particular its height, $\Delta r_\parallel$. The size needs to be determined by taking into account the number density of the given galaxy sample as well as properties of the galaxies. 
In order to see the sensitivity of the choice of $\Delta r_\parallel$ on the final result, 
we repeated the same analysis by adopting the size much larger than the RSD displacement scale, 
$\Delta r_\parallel=30\himpc$ in Appendix \ref{sec:changing_pi}.
By comparing the right panel of Fig. \ref{fig:pk_s_9.2e-5_corr_f} and Fig. \ref{fig:p2_s_9.2e-5_corr_f_pi30}, 
one can see that the accuracy of our model is not very sensitive to the choice of the cylinder size. 
Nevertheless, we will use mock catalogs to calibrate the appropriate cylinder size when we analyze the real data.


\section{Discussion and conclusion} \label{sec:conclusion}

In this paper we have developed a novel method for reconstructing the
power spectrum of dark matter halos from the observed galaxy
distribution in redshift space. The central tenet of our method is
that satellite galaxies are a nuisance to infer cosmological
parameters from galaxy clustering. Therefore, we remove satellite
galaxies from the sample used to measure the anisotropic clustering of
galaxies. This minimizes the 1-halo term contribution as well as
the Finger-of-God (FoG) effect due to virial
motions of satellite galaxies in massive halos, and makes the measured
clustering signal more sensitive to the typical halos hosting the
central galaxies of the sample.

Our method relies on the cylinder-grouping (CG) method, which sorts
galaxies in a given catalog in descending order of their luminosity or
stellar masses, places a cylinder around each galaxy from the top of
the sorted list and regards all the galaxies inside the cylinder as
member galaxies of a halo.  This is similar but simpler than the
redshift space FOF method used by \cite{Reid:2009}.
Since the line-of-sight length of cylinder has to be long enough to
cover a possible large displacement due to the virial motions of
galaxies, our fiducial choice of $\Delta r_\parallel=15~\himpc$ seems
sufficient to remove most of the satellite galaxies around massive
halos. However, we found that our method generates an apparent
anisotropy in the distribution of CGM-selected halos due to the
anisotropy of the cylinder used to identify satellite galaxies.
We showed that this apparent anisotropy can be corrected for
(equation~\ref{eq:baldauf_mod}), using the cylindrical window function
on small scales as well as the linear matter power spectrum with
coefficients up to the order $\mu^2$ on large scales.

We tested the accuracy of our method using mock galaxy catalogs at
$z=0.5$, which are constructed from $N$-body simulations to mimic the
SDSS-III BOSS survey.  We first studied the quadrupole power spectrum
of CGM halos in {\em real space}, which is caused by the apparent
anisotropic clustering that we discussed above. We showed that the
correction terms involving the cylindrical window can explain most of
the quadrupole power at $k\simgt 0.1~\hmpci$. On the other hand, the
apparent clustering at smaller $k$ can be explained by our model, by
adjusting the model parameter ($\alpha_{\mu^2}$ for the quadrupole
power spectrum), as shown in Figs.~\ref{fig:pk_quad_window_r_9.2e-5}.
The sum of these correction terms can fairly well reproduce the
apparent quadrupole power. Next, we showed that our model can also
reproduce the redshift-space quadrupole power spectrum of the
underlying halos, to an accuracy to within $5\%$ in the power up to
$k\simeq 0.5 \hmpci$. To arrive at this result, we did not introduce
any additional parameter: we assumed that the large-scale apparent
clustering is given by the above correction function in real space.
Even if we allow model parameters ($\alpha_{\mu^0}$ and
$\alpha_{\mu^2}$) to freely vary, we found a similar-level agreement.
The correction term is about 10\% in the quadrupole power at $k\sim
0.1~\hmpci$, while the correction terms involving the cylindrical
window give about 40\% at $k\sim 0.5~\hmpci$, for a galaxy mock
catalog (with number density of $\bar{n}_g\sim 10^{-4}~(\hmpci)^3$,
similar to LRGs or massive CMASS galaxies). If we use a galaxy sample
with a higher number density, the contributions of the correction
terms are greater.

The agreement of our model with the underlying halo power spectrum is
encouraging. If we can use the measured quadrupole spectrum up to the
weakly nonlinear regime $k\simgt 0.1\hmpci$, it promises a significant
improvement in the RSD measurement: e.g., the clustering information
up to maximum wavenumber $k=0.2\hmpci$ is equivalent to a galaxy
survey with about factor of 8 wider area using the information up to
$k=0.1\hmpci$. However, in the weakly nonlinear regime the Kaiser
factor is no longer valid, and the RSD effect becomes scale-dependent
due to nonlinear mode coupling \citep{Scoccimarro:2004}. Nevertheless,
the halo power spectrum, if recovered, enables a direct comparison
with model halo power spectrum estimated from a suite of $N$-body
simulations. For the cosmological analysis, we need to construct a
``cosmic emulator'' of redshift-space power spectra of halos given as
a function of halo mass, redshift and cosmological models (Nishimichi
et al. in preparation). If the satellite galaxy contribution can be
shown to be negligible and robust to a variety of methods that
populate satellite galaxies in halos, one could even build emulator of
redshift space power spectra of halos, where these halos are filtered
using cylinders of the same size as used in data, allowing a direct
comparison to observations (and bypassing the need for an accurate
empirical model as constructed in this paper). The ratio of the
monopole and quadrupole spectra of halos contains a wealth of
information about the RSD effect, thereby allowing us to improve the
constraints on the growth of large-scale structure. In addition, the
scale-dependent RSD effects can be used to test theories of gravity on
these scales. We will explore these avenues in future work.

The inability of redshift surveys to obtain redshifts of closely
spaced galaxies due to fiber collisions is one of the biggest
hindrance to use RSDs for precision cosmology \citep{Lin:1996,Reid:2014}. Fiber
collisions become more important for future high-$z$ surveys because a
fixed angular scale corresponds to a larger scale at higher redshifts.
Since our CGM halo reconstruction method needs to keep only one galaxy
out of the host halo, our model is less affected by fiber collisions.

There are several effects that we ignored in this paper that should be
taken into account for an implementation of this method for real data.
We have not discussed the mis-centering of (brightest or most massive)
galaxies in their halos \citep{Hikage:2013a} \citep[see
also][]{Skibba:2011,Hikage:2012,Masaki:2013,Hoshino:2015,Saito:2016}.
This becomes especially true if the spectroscopic target galaxies are
selected based on cuts in magnitude and color space from the imaging
data. This is indeed the case for all spectroscopic surveys, and the
true central galaxies may not necessarily make the targeting cut.
The off-centered galaxies can in principle be constrained by combining
the redshift-space clustering of CGM halos with the weak gravitational
lensing signal (or cross-correlation with photometric galaxies).
Gravitational lensing can probe the matter distribution around the
true center, as proposed in \citet{Hikage:2012} \citep[also
see][]{Oguri:2011,Hikage:2013a}.

The next logical step is to apply this method to existing data from
galaxy surveys, such as the CMASS sample of BOSS survey. In order to
stress our method, we need to use a more accurate mock catalog of
the CMASS sample to calibrate possible residual systematic errors. The
mock catalog can be used to add theory priors on model parameters in
our method (the coefficient parameters to model the large-scale
apparent clustering). In particular, the footprint of the ongoing
Subaru Hyper Suprime-Cam (HSC) survey \citep{Miyazaki:2012} fully
overlaps with the BOSS survey. This will allow us to combine the
redshift-space power spectrum of CGM halos, reconstructed from the
CMASS galaxies, with the galaxy-galaxy weak lensing, as we discussed
above.  Ultimately, our goal is to apply our method to even larger,
higher-redshift galaxy surveys such as the Subaru Prime Focus
Spectrograph (PFS) survey \citep{Takada:2014}.  Since the PFS combined
with HSC promises to provide very tight constraints on cosmological
models, we need to carefully test whether the accuracy we have
obtained in this paper for the theoretical modeling scheme is
sufficient given the statistical power of these surveys. This will be
addressed in future work.

\section*{Acknowledgments}
We thank Beth Reid and Uro\v{s} Seljak for useful discussion at the
early stage of this work.  T.O. is supported by JSPS KAKENHI Grant
Number JP26887012.  This work is in part supported by  
Grant-in-Aid for Scientific Research from the JSPS Promotion of Science
(No.~23340061 and 26610058), MEXT Grant-in-Aid for Scientific Research
on Innovative Areas (No.  15H05893, 15K21733, 15H05892)
and JSPS
Program for Advancing Strategic International Networks to Accelerate the
Circulation of Talented Researchers.

\appendix

\section{Changing the cylinder size}\label{sec:changing_pi}

\begin{figure}
\begin{center}
\includegraphics[width=85mm,bb=0 0 610 570]{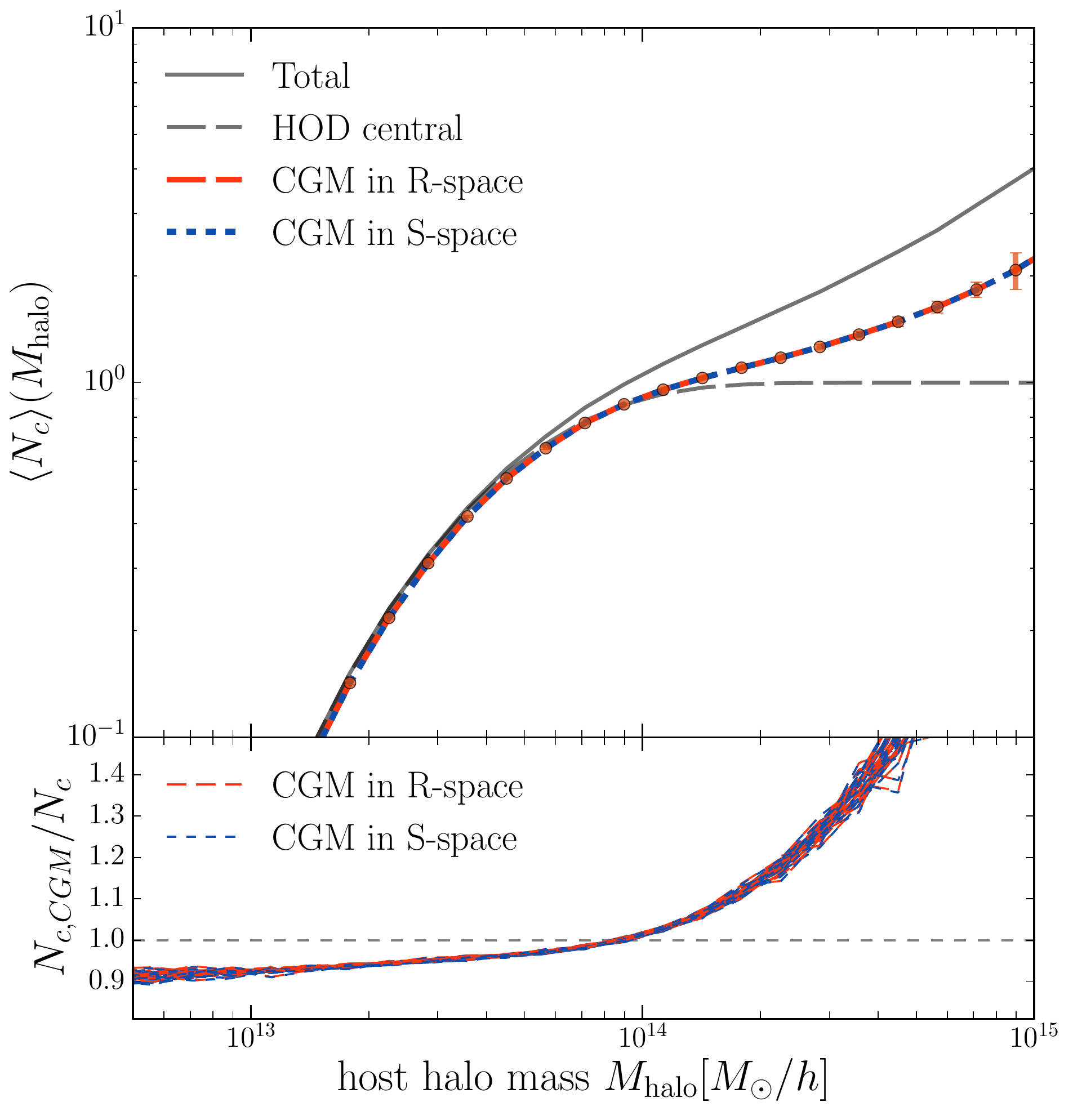}
\caption{Same as Fig.\ref{fig:hod_cic} but with the cylinder size $(\Delta r_\perp, \Delta r_\parallel)=(1.5,30)\himpc$. }
\label{fig:hod_cic_pi30}
\end{center}
\end{figure}

\begin{figure}
\begin{center}
\includegraphics[width=85mm,bb=0 0 610 460]{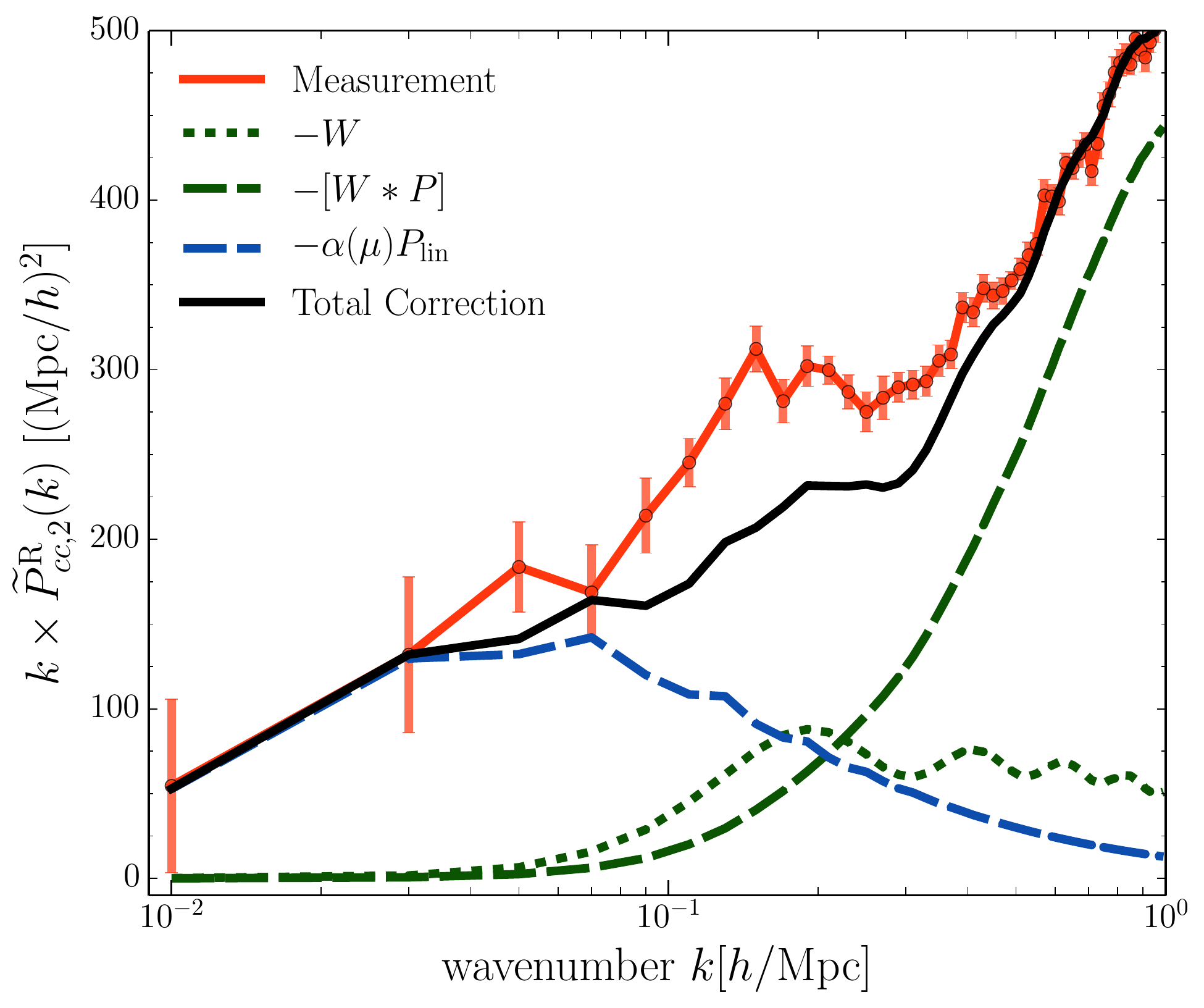}
\caption{Same as Fig.~\ref{fig:pk_quad_window_r_9.2e-5} but for the different cylinder size, $\Delta r_\parallel = 30\himpc$. }
\label{fig:pk_quad_window_r_9.2e-5_pi30}
\end{center}
\end{figure}

The size of the cylinder, namely the radius ($\Delta r_\perp$) and
height (2$\Delta r_\parallel$), is arbitrary, and depends upon the
galaxy sample being considered. These should be chosen so that the
majority of the multiple systems are properly classified as centrals
and satellites but to be as small as possible in order to avoid larger
corrections. Especially, the choice of $\Delta r_\parallel$ is
crucially important since this value determines how well we can
eliminate the FoG contamination.  Our default values, $(\Delta
r_\perp,\Delta r_\parallel)=(1.5,15)[\himpc]$, are such an optimal
choice. On the other hand, a galaxy pair in the same halo but
separated along the transverse direction has a small FoG effect, and
so imperfect grouping for such satellites does not affect the
measurements by a large amount.

Fig.~\ref{fig:dist_cylinder_galaxy} implies that there exist galaxy
pairs in the same massive halo are not identified due to a small
$\Delta r_\perp$, rather than $\Delta r_\parallel$. Thus the failure
of identifying the members of the massive halos seen in
Fig.\ref{fig:hod_cic} is not expected to cause a serious problem. In
order to explore the effects of a larger cylinder in the redshift
direction, we adopt a cylinder height $\Delta r_\parallel=30\himpc$,
and repeat the same analysis for the low-density galaxy sample.
Fig.~\ref{fig:hod_cic_pi30} is the same as Fig.~\ref{fig:hod_cic} but
with $\Delta r_\parallel=30\himpc$. There is almost no difference
between the two cases, implying that $\Delta r_\parallel=15\himpc$ is
large enough to identify the galaxies with the radial displacements by
the FoG and the the behavior of $N_{\rm c,CGM}/N_{\rm c}>1$ for large
$M_{\rm halo}$ is due to the galaxy pairs along the tangential
direction.

\begin{figure}
\begin{center}
\includegraphics[width=85mm,bb=0 0 610 570]{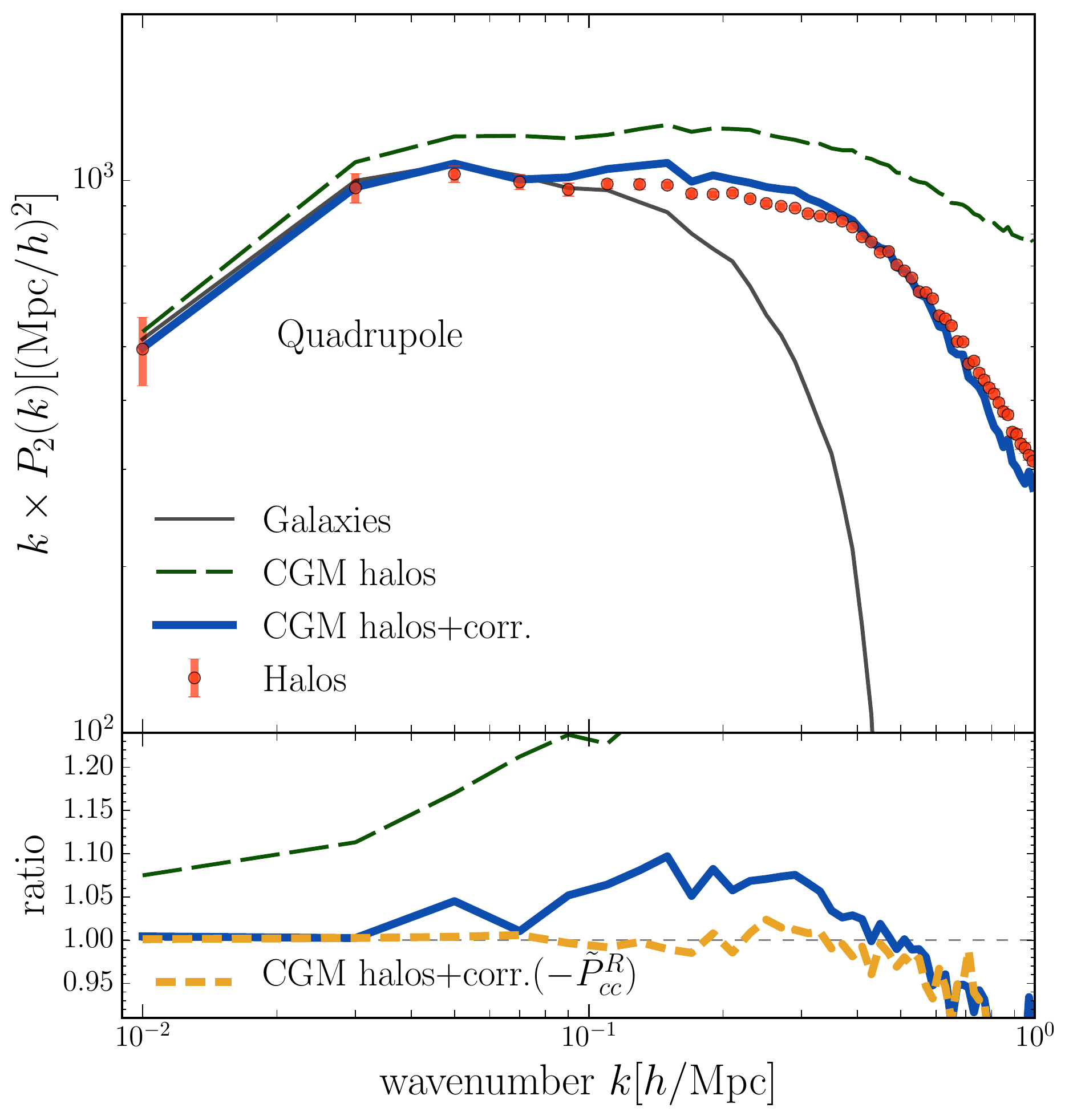}
\caption{Same as Fig.~\ref{fig:pk_s_9.2e-5_corr_f} but for the different cylinder size, $\Delta r_\parallel = 30\himpc$. }
\label{fig:p2_s_9.2e-5_corr_f_pi30}
\end{center}
\end{figure}

Fig.~\ref{fig:pk_quad_window_r_9.2e-5_pi30} plots the real-space
quadrupole power spectrum, the same as
Fig.~\ref{fig:pk_quad_window_r_9.2e-5} but with $\Delta
r_\parallel=30\himpc$. Due to the larger cylinder height, the
exclusion effect is more significant. Just like
Fig.~\ref{fig:pk_quad_window_r_9.2e-5}, the sum of the window function
$W$ and its convolution $[W*\wt{P}]$ as depicted by the solid black
line captures the measured real-space quadrupole while there is a
systematic offset at $k>0.1\hmpci$. The final result of our correction
model for the redshift-space quadrupole is shown in
Fig.~\ref{fig:p2_s_9.2e-5_corr_f_pi30}. As seen in the bottom panel,
our model presented by the solid blue line achieves the accuracy of
better than $5$-$7\%$ on scales $k<1\hmpci$ although we adopt such an
extremely large cylinder size. Furthermore, the model with the
correction terms directly determined from the real-space measurement,
shown as the yellow dashed line, has almost the same accuracy as the
result with $\Delta r_\parallel = 15\himpc$.

\section{Results for high-density sample}\label{sec:high_density}
\begin{figure*}
\begin{center}
\includegraphics[width=85mm,bb=0 0 610 570]{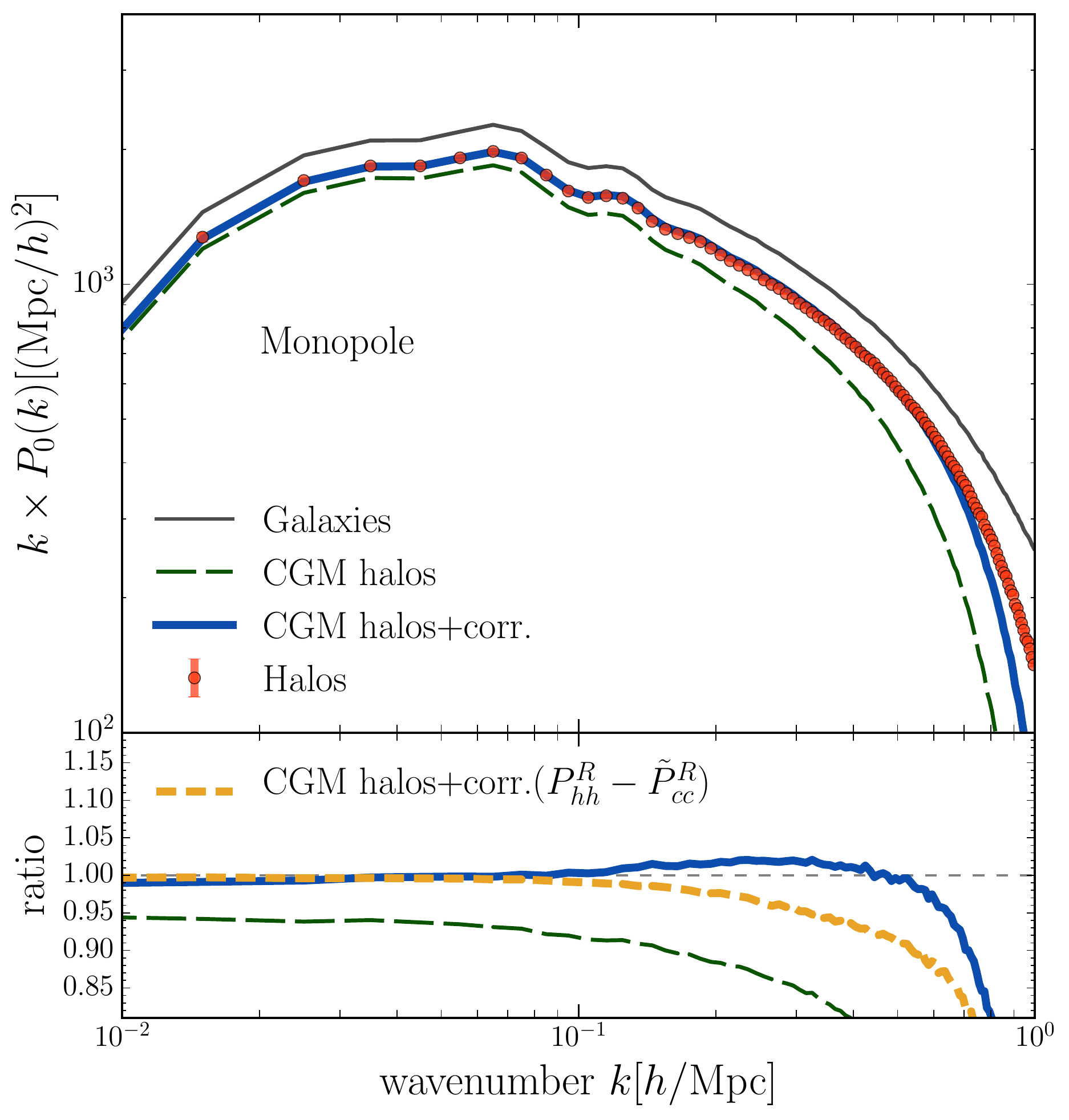}
\includegraphics[width=85mm,bb=0 0 610 570]{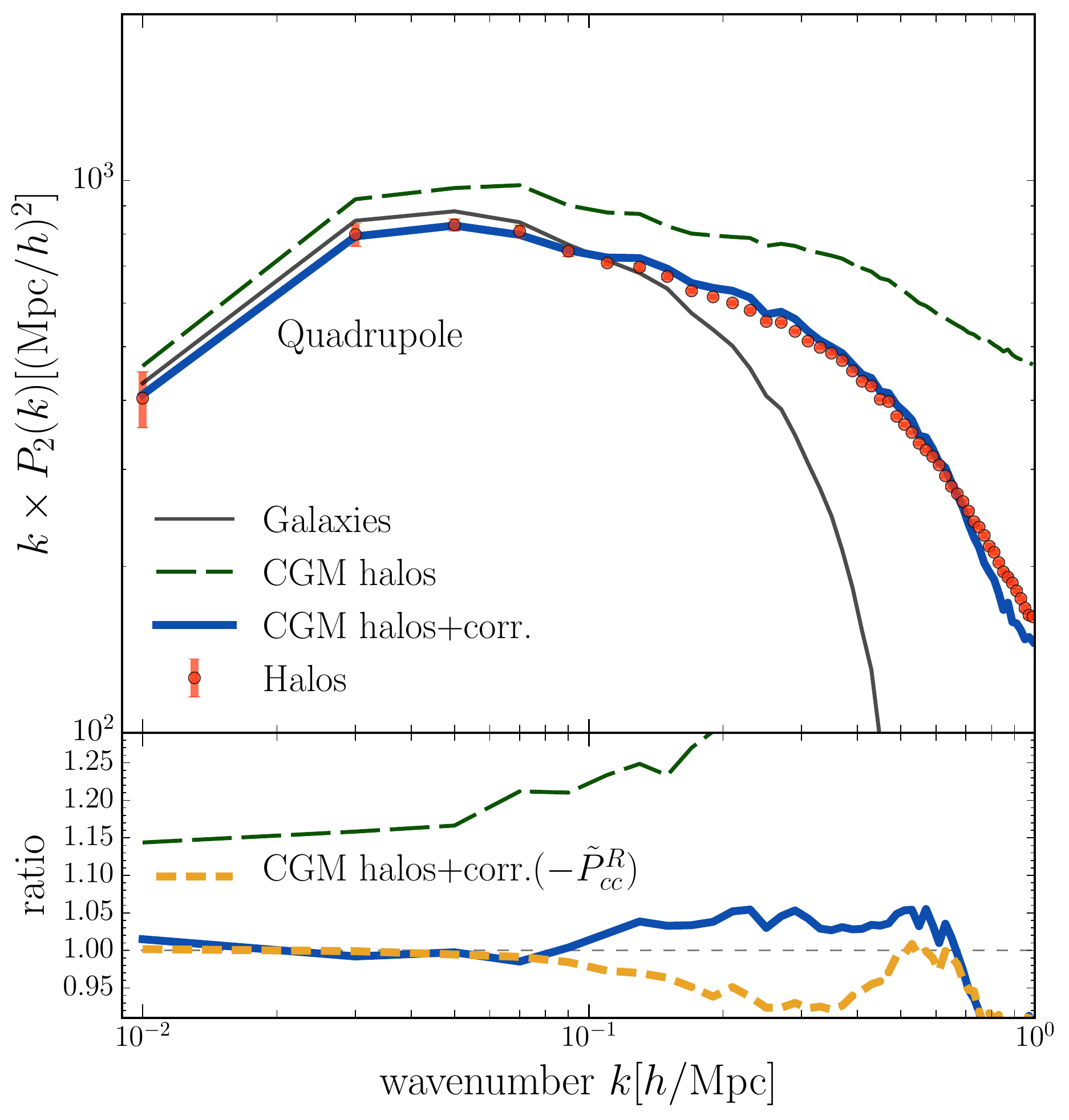}
\caption{Same as \ref{fig:pk_s_9.2e-5_corr_f}, but for the higher density sample.}
\label{fig:pk_s_3.5e-4_corr_f}
\end{center}
\end{figure*}

The performance of our correction model depends on the number density of
the galaxy sample.  As we have seen in Section~\ref{sec:cgm_rec},
larger corrections are required for a higher-density sample.  In this
appendix we present the results when we apply the same technique to the
high-density mock sample with $\bar{n}=3.4 \times 10 ^{-4} (\hmpci)^3$,
roughly corresponding to the full BOSS CMASS galaxy sample without any
stellar mass cut.

The results for the high-density sample are shown for the monopole
(left) and quadrupole (right) spectra in
Fig.~\ref{fig:pk_s_3.5e-4_corr_f}.  For the monopole, the CGM halo
spectrum without the correction terms has $>5\%$ offset from the true
halo spectrum even on large scales.  However, this can be corrected
for by our model to within $5\%$ accuracy up to $k\sim 0.3\hmpci$ as
shown in the bottom left panel.
The coefficients of equations (\ref{eq:Delta_p0r}) and
(\ref{eq:Delta_p2r}) for the high-density sample are,
$\alpha_{\mu^0}=0.399$ and $\alpha_{\mu^2}=0.594$. For the
quadrupole, there is $\sim 20\% $ offset of the CGM halo spectrum
without correction from the true halo spectrum at $k<0.1\hmpci$ and it
blows up to $30 \%$ at $k\sim 0.2\hmpci$.  Our correction model can
improve the results to a large extent, but a discrepancy of $\sim 10\%$
remains if we adopt the non-parametric model for the large-scale correction, 
as shown by the gray dashed line at the right bottom panel
of Fig.~\ref{fig:pk_s_3.5e-4_corr_f}.  
In comparison to Fig.~\ref{fig:pk_s_9.2e-5_corr_f},
one can see that the accuracy of our correction model certainly
depends on the number density of the sample and the lower-density
sample gives more precise results as expected.  

\section{Effect of off-centered galaxies}\label{sec:off-centering}

\begin{figure*}
\begin{center}
\includegraphics[width=85mm,bb=0 0 610 570]{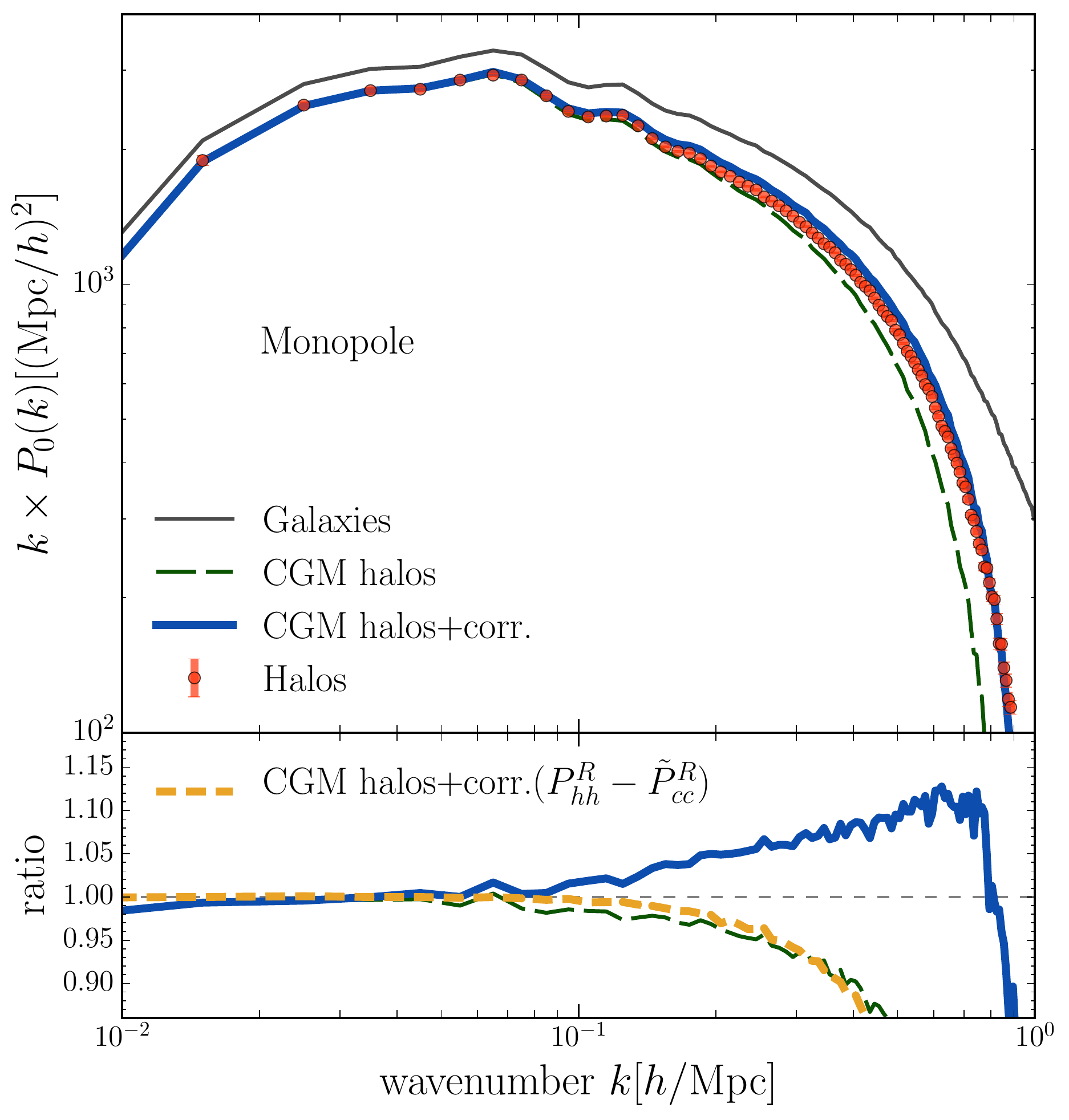}
\includegraphics[width=85mm,bb=0 0 610 570]{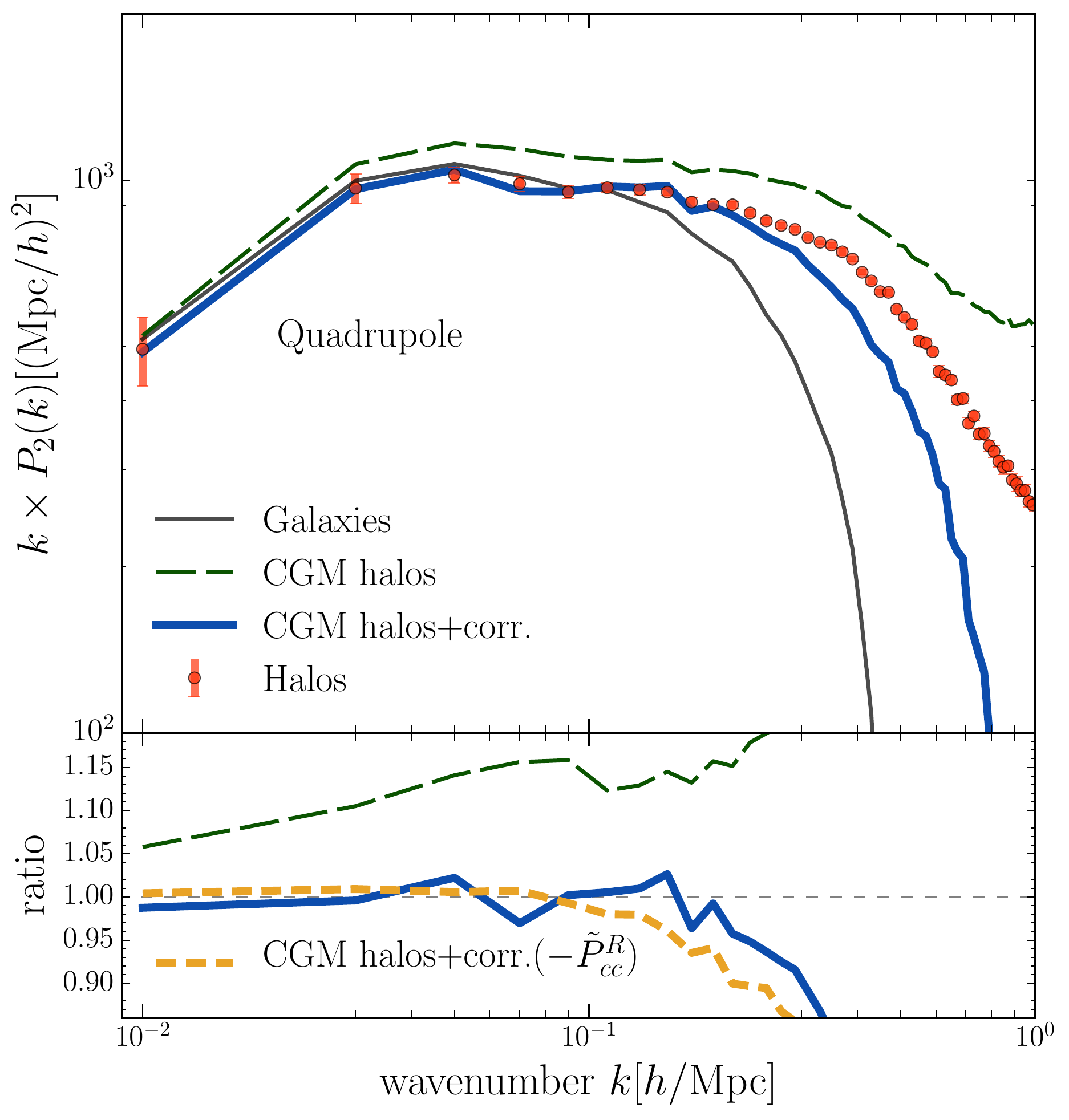}
\caption{Same as Fig.~\ref{fig:pk_s_9.2e-5_corr_f} but the off-centering effect of central galaxies are not corrected and the results compared to the results of central galaxies. As a fair comparison, the red points at the top panels here include the residual FoG effect from the central galaxies which have offset. }
\label{fig:pk_s_corr_fog_f}
\end{center}
\end{figure*}

The main results presented in Fig.~\ref{fig:pk_s_9.2e-5_corr_f}
assumed that the off-centering effect of central galaxies can be
perfectly corrected for using the galaxy-galaxy lensing. In this
appendix we focus on the situation that we analyze only the galaxy
clustering data, i.e., without any knowledge or constraints on the
off-centering effect.

The results for the monopole and quadrupole without correction for the
off-centering effect are shown in the top panels of
Fig.~\ref{fig:pk_s_corr_fog_f}.  The thin gray lines are the original
galaxy power spectra, the same as those presented in
Fig.~\ref{fig:pk_s_9.2e-5_corr_f}. The red points are the power
spectrum of halos, but to make a fair comparison, this power spectrum
includes the residual FoG effect due to miscentering unlike that in
Fig.~\ref{fig:pk_s_corr_fog_f}. Thus it is equivalent to the power
spectrum of central galaxies $P_{\rm cc}^S$, a small fraction of which
are actually not at the center of the halos. The green dashed lines
are the multipoles of the CGM halos, which are slightly suppressed at
small scales relative to the thin lines in
Fig.~\ref{fig:pk_s_corr_fog_f} due to the residual FoG effect. Thus,
the reconstructed halo power spectra are suppressed by a similar
amount, as shown as the thick dashed lines. Here the large-scale
correction $\alpha(\mu)P^R_{\rm lin}(k)$ is the same and only the
differences arise from the CGM window and its convolution with the
underlying power spectrum.

The bottom panels of Fig.~\ref{fig:pk_s_corr_fog_f} show the ratios of
the reconstructed spectra of halos to those of central galaxies.  The
denominators include the FoG from off-centered galaxies, while the
numerators have both the FoG effect as well as the residual FoG from the
satellites which were failed to captured by the CGM. Thus the ratios
are suppressed relative to the case where the FoG effect is assumed to
be perfectly corrected for using galaxy-galaxy lensing as shown at the
bottom panels of Fig.~\ref{fig:pk_s_9.2e-5_corr_f}. Even though we do
not correct for the FoG, a similar precision can obtained for the halo
power spectrum reconstruction. 

\section{Velocity bias arising from nonlinear transformation}\label{sec:velocity_bias}

\citet{Seljak:2012} showed that any nonlinear transformation of data in redshift space would create
velocity bias. This is exactly the case for our model prescription since placing cylinders on redshift-space galaxy distribution to identify CGM halos is also a nonlinear transformation, $\delta_{\rm g}^S \to \wt{\delta}_{\rm h}^S =F (\delta_{\rm g}^S ) $.
To see if the CG method produces the velocity bias, we consider to modify Eq. (\ref{eq:baldauf_mod}) and repeat the analysis as follows. 
First, the cylinders are placed at the positions of galaxies in redshift space. 
We then measure $\wt{P}_{cc}^S$ and 
replace the convolution term in equation (\ref{eq:phh_decom_cgm}) by $\left[W(\vk)*\wt{P}_{cc,0}^S(k)\right]$.
We determine the large-scale coefficient parameters $\alpha_{\mu^0}$ and $\alpha_{\mu^2}$ by comparing with the real-space multipoles of the CGM halos
(note again that the cylinders are identified in redshift space).
The determined coefficients are used to model the reconstructed halo power spectrum in redshift space. 
In this way, the reconstructed spectrum would inevitably contain the velocity bias according to \citet{Seljak:2012}, 
and the modeling result would be biased compared to the result shown in Section \ref{sec:cgm_halo_power}.

The coefficients are determined to exactly the same values as our default case, $\alpha_{\mu^0}=0.246$ and $\alpha_{\mu^2}=0.503$. We find that the accuracy of the reconstructed quadrupole spectrum of halos is within $6\%$, almost the same as the accuracy of $5\%$ for the default case. We thus conclude that the velocity bias arising from the CG method is negligibly small. However, we will investigate this effect in more detail in our future work. 

\bibliographystyle{mnras} 
\bibliography{ms.bbl}

\begin{thebibliography}{}
\makeatletter
\relax
\def\mn@urlcharsother{\let\do\@makeother \do\$\do\&\do\#\do\^\do\_\do\%\do\~}
\def\mn@doi{\begingroup\mn@urlcharsother \@ifnextchar [ {\mn@doi@}
  {\mn@doi@[]}}
\def\mn@doi@[#1]#2{\def\@tempa{#1}\ifx\@tempa\@empty \href
  {http://dx.doi.org/#2} {doi:#2}\else \href {http://dx.doi.org/#2} {#1}\fi
  \endgroup}
\def\mn@eprint#1#2{\mn@eprint@#1:#2::\@nil}
\def\mn@eprint@arXiv#1{\href {http://arxiv.org/abs/#1} {{\tt arXiv:#1}}}
\def\mn@eprint@dblp#1{\href {http://dblp.uni-trier.de/rec/bibtex/#1.xml}
  {dblp:#1}}
\def\mn@eprint@#1:#2:#3:#4\@nil{\def\@tempa {#1}\def\@tempb {#2}\def\@tempc
  {#3}\ifx \@tempc \@empty \let \@tempc \@tempb \let \@tempb \@tempa \fi \ifx
  \@tempb \@empty \def\@tempb {arXiv}\fi \@ifundefined
  {mn@eprint@\@tempb}{\@tempb:\@tempc}{\expandafter \expandafter \csname
  mn@eprint@\@tempb\endcsname \expandafter{\@tempc}}}

\bibitem[\protect\citeauthoryear{{Baldauf}, {Seljak}, {Smith}, {Hamaus}  \&
  {Desjacques}}{{Baldauf} et~al.}{2013}]{Baldauf:2013}
{Baldauf} T.,  {Seljak} U.,  {Smith} R.~E.,  {Hamaus} N.,   {Desjacques} V.,
  2013, \mn@doi [\prd] {10.1103/PhysRevD.88.083507}, \href
  {http://adsabs.harvard.edu/abs/2013PhRvD..88h3507B} {88, 083507}

\bibitem[\protect\citeauthoryear{{Baldauf}, {Codis}, {Desjacques}  \&
  {Pichon}}{{Baldauf} et~al.}{2016}]{Baldauf:2016}
{Baldauf} T.,  {Codis} S.,  {Desjacques} V.,   {Pichon} C.,  2016, \mn@doi
  [\mnras] {10.1093/mnras/stv2973}, \href
  {http://adsabs.harvard.edu/abs/2016MNRAS.456.3985B} {456, 3985}

\bibitem[\protect\citeauthoryear{{Behroozi}, {Wechsler}  \& {Wu}}{{Behroozi}
  et~al.}{2013}]{Behroozi:2013}
{Behroozi} P.~S.,  {Wechsler} R.~H.,   {Wu} H.-Y.,  2013, \mn@doi [\apj]
  {10.1088/0004-637X/762/2/109}, \href
  {http://adsabs.harvard.edu/abs/2013ApJ...762..109B} {762, 109}

\bibitem[\protect\citeauthoryear{{Berlind} et~al.,}{{Berlind}
  et~al.}{2006}]{Berlind:2006}
{Berlind} A.~A.,  et~al., 2006, \mn@doi [\apjs] {10.1086/508170}, \href
  {http://adsabs.harvard.edu/abs/2006ApJS..167....1B} {167, 1}

\bibitem[\protect\citeauthoryear{{Bernardeau}, {Colombi}, {Gazta{\~n}aga}  \&
  {Scoccimarro}}{{Bernardeau} et~al.}{2002}]{Bernardeau:2002}
{Bernardeau} F.,  {Colombi} S.,  {Gazta{\~n}aga} E.,   {Scoccimarro} R.,  2002,
  \mn@doi [\physrep] {10.1016/S0370-1573(02)00135-7}, \href
  {http://adsabs.harvard.edu/abs/2002PhR...367....1B} {367, 1}

\bibitem[\protect\citeauthoryear{{Beutler} et~al.,}{{Beutler}
  et~al.}{2014}]{Beutler:2014}
{Beutler} F.,  et~al., 2014, \mn@doi [\mnras] {10.1093/mnras/stu1051}, \href
  {http://adsabs.harvard.edu/abs/2014MNRAS.443.1065B} {443, 1065}

\bibitem[\protect\citeauthoryear{{Bryan} \& {Norman}}{{Bryan} \&
  {Norman}}{1998}]{Bryan:1998}
{Bryan} G.~L.,  {Norman} M.~L.,  1998, \mn@doi [\apj] {10.1086/305262}, \href
  {http://adsabs.harvard.edu/abs/1998ApJ...495...80B} {495, 80}

\bibitem[\protect\citeauthoryear{{Cole}, {Fisher}  \& {Weinberg}}{{Cole}
  et~al.}{1994}]{Cole:1994}
{Cole} S.,  {Fisher} K.~B.,   {Weinberg} D.~H.,  1994, \mnras, \href
  {http://adsabs.harvard.edu/abs/1994MNRAS.267..785C} {267, 785}

\bibitem[\protect\citeauthoryear{{Cooray} \& {Sheth}}{{Cooray} \&
  {Sheth}}{2002}]{Cooray:2002}
{Cooray} A.,  {Sheth} R.,  2002, \mn@doi [\physrep]
  {10.1016/S0370-1573(02)00276-4}, \href
  {http://adsabs.harvard.edu/abs/2002PhR...372....1C} {372, 1}

\bibitem[\protect\citeauthoryear{{Davis}, {Efstathiou}, {Frenk}  \&
  {White}}{{Davis} et~al.}{1985}]{Davis:1985}
{Davis} M.,  {Efstathiou} G.,  {Frenk} C.~S.,   {White} S.~D.~M.,  1985,
  \mn@doi [\apj] {10.1086/163168}, \href
  {http://adsabs.harvard.edu/abs/1985ApJ...292..371D} {292, 371}

\bibitem[\protect\citeauthoryear{{Eisenstein} et~al.,}{{Eisenstein}
  et~al.}{2005}]{Eisenstein:2005}
{Eisenstein} D.~J.,  et~al., 2005, \mn@doi [\apj] {10.1086/466512}, \href
  {http://adsabs.harvard.edu/abs/2005ApJ...633..560E} {633, 560}

\bibitem[\protect\citeauthoryear{{Eisenstein} et~al.,}{{Eisenstein}
  et~al.}{2011}]{Eisenstein:2011}
{Eisenstein} D.~J.,  et~al., 2011, \mn@doi [\aj] {10.1088/0004-6256/142/3/72},
  \href {http://adsabs.harvard.edu/abs/2011AJ....142...72E} {142, 72}

\bibitem[\protect\citeauthoryear{{Hamilton}}{{Hamilton}}{1992}]{Hamilton:1992}
{Hamilton} A.~J.~S.,  1992, \mn@doi [\apjl] {10.1086/186264}, \href
  {http://adsabs.harvard.edu/abs/1992ApJ...385L...5H} {385, L5}

\bibitem[\protect\citeauthoryear{{Hamilton}}{{Hamilton}}{1998}]{Hamilton:1998}
{Hamilton} A.~J.~S.,  1998, in {D.~Hamilton} ed.,  Astrophysics and Space
  Science Library Vol. 231, The Evolving Universe. pp 185--+ (\mn@eprint {}
  {arXiv:astro-ph/9708102})

\bibitem[\protect\citeauthoryear{{Hikage} \& {Yamamoto}}{{Hikage} \&
  {Yamamoto}}{2013}]{Hikage:2013}
{Hikage} C.,  {Yamamoto} K.,  2013, \mn@doi [\jcap]
  {10.1088/1475-7516/2013/08/019}, \href
  {http://adsabs.harvard.edu/abs/2013JCAP...08..019H} {8, 19}

\bibitem[\protect\citeauthoryear{{Hikage}, {Takada}  \& {Spergel}}{{Hikage}
  et~al.}{2012}]{Hikage:2012}
{Hikage} C.,  {Takada} M.,   {Spergel} D.~N.,  2012, \mn@doi [\mnras]
  {10.1111/j.1365-2966.2011.19987.x}, \href
  {http://adsabs.harvard.edu/abs/2012MNRAS.419.3457H} {419, 3457}

\bibitem[\protect\citeauthoryear{{Hikage}, {Mandelbaum}, {Takada}  \&
  {Spergel}}{{Hikage} et~al.}{2013}]{Hikage:2013a}
{Hikage} C.,  {Mandelbaum} R.,  {Takada} M.,   {Spergel} D.~N.,  2013, \mn@doi
  [\mnras] {10.1093/mnras/stt1446}, \href
  {http://adsabs.harvard.edu/abs/2013MNRAS.435.2345H} {435, 2345}

\bibitem[\protect\citeauthoryear{{Hoshino} et~al.,}{{Hoshino}
  et~al.}{2015}]{Hoshino:2015}
{Hoshino} H.,  et~al., 2015, \mn@doi [\mnras] {10.1093/mnras/stv1271}, \href
  {http://adsabs.harvard.edu/abs/2015MNRAS.452..998H} {452, 998}

\bibitem[\protect\citeauthoryear{{Jackson}}{{Jackson}}{1972}]{Jackson:1972}
{Jackson} J.~C.,  1972, \mnras, \href
  {http://adsabs.harvard.edu/abs/1972MNRAS.156P...1J} {156, 1P}

\bibitem[\protect\citeauthoryear{{Jennings}, {Baugh}  \& {Pascoli}}{{Jennings}
  et~al.}{2011}]{Jennings:2011}
{Jennings} E.,  {Baugh} C.~M.,   {Pascoli} S.,  2011, \mn@doi [\mnras]
  {10.1111/j.1365-2966.2010.17581.x}, \href
  {http://adsabs.harvard.edu/abs/2011MNRAS.410.2081J} {410, 2081}

\bibitem[\protect\citeauthoryear{{Jing}}{{Jing}}{2005}]{Jing:2005}
{Jing} Y.~P.,  2005, \mn@doi [\apj] {10.1086/427087}, \href
  {http://adsabs.harvard.edu/abs/2005ApJ...620..559J} {620, 559}

\bibitem[\protect\citeauthoryear{{Kaiser}}{{Kaiser}}{1987}]{Kaiser:1987}
{Kaiser} N.,  1987, \mnras, \href
  {http://adsabs.harvard.edu/abs/1987MNRAS.227....1K} {227, 1}

\bibitem[\protect\citeauthoryear{{Komatsu} et~al.,}{{Komatsu}
  et~al.}{2009}]{Komatsu:2009}
{Komatsu} E.,  et~al., 2009, \mn@doi [\apjs] {10.1088/0067-0049/180/2/330},
  \href {http://adsabs.harvard.edu/abs/2009ApJS..180..330K} {180, 330}

\bibitem[\protect\citeauthoryear{{Lewis}, {Challinor}  \& {Lasenby}}{{Lewis}
  et~al.}{2000}]{Lewis:2000}
{Lewis} A.,  {Challinor} A.,   {Lasenby} A.,  2000, \mn@doi [\apj]
  {10.1086/309179}, \href {http://adsabs.harvard.edu/abs/2000ApJ...538..473L}
  {538, 473}

\bibitem[\protect\citeauthoryear{{Lin}, {Kirshner}, {Shectman}, {Landy},
  {Oemler}, {Tucker}  \& {Schechter}}{{Lin} et~al.}{1996}]{Lin:1996}
{Lin} H.,  {Kirshner} R.~P.,  {Shectman} S.~A.,  {Landy} S.~D.,  {Oemler} A.,
  {Tucker} D.~L.,   {Schechter} P.~L.,  1996, \mn@doi [\apj] {10.1086/177993},
  \href {http://adsabs.harvard.edu/abs/1996ApJ...471..617L} {471, 617}

\bibitem[\protect\citeauthoryear{{Masaki}, {Hikage}, {Takada}, {Spergel}  \&
  {Sugiyama}}{{Masaki} et~al.}{2013}]{Masaki:2013}
{Masaki} S.,  {Hikage} C.,  {Takada} M.,  {Spergel} D.~N.,   {Sugiyama} N.,
  2013, \mn@doi [\mnras] {10.1093/mnras/stt981}, \href
  {http://adsabs.harvard.edu/abs/2013MNRAS.433.3506M} {433, 3506}

\bibitem[\protect\citeauthoryear{{Matsubara}}{{Matsubara}}{2008}]{Matsubara:2008a}
{Matsubara} T.,  2008, \mn@doi [\prd] {10.1103/PhysRevD.78.083519}, \href
  {http://adsabs.harvard.edu/abs/2008PhRvD..78h3519M} {78, 083519}

\bibitem[\protect\citeauthoryear{{Miyazaki} et~al.,}{{Miyazaki}
  et~al.}{2012}]{Miyazaki:2012}
{Miyazaki} S.,  et~al., 2012, in Ground-based and Airborne Instrumentation for
  Astronomy IV. p. 84460Z, \mn@doi{10.1117/12.926844}

\bibitem[\protect\citeauthoryear{{Mohammed} \& {Seljak}}{{Mohammed} \&
  {Seljak}}{2014}]{Mohammed:2014}
{Mohammed} I.,  {Seljak} U.,  2014, \mn@doi [\mnras] {10.1093/mnras/stu1972},
  \href {http://adsabs.harvard.edu/abs/2014MNRAS.445.3382M} {445, 3382}

\bibitem[\protect\citeauthoryear{{More}, {Miyatake}, {Mandelbaum}, {Takada},
  {Spergel}, {Brownstein}  \& {Schneider}}{{More} et~al.}{2015}]{More:2015}
{More} S.,  {Miyatake} H.,  {Mandelbaum} R.,  {Takada} M.,  {Spergel} D.~N.,
  {Brownstein} J.~R.,   {Schneider} D.~P.,  2015, \mn@doi [\apj]
  {10.1088/0004-637X/806/1/2}, \href
  {http://adsabs.harvard.edu/abs/2015ApJ...806....2M} {806, 2}

\bibitem[\protect\citeauthoryear{{Nishimichi} \& {Taruya}}{{Nishimichi} \&
  {Taruya}}{2011}]{Nishimichi:2011}
{Nishimichi} T.,  {Taruya} A.,  2011, \mn@doi [\prd]
  {10.1103/PhysRevD.84.043526}, \href
  {http://adsabs.harvard.edu/abs/2011PhRvD..84d3526N} {84, 043526}

\bibitem[\protect\citeauthoryear{{Nishizawa}, {Takada}  \&
  {Nishimichi}}{{Nishizawa} et~al.}{2013}]{Nishizawa:2013}
{Nishizawa} A.~J.,  {Takada} M.,   {Nishimichi} T.,  2013, \mn@doi [\mnras]
  {10.1093/mnras/stt716}, \href
  {http://adsabs.harvard.edu/abs/2013MNRAS.433..209N} {433, 209}

\bibitem[\protect\citeauthoryear{{Oguri} \& {Takada}}{{Oguri} \&
  {Takada}}{2011}]{Oguri:2011}
{Oguri} M.,  {Takada} M.,  2011, \mn@doi [\prd] {10.1103/PhysRevD.83.023008},
  \href {http://adsabs.harvard.edu/abs/2011PhRvD..83b3008O} {83, 023008}

\bibitem[\protect\citeauthoryear{{Okumura} \& {Jing}}{{Okumura} \&
  {Jing}}{2011}]{Okumura:2011}
{Okumura} T.,  {Jing} Y.~P.,  2011, \mn@doi [\apj] {10.1088/0004-637X/726/1/5},
  \href {http://adsabs.harvard.edu/abs/2011ApJ...726....5O} {726, 5}

\bibitem[\protect\citeauthoryear{{Okumura}, {Seljak}, {McDonald}  \&
  {Desjacques}}{{Okumura} et~al.}{2012a}]{Okumura:2012}
{Okumura} T.,  {Seljak} U.,  {McDonald} P.,   {Desjacques} V.,  2012a, \mn@doi
  [\jcap] {10.1088/1475-7516/2012/02/010}, \href
  {http://adsabs.harvard.edu/abs/2012JCAP...02..010O} {2, 10}

\bibitem[\protect\citeauthoryear{{Okumura}, {Seljak}  \&
  {Desjacques}}{{Okumura} et~al.}{2012b}]{Okumura:2012b}
{Okumura} T.,  {Seljak} U.,   {Desjacques} V.,  2012b, \mn@doi [\jcap]
  {10.1088/1475-7516/2012/11/014}, \href
  {http://adsabs.harvard.edu/abs/2012JCAP...11..014O} {11, 14}

\bibitem[\protect\citeauthoryear{{Okumura}, {Hand}, {Seljak}, {Vlah}  \&
  {Desjacques}}{{Okumura} et~al.}{2015}]{Okumura:2015}
{Okumura} T.,  {Hand} N.,  {Seljak} U.,  {Vlah} Z.,   {Desjacques} V.,  2015,
  \mn@doi [\prd] {10.1103/PhysRevD.92.103516}, \href
  {http://adsabs.harvard.edu/abs/2015PhRvD..92j3516O} {92, 103516}

\bibitem[\protect\citeauthoryear{{Peacock} \& {Dodds}}{{Peacock} \&
  {Dodds}}{1994}]{Peacock:1994}
{Peacock} J.~A.,  {Dodds} S.~J.,  1994, \mnras, \href
  {http://adsabs.harvard.edu/abs/1994MNRAS.267.1020P} {267, 1020}

\bibitem[\protect\citeauthoryear{{Peebles}}{{Peebles}}{1980}]{Peebles:1980}
{Peebles} P.~J.~E.,  1980, {The large-scale structure of the universe}.
Princeton, N.J., Princeton Univ. Press

\bibitem[\protect\citeauthoryear{{Reid} \& {Spergel}}{{Reid} \&
  {Spergel}}{2009}]{Reid:2009}
{Reid} B.~A.,  {Spergel} D.~N.,  2009, \mn@doi [\apj]
  {10.1088/0004-637X/698/1/143}, \href
  {http://adsabs.harvard.edu/abs/2009ApJ...698..143R} {698, 143}

\bibitem[\protect\citeauthoryear{{Reid} \& {White}}{{Reid} \&
  {White}}{2011}]{Reid:2011}
{Reid} B.~A.,  {White} M.,  2011, \mn@doi [\mnras]
  {10.1111/j.1365-2966.2011.19379.x}, \href
  {http://adsabs.harvard.edu/abs/2011MNRAS.417.1913R} {417, 1913}

\bibitem[\protect\citeauthoryear{{Reid} et~al.,}{{Reid}
  et~al.}{2010}]{Reid:2010}
{Reid} B.~A.,  et~al., 2010, \mn@doi [\mnras]
  {10.1111/j.1365-2966.2010.16276.x}, \href
  {http://adsabs.harvard.edu/abs/2010MNRAS.404...60R} {404, 60}

\bibitem[\protect\citeauthoryear{{Reid}, {Seo}, {Leauthaud}, {Tinker}  \&
  {White}}{{Reid} et~al.}{2014}]{Reid:2014}
{Reid} B.~A.,  {Seo} H.-J.,  {Leauthaud} A.,  {Tinker} J.~L.,   {White} M.,
  2014, \mn@doi [\mnras] {10.1093/mnras/stu1391}, \href
  {http://adsabs.harvard.edu/abs/2014MNRAS.444..476R} {444, 476}

\bibitem[\protect\citeauthoryear{{Saito}, {Baldauf}, {Vlah}, {Seljak},
  {Okumura}  \& {McDonald}}{{Saito} et~al.}{2014}]{Saito:2014}
{Saito} S.,  {Baldauf} T.,  {Vlah} Z.,  {Seljak} U.,  {Okumura} T.,
  {McDonald} P.,  2014, \mn@doi [\prd] {10.1103/PhysRevD.90.123522}, \href
  {http://adsabs.harvard.edu/abs/2014PhRvD..90l3522S} {90, 123522}

\bibitem[\protect\citeauthoryear{{Saito} et~al.,}{{Saito}
  et~al.}{2016}]{Saito:2016}
{Saito} S.,  et~al., 2016, \mn@doi [\mnras] {10.1093/mnras/stw1080}, \href
  {http://adsabs.harvard.edu/abs/2016MNRAS.460.1457S} {460, 1457}

\bibitem[\protect\citeauthoryear{{Sato} \& {Matsubara}}{{Sato} \&
  {Matsubara}}{2011}]{Sato:2011}
{Sato} M.,  {Matsubara} T.,  2011, \mn@doi [\prd] {10.1103/PhysRevD.84.043501},
  \href {http://adsabs.harvard.edu/abs/2011PhRvD..84d3501S} {84, 043501}

\bibitem[\protect\citeauthoryear{{Scoccimarro}}{{Scoccimarro}}{2004}]{Scoccimarro:2004}
{Scoccimarro} R.,  2004, \mn@doi [\prd] {10.1103/PhysRevD.70.083007}, \href
  {http://adsabs.harvard.edu/abs/2004PhRvD..70h3007S} {70, 083007}

\bibitem[\protect\citeauthoryear{{Seljak}}{{Seljak}}{2000}]{Seljak:2000}
{Seljak} U.,  2000, \mn@doi [\mnras] {10.1046/j.1365-8711.2000.03715.x}, \href
  {http://adsabs.harvard.edu/abs/2000MNRAS.318..203S} {318, 203}

\bibitem[\protect\citeauthoryear{{Seljak}}{{Seljak}}{2001}]{Seljak:2001}
{Seljak} U.,  2001, \mn@doi [\mnras] {10.1046/j.1365-8711.2001.04508.x}, \href
  {http://adsabs.harvard.edu/abs/2001MNRAS.325.1359S} {325, 1359}

\bibitem[\protect\citeauthoryear{{Seljak}}{{Seljak}}{2012}]{Seljak:2012}
{Seljak} U.,  2012, \mn@doi [\jcap] {10.1088/1475-7516/2012/03/004}, \href
  {http://adsabs.harvard.edu/abs/2012JCAP...03..004S} {3, 004}

\bibitem[\protect\citeauthoryear{{Seljak} \& {Vlah}}{{Seljak} \&
  {Vlah}}{2015}]{Seljak:2015}
{Seljak} U.,  {Vlah} Z.,  2015, \mn@doi [\prd] {10.1103/PhysRevD.91.123516},
  \href {http://adsabs.harvard.edu/abs/2015PhRvD..91l3516S} {91, 123516}

\bibitem[\protect\citeauthoryear{{Seljak} \& {Zaldarriaga}}{{Seljak} \&
  {Zaldarriaga}}{1996}]{Seljak:1996}
{Seljak} U.,  {Zaldarriaga} M.,  1996, \mn@doi [\apj] {10.1086/177793}, \href
  {http://adsabs.harvard.edu/abs/1996ApJ...469..437S} {469, 437}

\bibitem[\protect\citeauthoryear{{Simpson}, {Heavens}  \& {Heymans}}{{Simpson}
  et~al.}{2013}]{Simpson:2013}
{Simpson} F.,  {Heavens} A.~F.,   {Heymans} C.,  2013, \mn@doi [\prd]
  {10.1103/PhysRevD.88.083510}, \href
  {http://adsabs.harvard.edu/abs/2013PhRvD..88h3510S} {88, 083510}

\bibitem[\protect\citeauthoryear{{Skibba}, {van den Bosch}, {Yang}, {More},
  {Mo}  \& {Fontanot}}{{Skibba} et~al.}{2011}]{Skibba:2011}
{Skibba} R.~A.,  {van den Bosch} F.~C.,  {Yang} X.,  {More} S.,  {Mo} H.,
  {Fontanot} F.,  2011, \mn@doi [\mnras] {10.1111/j.1365-2966.2010.17452.x},
  \href {http://adsabs.harvard.edu/abs/2011MNRAS.410..417S} {410, 417}

\bibitem[\protect\citeauthoryear{{Smith}, {Scoccimarro}  \& {Sheth}}{{Smith}
  et~al.}{2007}]{Smith:2007}
{Smith} R.~E.,  {Scoccimarro} R.,   {Sheth} R.~K.,  2007, \mn@doi [\prd]
  {10.1103/PhysRevD.75.063512}, \href
  {http://adsabs.harvard.edu/abs/2007PhRvD..75f3512S} {75, 063512}

\bibitem[\protect\citeauthoryear{{Springel}, {White}, {Tormen}  \&
  {Kauffmann}}{{Springel} et~al.}{2001}]{Springel:2001}
{Springel} V.,  {White} S.~D.~M.,  {Tormen} G.,   {Kauffmann} G.,  2001,
  \mn@doi [\mnras] {10.1046/j.1365-8711.2001.04912.x}, \href
  {http://adsabs.harvard.edu/abs/2001MNRAS.328..726S} {328, 726}

\bibitem[\protect\citeauthoryear{{Takada} et~al.,}{{Takada}
  et~al.}{2014}]{Takada:2014}
{Takada} M.,  et~al., 2014, \mn@doi [\pasj] {10.1093/pasj/pst019}, \href
  {http://adsabs.harvard.edu/abs/2014PASJ...66R...1T} {66, 1}

\bibitem[\protect\citeauthoryear{{Taruya}, {Nishimichi}  \& {Saito}}{{Taruya}
  et~al.}{2010}]{Taruya:2010}
{Taruya} A.,  {Nishimichi} T.,   {Saito} S.,  2010, \mn@doi [\prd]
  {10.1103/PhysRevD.82.063522}, \href
  {http://adsabs.harvard.edu/abs/2010PhRvD..82f3522T} {82, 063522}

\bibitem[\protect\citeauthoryear{{Tegmark} et~al.,}{{Tegmark}
  et~al.}{2006}]{Tegmark:2006}
{Tegmark} M.,  et~al., 2006, \mn@doi [\prd] {10.1103/PhysRevD.74.123507}, \href
  {http://adsabs.harvard.edu/abs/2006PhRvD..74l3507T} {74, 123507}

\bibitem[\protect\citeauthoryear{{Tinker}, {Weinberg}  \& {Zheng}}{{Tinker}
  et~al.}{2006}]{Tinker:2006}
{Tinker} J.~L.,  {Weinberg} D.~H.,   {Zheng} Z.,  2006, \mn@doi [\mnras]
  {10.1111/j.1365-2966.2006.10114.x}, \href
  {http://adsabs.harvard.edu/abs/2006MNRAS.368...85T} {368, 85}

\bibitem[\protect\citeauthoryear{{Vlah}, {Seljak}, {Okumura}  \&
  {Desjacques}}{{Vlah} et~al.}{2013}]{Vlah:2013}
{Vlah} Z.,  {Seljak} U.,  {Okumura} T.,   {Desjacques} V.,  2013, \mn@doi
  [\jcap] {10.1088/1475-7516/2013/10/053}, \href
  {http://adsabs.harvard.edu/abs/2013JCAP...10..053V} {10, 53}

\bibitem[\protect\citeauthoryear{{Weinberg}, {Mortonson}, {Eisenstein},
  {Hirata}, {Riess}  \& {Rozo}}{{Weinberg} et~al.}{2013}]{Weinberg:2013}
{Weinberg} D.~H.,  {Mortonson} M.~J.,  {Eisenstein} D.~J.,  {Hirata} C.,
  {Riess} A.~G.,   {Rozo} E.,  2013, \mn@doi [\physrep]
  {10.1016/j.physrep.2013.05.001}, \href
  {http://adsabs.harvard.edu/abs/2013PhR...530...87W} {530, 87}

\bibitem[\protect\citeauthoryear{{White}}{{White}}{2001}]{White:2001}
{White} M.,  2001, \mn@doi [\mnras] {10.1046/j.1365-8711.2001.03956.x}, \href
  {http://adsabs.harvard.edu/abs/2001MNRAS.321....1W} {321, 1}

\bibitem[\protect\citeauthoryear{{White} et~al.,}{{White}
  et~al.}{2011}]{White:2011}
{White} M.,  et~al., 2011, \mn@doi [\apj] {10.1088/0004-637X/728/2/126}, \href
  {http://adsabs.harvard.edu/abs/2011ApJ...728..126W} {728, 126}

\bibitem[\protect\citeauthoryear{{Zheng} \& {Song}}{{Zheng} \&
  {Song}}{2016}]{Zheng:2016}
{Zheng} Y.,  {Song} Y.-S.,  2016, \mn@doi [\jcap]
  {10.1088/1475-7516/2016/08/050}, \href
  {http://adsabs.harvard.edu/abs/2016JCAP...08..050Z} {8, 050}

\bibitem[\protect\citeauthoryear{{Zheng} et~al.,}{{Zheng}
  et~al.}{2005}]{Zheng:2005}
{Zheng} Z.,  et~al., 2005, \mn@doi [\apj] {10.1086/466510}, \href
  {http://adsabs.harvard.edu/abs/2005ApJ...633..791Z} {633, 791}

\bibitem[\protect\citeauthoryear{{van den Bosch}, {More}, {Cacciato}, {Mo}  \&
  {Yang}}{{van den Bosch} et~al.}{2013}]{van-den-Bosch:2013}
{van den Bosch} F.~C.,  {More} S.,  {Cacciato} M.,  {Mo} H.,   {Yang} X.,
  2013, \mn@doi [\mnras] {10.1093/mnras/sts006}, \href
  {http://adsabs.harvard.edu/abs/2013MNRAS.430..725V} {430, 725}

\makeatother
\end{thebibliography}

\label{lastpage}

\end{document}